\documentclass[onecolumn,authoryear]{els-mrw} 

\usepackage{amsmath,amssymb,amsfonts,amsthm,makeidx,graphicx}
\usepackage{txfonts}
\usepackage{helvet}
\usepackage{siunitx} 

\usepackage
{hyperref}
\newcommand{\doi}[1]{\href{https://doi.org/#1}{doi:\nolinkurl{#1}}}
\newcommand{\eprint}[2][arXiv]{\href{https://arxiv.org/abs/#2}{#1:\nolinkurl{#2}}}

\newcommand{\1}
{\large }

\usepackage[dvipsnames]{xcolor}
\definecolor{mydarkgreen}{RGB}{0,100,0}        

\newcommand{\be}{\begin{equation}}
\newcommand{\ee}{\end{equation}}
\newcommand{\bea}{\begin{eqnarray}}
\newcommand{\eea}{\end{eqnarray}}

\newcommand{\6}{\partial }
\newcommand{\J}{\mathcal{J}}
\newcommand{\WSS}{WSS } 

\newcommand{\FYM}{F_{\rm YM}}
\newcommand{\Nc}{N_{c}}

\newcommand{\SU}{\mathrm{SU}}
\newcommand{\U}{\mathrm{U}}
\newcommand{\SO}{\mathrm{SO}}
\newcommand{\gym}{g_{\mathrm{YM}}}
\newcommand{\ukk}{u_{\mathrm{KK}}}
\newcommand{\mkk}{M_{\mathrm{KK}}}
\newcommand{\MKK}{M_{\mathrm{KK}}}
\newcommand{\AdS}{\mathrm{AdS}}
\newcommand{\Dbar}{$\overline{\mathrm{D8}}$}
\newcommand{\AD}{$\overline{\mathrm{D8}}$}
\newcommand{\ra}{\rightarrow}
\newcommand{\cN}{{\cal N}}
\newcommand{\cO}{{\cal O}}

\begin{document}


\chapter{Holographic Approaches to Hadron Structure}
\label{chap:holographicapproaches}


\author[1,2]{Luigi Cappiello}%
\author[3]{Anton Rebhan}%
\author[4,5,6]{Shigeki Sugimoto}%

\address[1]{\orgname{Universit\`a di Napoli  Federico II}, \orgdiv{Dipartimento di Fisica \lq\lq E. Pancini\rq\rq}, \orgaddress{via Cintia, I-80126 Naples, Italy}}
\address[2]{\orgname{INFN}, \orgdiv{Sezione di Napoli}, \orgaddress{C.U. di Monte Sant'Angelo, via Cintia Edificio 6, I-80126 Naples, Italy}}
\address[3]{\orgname{Technische Universit\"at Wien}, \orgdiv{Institute for Theoretical Physics}, \orgaddress{Wiedner Hauptstr. 8-10, A-1040 Vienna, Austria}}
\address[4]{\orgname{Kyoto University}, \orgdiv{Department of Physics}, \orgaddress{Kyoto 606-8502, Japan}}
\address[5]{\orgname{Kyoto University}, \orgdiv{Center for Gravitational Physics and Quantum Information, YITP}, \orgaddress{Kyoto 606-8502, Japan}}
\address[6]{\orgname{The University of Tokyo}, \orgdiv{Kavli IPMU (WPI), UTIAS, Kashiwa, Chiba 277-8583, Japan}} 


\maketitle

\begin{abstract}[Abstract]
We give an introduction to and an overview of holographic approaches to QCD and hadron structure. 
After an introduction to the string-theoretic foundations of holographic QCD, we review the predictions of top-down and bottom-up models
for the phenomenology of mesons, baryons, and glueballs.
We focus on the top-down model following from probe flavor branes
in the Witten model of low-energy Yang-Mills theory and, in the
more phenomenological approach of bottom-up models, mostly on minimal
models based on five-dimensional anti-de Sitter space,
which also implement global symmetries and their anomalies consistently, but allow the
high-energy behavior to better match real QCD.
\end{abstract}

\begin{keywords}
 	gauge-gravity duality \sep holography \sep QCD \sep hadrons \sep mesons \sep baryons \sep glueballs
\end{keywords}

\begin{glossary}[Nomenclature]
	\begin{tabular}{@{}lp{34pc}@{}}
        AdS & anti-de Sitter\\
        CFT & conformal field theory\\
        CS & Chern-Simons\\
        DBI & Dirac-Born-Infeld\\
        GOR & Gell-Mann-Oakes-Renner\\
        HLbL & hadronic light-by-light scattering\\
        hQCD & holographic QCD\\
        HVP & hadronic vacuum polarization\\
        HW & hard wall\\
        IR & infrared\\
        KK & Kaluza-Klein\\
        NS & Neveu-Schwarz\\
        $N_c$ & number of colors\\
        $N_f$ & number of flavors\\
        OPE & operator product expansion\\
        pQCD & perturbative QCD\\
        QCD & quantum chromodynamics\\
        RR & Ramond-Ramond\\
        SW & soft wall\\
        SYM & supersymmetric Yang-Mills theory\\
        UV & ultraviolet\\
		VMD & vector meson dominance\\
        (W)SS & (Witten-)Sakai-Sugimoto (D4-D8 model)\\
        WZW & Wess-Zumino-Witten\\
        YM & Yang-Mills\\
	\end{tabular}
\end{glossary}

\newpage

\section*{Key points}
\begin{itemize}
\item Holographic QCD provides a complementary approach to nonperturbative hadron physics, connecting hadronic observables to a higher-dimensional description of strongly coupled gauge theories, specifically in the limit of large color number.
\item Confinement and chiral symmetry breaking are given a geometric interpretation, particularly in top-down models derived from string theory and D-brane constructions, where global symmetries and their anomalies are consistently encoded.
\item Bottom-up models keep essential features of the top-down models with a string-theoretic foundation, but allow greater flexibility to adjust the high-energy behavior to become closer to real QCD.
\item The infinite towers of hadronic resonances of holographic QCD can reconcile hadronic descriptions with QCD short-distance constraints where conventional hadronic models fail, offering important advantages, e.g., for transition form factors and hadronic light-by-light scattering.
\item Holographic approaches provide not only qualitative but often quantitative predictions which can be used to assess their virtues and limitations with experimental data or lattice QCD; to a corresponding extent, they promise new insights in difficult areas such as glueball interactions, gravitational form factors, or strongly coupled QCD matter at finite baryon density.
\end{itemize}

\1

\section{Introduction}
\label{sec:Introduction}

Quantum chromodynamics (QCD) provides the fundamental theory of the strong interaction, yet understanding how its quark and gluon degrees of freedom give rise to the spectrum, interactions, and internal structure of hadrons remains the central challenge in nuclear and particle physics. At low energies, the strong coupling prevents a straightforward perturbative treatment, because the characteristic phenomena of hadron physics such as confinement and spontaneous chiral symmetry breaking are intrinsically nonperturbative. Developing tractable descriptions of these phenomena, while maintaining a connection to the underlying theory, is therefore essential for understanding the observable properties of hadrons.

Several complementary approaches have been developed to meet this challenge. Phenomenological hadro\-nic models provide powerful tools for describing low-energy processes in terms of mesons, baryons, and their interactions. Lattice QCD, on the other hand, offers a first-principles numerical approach to nonperturbative strong-interaction physics and has become a powerful source of quantitative predictions. Nevertheless, complementary analytical frameworks remain valuable, especially when one seeks to understand how hadronic resonances, effective interactions, symmetry principles, and short-distance constraints fit together within a common description.

Holographic QCD (hQCD), a spin-off of the famous and well-established AdS/CFT correspondence \citep{Aharony:1999ti}, provides such a complementary perspective. Relying on a largely hypothetical general gauge/gra\-vity duality, it represents the dynamics of a strongly coupled four-dimensional quantum gauge theory through a higher-dimensional theory of classical fields with gravity. The additional dimension represents the energy scale and provides a framework for organizing hadronic excitations and their interactions; a holographic dictionary relates fields in the higher-dimensional description to operators and observables in the gauge theory living on the conformal boundary. Through the extra dimension, infinite towers of hadronic resonances arise naturally, allowing certain aspects of the short-distance behavior of QCD to be incorporated alongside the phenomenology of individual hadrons. Holographic QCD is thus useful not only as a model for particular observables, but also as a framework in which to explore how confinement, chiral symmetry breaking, resonance dynamics, and hadron structure may be related.

Two broad strategies have emerged. The \emph{top-down approach} starts from a string-theoretic construction and derives its holographic description, offering a geometric interpretation of certain features of low-energy QCD. A prominent example is the Witten–Sakai–Sugimoto (WSS) \citep{Witten:1998zw,Sakai:2004cn,Sakai:2005yt} model, in which the flavor-brane configuration realizes chiral symmetry and its spontaneous breaking, while mesons, baryons, and glueballs admit natural interpretations in terms of string and brane excitations. On the other hand, the \emph{bottom-up approach}, while inspired by the lessons learned from the top-down approach, begins with a five-dimensional effective action that is more freely designed to reproduce relevant properties of QCD.
The simplest models (to which this review will limit itself mostly)
employ pure 5-dimensional anti-de Sitter (AdS) space,
which provides the arena for the original AdS/CFT duality of Maldacena 
involving 4-dimensional maximally supersymmetric Yang-Mills (SYM) theory, and introduce
confinement by either a hard-wall \citep{Erlich:2005qh,DaRold:2005mxj,Hirn:2005nr} or soft-wall cutoff \citep{Karch:2006pv} (the latter by
adding a suitable dilaton field).
Such models, and in particular their improved extensions, allow different aspects of the hadronic spectrum, chiral dynamics, and short-distance behavior to be investigated with greater phenomenological flexibility. These approaches embody a useful tension between theoretical structure and quantitative adaptability: the top-down approach seeks to derive hadronic dynamics from a fundamental construction, whereas the bottom-up approach allows the framework to be tailored more directly to the physics one wishes to describe.

The complementarity of holographic QCD is particularly evident in observables that are sensitive to both resonance dynamics and QCD short-distance constraints. Hadronic light-by-light scattering, for example, contributes to the anomalous magnetic moment of the muon and probes the transition form factors of pseudoscalar, axial-vector, and tensor mesons. In this setting, holographic models can incorporate the effects of entire resonance towers and, in suitable bottom-up models, reproduce short-distance behavior that cannot generally be captured by conventional vector-meson-dominance (VMD) models containing only finitely many resonances. Moreover, the holographic framework provides a route to calculating hadronic form factors and studying the internal distributions of mass, angular momentum, pressure, and shear encoded in gravitational form factors. Comparison with experimental measurements, dispersive analyses, and lattice calculations then provides a means of testing the physical content of the models.

The limitations of this framework should nevertheless be kept in mind. Existing holographic constructions are not exact dual descriptions of real-world QCD---at best only in certain regimes (large coupling, large $N_c$, low energy). In particular, top-down models typically retain additional degrees of freedom absent from QCD, while bottom-up models necessarily introduce vastly simplifying assumptions about the effective action and background geometry. Their predictions must therefore be evaluated in the light of the approximations employed and the observables under consideration. The aim is not to replace established approaches, but to develop and test a complementary description that can connect qualitative physical mechanisms with quantitative hadron phenomenology.

This review introduces the foundations of holographic QCD and surveys its applications to hadron structure. We first present the string-theoretic foundations of holographic duality (where the reader is invited to
just glance over the technical details imported from string theory) and display the construction of the top-down and bottom-up models. We then examine mesons, including their spectra, interactions, and chiral properties, before turning to baryons and glueballs. Finally, we review further developments in finite-temperature and finite-density physics, higher-spin mesons, hadronic light-by-light scattering, and gravitational form factors. Taken together, these applications illustrate both the explanatory appeal and the present limitations of holographic QCD, and show how comparisons with experiment and other nonperturbative methods can sharpen our understanding of the strong interaction.

\section{Holographic dualities}

\subsection{Brief introduction to superstring theory and D-branes}
\label{briefintro}
String theory is a theoretical framework in which elementary particles are described as tiny strings.
It originally emerged as a theory of hadrons around 1970. However, it was not successful in that direction at that time. One of the reasons is that the spectrum of string theory did not agree with that of hadrons observed in experiments. For example, string theory predicts massless spin-2 or spin-1 states, while hadrons of spin-2 or spin-1 are all heavy. In the mid-1970s, these massless spin-2 and spin-1 states were interpreted as gravitons and gauge particles, respectively, and string theory evolved into a candidate for a unified theory of all the elementary forces of nature, including gravity. This interpretation of string theory was further developed in the 1980s along with many important theoretical discoveries. This period is called the first string revolution. 
In the mid-1990s, the second string revolution began. It was triggered by Polchinski’s rediscovery of D-branes \citep{Polchinski:1995mt}. D-branes are a kind of solitonic objects in string theory, and they play an important role in understanding the non-perturbative properties of the theory. They also led to the discovery of holographic duality, which will be reviewed in the next subsection. Holographic QCD was born as an application of the idea of holography to QCD. As we shall see in the following sections, it suggests that string theory can indeed serve as a theory of hadrons as was originally envisioned.

Since D-branes are one of the key ingredients of this section, we now elaborate on them in a little more detail.
The types of string theory considered in this section are Type IIA and Type IIB superstring theories. Both of them are defined in ten-dimensional spacetime and have $\cN=2$ spacetime supersymmetry. The massless bosonic fields in these theories are listed in Table \ref{tab:typeII};
\begin{table}[h]
\centering
    \begin{tabular}{|c|c|c|}
    \hline
    & NSNS sector    & RR sector \\
    \hline
    Type IIA & $g_{mn}$, $B_2$, $\phi$& $C_1$, $C_3$\\
    Type IIB &$g_{mn}$, $B_2$, $\phi$& $C_0$, $C_2$, $C_4$\\
    \hline
    \end{tabular}
    \caption{Massless bosonic fields in Type IIA and Type IIB superstring theories.}
    \label{tab:typeII}
\end{table}
$g_{mn}$ ($m,n=0,1,\cdots,9$) is the metric of the ten-dimensional spacetime corresponding to the graviton, $B_2$ is a 2-form field called the Kalb-Ramond field or simply $B$-field, $\phi$ is a scalar field called the dilaton field, and $C_n$ are $n$-form fields called Ramond-Ramond (RR) fields.\footnote{$C_4$ is subject to the self-duality condition $dC_4=*dC_4$, where $*$ is the Hodge dual operation.} 

A D$p$-brane is defined as a $(p+1)$-dimensional object on which open strings can end. It has been shown that there exist D$p$-branes with even and odd $p$ for Type IIA and Type IIB superstring theory, respectively, which are charged under the RR-fields. For example, consider D$p$-branes with $p=3,4,\cdots,7$ extended along $x^{0,1,\cdots,p}$ directions in the flat ten-dimensional Minkowski spacetime. The transverse space is a $(9-p)$-dimensional plane parametrized by $x^{p+1,p+2,\cdots,9}$. Then, integrating the field strength of the RR $(7-p)$-form field over an $(8-p)$-dimensional sphere $S^{8-p}$ in the transverse space, we get
\be
\frac{1}{(2\pi\ell_s)^{7-p}}\int_{S^{8-p}}dC_{7-p}=N\ ,
\label{RRcharge}
\ee
where $N$ is the number of the D$p$-branes placed inside the sphere.\footnote{This relation \eqref{RRcharge} can be extended to the cases with $p\le 2$ by introducing $C_n$ with $n\ge 5$ satisfying the relation $dC_n\propto *dC_{8-n}$ (for $B_2=0$).}

One of the important properties of D-branes is that the spectrum of an open string ending on the D-brane contains a massless gauge particle. When $N$ D$p$-branes are stacked on top of each other, the gauge field arising from open strings stretched between the $i$-th and $j$-th D-branes ($i,j=1,2,\cdots,N$) carries two indices $i$ and $j$ associated with the D-branes. They form an $N\times N$ matrix corresponding to a $\U(N)$ gauge field. Therefore, a $(p+1)$-dimensional $\U(N)$ gauge theory is realized on the D$p$-brane world-volume. In this way, it is possible to construct non-Abelian gauge theories in string theory quite easily using D-branes. Owing to this property, there has been a fruitful interplay between string theory and gauge theory, which was one of the main driving forces behind the second string revolution.



\subsection{AdS/CFT correspondence}
\label{AdS/CFT}

In 1997, 
\citet{Maldacena:1997re} made a striking conjecture that four-dimensional ${\cal N}=4$ supersymmetric Yang-Mills (SYM) theory is equivalent to type IIB superstring theory on AdS$_5\times S^5$.  This is the first and the most concrete example of the so-called AdS/CFT correspondence, which asserts an equivalence between a $d$-dimensional conformal field theory (CFT) and a gravitational theory defined on a spacetime that is the product of $(d+1)$-dimensional AdS space and an internal manifold. It was soon generalized to non-supersymmetric and non-conformal systems. One of the surprising features of this duality is that the spacetime dimensions of the two descriptions are different. By analogy with optical holography, in which a three-dimensional image is encoded on a two-dimensional surface, this type of duality is called a holographic duality. It is also referred to as gauge/string duality or gauge/gravity duality, as it relates a gauge theory to a string theory, or more generally, to a theory of quantum gravity.

To explain the basic idea, let us consider $N_c$ D3-branes in type IIB superstring theory. As mentioned in section \ref{briefintro}, we obtain a four-dimensional $\U(N_c)$ gauge theory on the world-volume.
In fact, the massless fields on the D3-branes turn out to be the same as those of the ${\cal N}=4$ SYM: a gauge field, 6 scalar fields and 4 Weyl fermions in the adjoint representation of the gauge group. In addition to the massless modes, there are infinitely many massive modes in the spectrum. Their masses are proportional to $1/\ell_s$, where $\ell_s$ is a parameter called string length. In order to get precisely ${\cal N}=4$ SYM, we take a decoupling limit $\ell_s\rightarrow 0$. The couplings to closed strings, including the gravitational interaction, as well as other $1/\ell_s$ suppressed interactions, vanish in this limit. 

On the other hand, the D3-branes (before taking the $\ell_s\ra 0$ limit) are described in supergravity as a curved background corresponding to the D3-branes, which is obtained as a solution of the equations of motion of type IIB supergravity
sourced by the $N_c$ D3-branes. The metric\footnote{In this review, we use the mostly minus convention for the metric.} for the solution corresponding to $N_c$ D3-branes extended along $x^{0,1,2,3}$ directions and localized at $x^4=x^5=\cdots=x^9=0$ is
\be
ds^2=H(u)^{-1/2}\eta_{\mu\nu}dx^\mu dx^\nu-H(u)^{1/2}\left(du^2+u^2 d\Omega_5^2\right)
\label{metric1}
\ee
with
\be
H(u)=1+\frac{4\pi g_s N_c\ell_s^4}{u^4}\ ,
\ee
where $u$ is the radial coordinate in the $x^{4,5,\cdots,9}$-plane,
$\eta_{\mu\nu}$ ($\mu,\nu=0,1,2,3$) is the four-dimensional Minkowski metric and $d\Omega_5^2$ is the line element of the unit $S^5$. 
The strength of interactions in string theory is determined by the string coupling $g_s$, and the
gauge coupling $\gym$ of the gauge theory on the D3-brane is related to $g_s$ by $\gym^2=2\pi g_s$. Since the D3-branes couple to the RR four-form field $C_4$, the background has a non-trivial RR flux satisfying
\be
\frac{1}{(2\pi\ell_s)^4}\int_{S^5}dC_4=N_c\ .
\ee

As explained above, we should take a decoupling limit $\ell_s\ra 0$ to describe ${\cal N}=4$ SYM. To this end, we focus on the region $u\ra0$, while keeping $u/\ell_s^2$ finite. To see this, consider a configuration with $(N_c-1)$ D3-branes at $u=0$ and one D3-brane at $u=u_1$. This corresponds to the case that $\U(N_c)$ gauge symmetry is spontaneously broken to $\U(N_c-1)\times \U(1)$ by giving a non-trivial vacuum expectation value to a scalar field in ${\cal N}=4$ SYM. A W-boson corresponds to an open string stretched between $u=0$ and $u=u_1$. Then, the mass scale of the W-boson is $u_1/\ell_s^2$, which should be kept finite in the limit $\ell_s\ra 0$. It is then convenient to introduce a new coordinate $z:=(2\lambda)^{1/2}\ell_s^2/u$, where $\lambda=\gym^2N_c$ is the 't Hooft coupling. In the $\ell_s\ra 0$ limit, the line element \eqref{metric1} becomes
\be
ds^2=\frac{\ell_s^2}{\tilde\ell_s^2} d\tilde s^2\ ,~~~
d\tilde s^2=-L^2\left(z^{-2}(-\eta_{\mu\nu}dx^\mu dx^\nu+dz^2)+d\Omega_5^2\right)\ ,
\label{metric2}
\ee
where $\tilde\ell_s:=L/(2\lambda)^{1/4}$ and $L$ is an arbitrary parameter with the dimension of length. The line element $d\tilde s^2$ is that of the AdS$_5\times S^5$ with radius $L$.
The string world-sheet action (Nambu-Goto action) in this background is
\be
S_{\rm NG}=T_s\int d^2\sigma\sqrt{-\det g_{ij}}
=\tilde T_s\int d^2\sigma\sqrt{-\det \tilde g_{ij}}\ ,
\label{S_NG}
\ee
where $g_{ij}$ and $\tilde g_{ij}$ are the induced metric on the string world-sheet defined from the background metric given by $ds^2$ and $d\tilde s^2$ in \eqref{metric2}, respectively, and
\be
T_s:=\frac{1}{2\pi\ell_s^2}\ ,~~~\tilde T_s:=\frac{1}{2\pi\tilde\ell_s^2}
\ee
are the string tensions.
It is important to note that $\ell_s$ has disappeared in the last expression of the action \eqref{S_NG} and we can safely take the $\ell_s\ra0$ limit. This is an action for a string with tension given by $\tilde T_s$ embedded in AdS$_5\times S^5$ spacetime with radius $L$. Although the original metric given by $ds^2$ is singular in the $\ell_s\ra0$ limit, the rescaled metric given by $d\tilde s^2$ and the string tension $\tilde T_s$ remain finite. In this way, we find that ${\cal N}=4$ SYM corresponds to string theory in AdS$_5\times S^5$.
As an immediate consistency check, one can see that the conformal symmetry and R-symmetry, $\SO(2,4)\times\SO(6)$, of ${\cal N}=4$ SYM match the isometries of AdS$_5\times S^5$.

The correspondence between parameters in string theory and ${\cal N}=4$ SYM, up to numerical factors, is summarized as
\be
g_s\longleftrightarrow\frac{\lambda}{N_c}\ ,~~~~~\frac{\tilde\ell_s}{L}\longleftrightarrow\frac{1}{\lambda^{1/4}}\ .
\ee
From this we see that the perturbative expansion (expansion in powers of $g_s$) and the $\alpha'$ expansion (expansion with respect to $\tilde\ell_s/L$)
in string theory correspond to 't Hooft's $1/N_c$ expansion and strong coupling expansion (expansion in negative powers of $\lambda$) in ${\cal N}=4$ SYM, respectively. The supergravity approximation is reliable when $g_s\ll 1$ and $\tilde\ell_s/L\ll 1$, which corresponds to the regime $1\ll\lambda\ll N_c$. Therefore, using the AdS/CFT correspondence, large $N_c$, strongly coupled ${\cal N}=4$ SYM (at leading order in the $\lambda/N_c$ and $1/\lambda$ expansions) can be analyzed by classical supergravity in a weakly curved background. Quantum gravity effects are encoded in the $1/N_c$ corrections of ${\cal N}=4$ SYM.

The above argument provides an intuitive understanding of the equivalence between ${\cal N}=4$ SYM and type IIB superstring theory on AdS$_5\times S^5$.
However, it is not regarded as a proof, because the two descriptions are reliable in different regions of parameter space. For this reason, extensive efforts have been made to test this duality, yielding a vast amount of nontrivial supporting evidence, and it is now widely believed to hold. Note that the above argument does not rely on supersymmetry or conformal symmetry, though the original proposal of \citet{Maldacena:1997re} was made for superconformal field theories. 
It is then natural to ask whether the holographic duality holds for more realistic, non-supersymmetric field theories, which will be discussed in the following sections.

A more explicit correspondence between $d$-dimensional CFT and string theory in AdS$_{d+1}$ space was given by Gubser, Klebanov and Polyakov \citep{Gubser:1998bc}, and 
\citet{Witten:1998qj}. It was argued that there is a one-to-one correspondence between a gauge-invariant local operator $\cO(x^\mu)$ in CFT and a field $\phi(x^\mu,z)$ in AdS space, and the correlation functions of the operator $\cO(x^\mu)$ in CFT can be calculated by a formula called GKP-Witten relation:
\be
\left\langle e^{\int d^d\! x\,\phi_0(x^\mu)\cO(x^\mu)}\right\rangle_{\rm CFT}=
Z_{\rm string}{[}\phi {]}\ .
\label{GKPW}
\ee
The left-hand side is the generating function of the correlation functions in CFT, and the right-hand side is the partition function of string theory in AdS space.
The source field $\phi_0(x^\mu)$ for the operator $\cO(x^\mu)$ in the left hand side is related to the field $\phi$ in the right hand side by the boundary condition:
\be
\phi(x^\mu,\epsilon)=\epsilon^{d-\Delta}\phi_0(x^\mu)\ ,
\label{bc}
\ee
where $\Delta$ is the scaling dimension of the operator $\cO$;
$\epsilon$ is a cut-off parameter, which will eventually be taken to zero. The boundary condition \eqref{bc} has to be consistent with the equations of motion, which imply a relation between the scaling dimension $\Delta$ and the mass $m$ as
\begin{align}
\Delta=
\frac{d}{2}+\sqrt{\frac{d^2}{4}+L^2 m^2}
\end{align}
for a scalar field of mass $m$.
In general, evaluating the string partition function on the right-hand side of \eqref{GKPW} is very difficult. However, when the supergravity approximation is valid, it can be approximated by
\be
Z_{\rm string}[\phi]\simeq
e^{-S_{\rm SUGRA}[\phi_{\rm cl}]}\ ,
\ee
where $S_{\rm SUGRA}$ is the supergravity action and $\phi_{\rm cl}$ is a classical solution of the equations of motion satisfying the boundary condition \eqref{bc}. This expression enables us to compute correlation functions of CFT using the classical supergravity action.

There is one subtle point that might be worth mentioning here. As pointed out by \citet{Witten:1998qj}, the gauge group of $\cN=4$ SYM dual to Type IIB superstring theory on AdS$_5\times S^5$ is $\SU(N_c)$, rather than $\U(N_c)$, because the operators associated with the $\U(1)$ factor of $\U(N_c)$ do not have counterparts in the supergravity analysis.\footnote{The $\U(1)$ degrees of freedom can be included in the supergravity description by modifying that boundary condition such that the boundary degrees of freedom called AdS singleton modes survive; see \citet{Maldacena:2001ss}.}

\subsection{Holographic dual of low-energy Yang-Mills theory}
\label{holoYM}

As a crucial step toward the construction of holographic QCD, we discuss Witten’s approach \citep{Witten:1998zw} to obtain a holographic description of the pure $\SU(N_c)$ Yang–Mills theory.\footnote{For more details, see section 6.2.2 of the well-known review \citep{Aharony:1999ti} of the AdS/CFT correspondence.} Consider type IIA superstring theory compactified on $S^1$. The $S^1$ is parametrized by $\tau=x^4$, and its radius is denoted as $\mkk^{-1}$ so that the Kaluza-Klein scale is given by $\mkk$. We place a stack of $N_c$ D4-branes extended along the $x^{0,1,2,3,4}$ directions and localized at $x^5=x^6=\cdots =x^9=0$. Since the D4-branes wrap $S^1$, the effective gauge theory realized on their world-volume becomes four-dimensional at low energies. In order to break supersymmetry, we impose an anti-periodic boundary condition for fermions along the $S^1$:
\be
\Psi(\tau+2\pi\mkk^{-1})=(-1)^F\Psi(\tau)\ ,
\label{apbc}
\ee
where $\Psi$ is a field in the system and $F$ is its fermion number.
Due to the boundary condition \eqref{apbc}, the fermions in the four-dimensional spacetime become massive and the supersymmetry is completely broken. The four-dimensional scalar fields also become massive due to quantum corrections.\footnote{
An explicit one-loop calculation shows that the quantum corrections to the mass squared for the scalar fields are positive. We expect them to be massive even in the strongly coupled regime. Note also that the scalar fields corresponding to the Nambu-Goldstone bosons associated with the spontaneous breaking of the translational symmetry in $x^{5,6,7,8,9}$ directions remain massless. They are gauge singlets and decoupled from the Yang-Mills field.}
Then, the low-energy effective theory realized on the four-dimensional spacetime becomes the pure Yang-Mills (YM) theory.

If we consider the Euclidean version of this setup and regard $\tau$ as the Euclidean time, the boundary condition (\ref{apbc}) corresponds to a thermal system with temperature $T=\mkk/(2\pi)$. Fortunately, the supergravity solution corresponding to this system is explicitly known. It is a Euclidean black D4-brane solution with Hawking temperature given by $T=\mkk/(2\pi)$. Then, the original Lorentzian theory is obtained by Wick rotating $x^0$. The metric is
\begin{align}
    ds^2=H(u)^{-1/2}\left(\eta_{\mu\nu}dx^\mu dx^\nu-f(u)d\tau^2\right)
    -H(u)^{1/2}\left(\frac{du^2}{f(u)}+u^2d\Omega_4^2\right)
    \label{D4metric}
\end{align}
with
\begin{align}
    H(u)=1+\frac{Q_4}{u^3}\ ,~~~f(u)=1-\frac{\ukk^3}{u^3}\ ,
\end{align}
where $\mu,\nu=0,1,2,3$. Here, the radial coordinate $u$ is restricted to $u\ge \ukk$, and the lower bound $u=\ukk$ corresponds to the horizon in the black D4-brane solution.
Note that the radius of $S^4$, which can be read off from the coefficient of $d\Omega_4^2$ in (\ref{D4metric}), remains finite everywhere. This fact will play an important role in section \ref{baryon_as_D-brane}.
The parameters $Q_4$ and $\ukk$ are given by
\begin{align}
    Q_4=\pi g_s N_c\ell_s^3=\frac{\lambda\ell_s^2}{2\mkk}\ ,~~~
    \ukk=\frac{4}{9}Q_4\mkk^2=\frac{2}{9}\lambda\mkk\ell_s^2\ ,
    \label{Qukk}
\end{align}
where $\lambda=\gym^2N_c=2\pi g_s\ell_s\mkk N_c$ is a parameter interpreted as the 't Hooft coupling of the four-dimensional gauge theory. The relation between $\ukk$ and $Q_4$ in \eqref{Qukk} is fixed by requiring the regularity of the geometry at $u=\ukk$.\footnote{In the second equation of (\ref{Qukk}), terms that vanish in the decoupling limit are neglected.}
In addition, the dilaton field $\phi$ and the RR 3-form field $C_3$ are non-trivial as
\begin{align}
    e^\phi=g_sH(u)^{-1/4}\ ,~~~
    \frac{1}{(2\pi\ell_s)^3}\int_{S^4}dC_3=N_c\ .
    \label{D4dilaton-RR}
\end{align}

As discussed in the previous subsection, we have to take a decoupling limit. In this case, it turns out that we should take a limit $\ell_s\ra 0$ while $\lambda$ and the coordinate $r$ ($r\ge 0$) defined by
\be
\frac{u^3}{\ukk^3}=1+r^2=:k(r)
\ee
are kept fixed. In this limit, the metric \eqref{D4metric} and the dilaton field \eqref{D4dilaton-RR} become
\begin{align}
&ds^2=\frac{\ell_s^2}{\tilde\ell_s^2}d\tilde s^2\ ,~~~
d\tilde s^2=k(r)^{1/2}\eta_{\mu\nu}dx^\mu dx^\nu-
\left(k(r)^{-5/6} dr^2+k(r)^{-1/2}r^2d\tau^2+\frac{9}{4}k(r)^{1/6}d\Omega_4^2
\right)\ ,
\label{D4metric2}\\
&e^\phi=\frac{\lambda^{3/2}}{3\sqrt{3}\pi N_c}k(r)^{1/4}\ ,
\label{D4dilaton2}
\end{align}
in units with $\mkk=1$, where we have defined $\tilde\ell_s^2=27/(4\lambda)$. Thus, we obtain type IIA superstring theory in the background with metric given by $d\tilde s^2$ in \eqref{D4metric2} and the string tension
\be
\tilde T_s=\frac{1}{2\pi\tilde\ell_s^2}=\frac{2\lambda}{27\pi}\mkk^2\ .
\label{tension}
\ee
This string theory configuration is supposed to be dual to the gauge theory realized on the D4-brane world-volume, which flows to the pure YM theory at low energies.

In principle, if one could take a limit where the radius of $S^1$ is sent to zero and thus $\mkk\rightarrow\infty$ (together with the decoupling limit $\ell_s\rightarrow 0$) with the dynamical QCD scale $\Lambda_{\rm QCD}$ kept fixed, we expect to obtain precisely four-dimensional pure YM theory. However, the asymptotic freedom of YM theory implies that $\lambda$ should be taken to zero in the $\mkk\ra\infty$ limit in order to keep $\Lambda_{\rm QCD}$ fixed. Because the supergravity approximation is valid when $\lambda\gg1$, it is not easy to take this limit in the holographic description. For this reason, we keep $\mkk$ finite in what follows. Thus, the theory realized on this brane configuration is the YM theory with additional matter fields of masses around $\mkk\approx\mbox{1 GeV}$ or higher. These additional massive fields are called KK artifacts. Therefore, roughly speaking, the system starts to deviate from the YM theory when the energy scale of interest exceeds $\mkk$. This feature is inherited in the top-down holographic QCD discussed in later sections, since this D4-brane system is used to introduce the gluon sector of QCD.
For low-energy quantities, it is reasonable to expect that the contributions from massless gluons and quarks are much larger than those from the massive modes. In fact, lattice QCD calculations reproduce the experimental results remarkably well even when the contribution from the charm quark, whose mass is around 1.3 GeV, is completely neglected. Therefore, we expect that holographic QCD can reproduce the low-energy properties of QCD.

\begin{figure}[h]
\begin{center}
\begin{picture}(300,60)(0,0)
\put(10,6){
\includegraphics[scale=0.1]{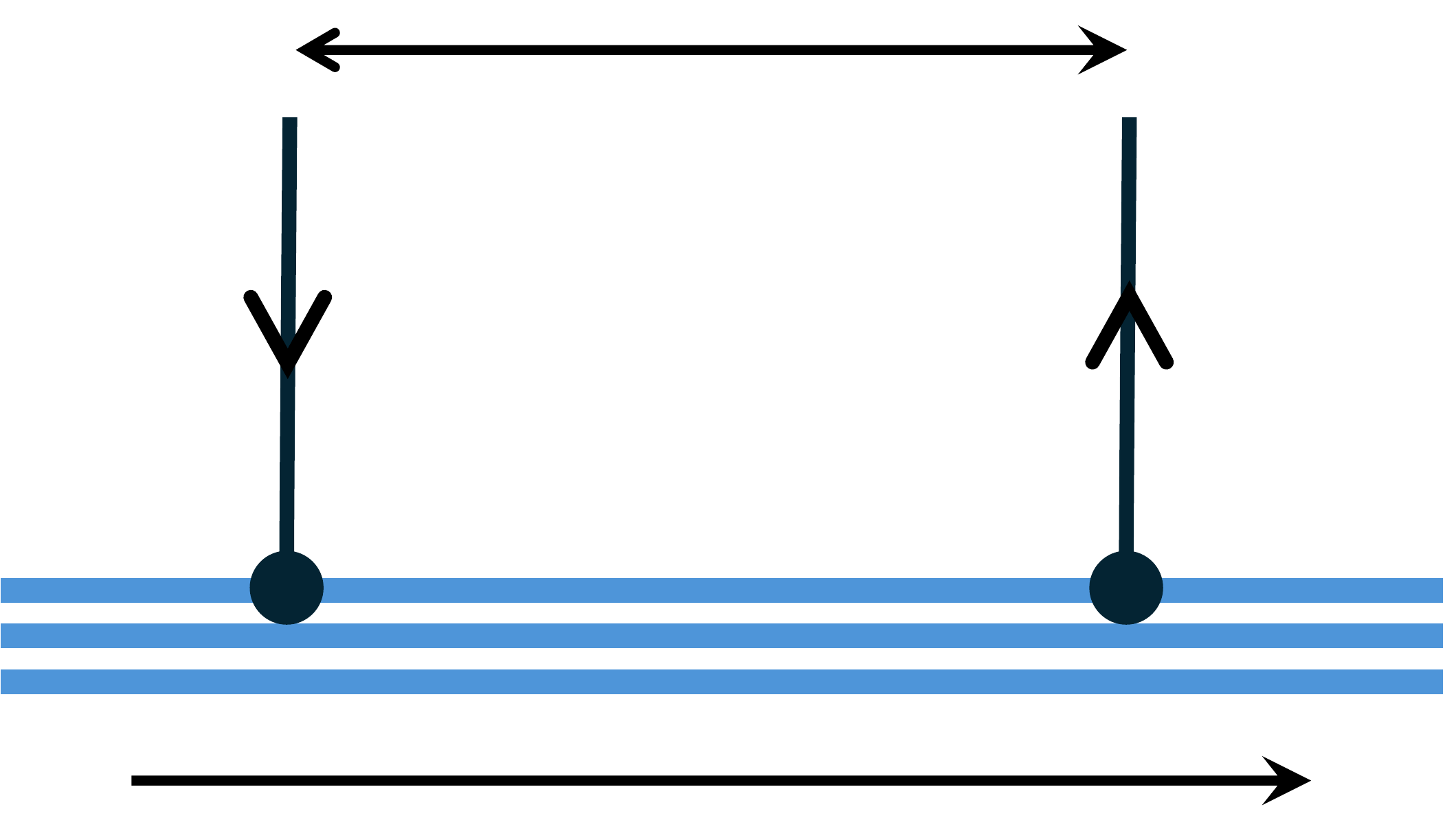}}
\put(3,15){\makebox(0,0){D4}}
\put(62,67){\makebox(0,0){$L$}}
\put(47,48){\makebox(0,0){string}}
\put(105,48){\makebox(0,0){string}}
\put(36,29){\makebox(0,0){$q$}}
\put(96,29){\makebox(0,0){$\bar q$}}
\put(110,10){\makebox(0,0){$x^1$}}
\put(150,40){\makebox(0,0){{\Large $\Longrightarrow$}}}
\put(200,6){
\includegraphics[scale=0.1]{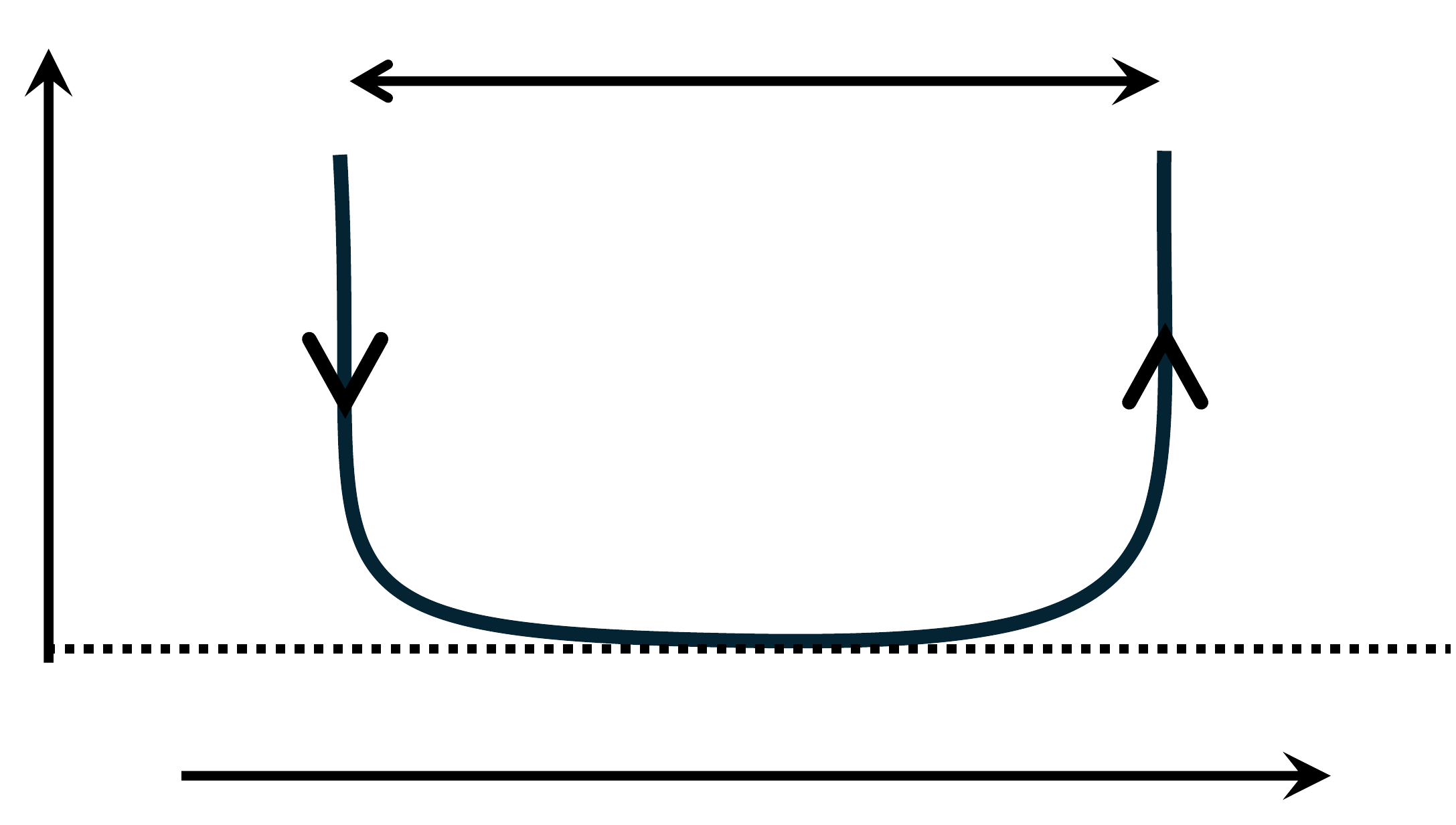}}
\put(200,63){\makebox(0,0){$r$}}
\put(299,48){\makebox(0,0){string}}
\put(240,48){\makebox(0,0){string}}
\put(255,67){\makebox(0,0){$L$}}
\put(305,10){\makebox(0,0){$x^1$}}
\put(180,19){\makebox(0,0){$r=0$}}
\put(193,19){\vector(1,0){9}}
\end{picture}
\caption{The left panel shows a pair of external quark and antiquark introduced by putting a pair of open strings attached to the D4-brane. The right panel is its holographic description. In the holographic description,  the strings in the left panel have to be connected, giving rise to a confining potential.
\label{qqbar}}
\end{center}
\end{figure}

As an application, the confinement of YM theory can be readily understood within the holographic description as follows. In the D4-brane configuration that describes YM theory, an external quark can be introduced by adding an open string with one endpoint attached to the D4-brane and extending along the $u$-direction.
Similarly, an antiquark can be introduced by putting an open string with opposite orientation. Consider a configuration with a pair of strings corresponding to an external quark-antiquark pair separated by a distance $L$ along the $x^1$-direction. (See the left panel of Fig.~\ref{qqbar}.) In the holographic description, the D4-branes are completely replaced by the corresponding curved background given in \eqref{D4metric2} and \eqref{D4dilaton2},
and there is no place where open strings can end. Therefore, the two strings have to be connected and, when $L$ is large, we obtain a string extending along the $x^1$-direction near $r=0$, which corresponds to the color flux tube in QCD. 
(See the right panel of Fig.~\ref{qqbar}.)
Note that the Nambu-Goto action (\ref{S_NG}) for a string stretched along the $x^1$-direction with fixed $r$ can be written as
\be
S_{\rm NG}=T_{\rm eff}(r)\int dx^0 dx^1
\ee
with the effective string tension
\be
T_{\rm eff}(r)=k(r)^{1/2}\tilde T_s\ .
\ee
Because the effective tension of the string at $r=0$ ($T_{\rm eff}(0)=\tilde T_s$) is non-zero and finite, the quark-antiquark potential is proportional to the distance $L$ for large $L$. This gives a confining force between the external quark-antiquark pair.

As can be seen from this example, confinement occurs when the radial coordinate has a lower bound where the string tension remains nonzero. This feature will be used to construct bottom-up models of holographic QCD.

\section{Holographic QCD}
There are two approaches to construct holographic models of QCD, called top-down and bottom-up approaches. 
The top-down approach starts with a setup in string theory that realizes QCD at low energies and then finds its holographic dual description following Maldacena's original idea of AdS/CFT correspondence, summarized in section \ref{AdS/CFT}.
The bottom-up approach is to find a phenomenological model of hadrons described by a 5-dimensional action, from which properties of hadrons are extracted by using the dictionary of holographic duality. In this section, we review the holographic models of QCD constructed in these two approaches in turn. The applications to the analysis of QCD and hadron physics will be discussed in the following sections.

\subsection{Top-down approach}
\label{topdown}
\subsubsection{General strategy}
\label{general}
Top-down holographic QCD has been constructed by adding additional D-branes, called flavor branes, to Witten's D4-brane system, which describes pure YM theory (see section \ref{holoYM}), in order to introduce quarks in the fundamental representation.%
\footnote{See \citet{Mazu:2007tp} and \citet{Saitou:2008xv} for other proposals to construct holographic QCD in string theory.} The idea is that open strings stretched between the D4 and the flavor branes induce matter fields in the fundamental representation of the $\SU(N_c)$ gauge group.\footnote{The idea of introducing matter fields in the fundamental representation by adding flavor branes in the context of AdS/CFT correspondence was first proposed by \citet{Karch:2002sh} and applied to supersymmetric systems such as the D3-D7 system.}
If spin-1/2 fermions are created from these open strings, they can be interpreted as fundamental quarks.

The holographic dual description of this system is obtained by replacing the D4-branes with the corresponding supergravity background (\ref{D4metric2})--(\ref{D4dilaton2}). The flavor branes are usually treated as probe branes \citep{Karch:2002sh} embedded in the D4-brane background. Namely, the backreaction to the geometry due to the existence of the flavor branes is neglected, which is justified as long as the $1/N_c$ expansion of large $N_c$ QCD with the number of flavors $N_f$ treated as order 1 is a good approximation.\footnote{The backreaction of the flavor branes in the \WSS model has been studied
by \citet{Burrington:2007qd} and \citet{Bigazzi:2014qsa}; a review of studies in other top-down models is given by \citet{Nunez:2010sf}.}

\subsubsection{D4-D6 model}
\label{D4D6}
The first proposal in line with the strategy described in the previous subsection was made by \citet{Kruczenski:2003uq}, in which D6-branes are used as the flavor branes. In this setup, there are $N_c$ D4-branes, called color branes, extended along $x^{0,1,2,3,4}$ directions and $N_f$ D6-branes extended along $x^{0,1,2,3,5,6,7}$ directions. As explained in section \ref{holoYM}, the $x^4$-direction is compactified to $S^1$ with the supersymmetry-breaking boundary condition and the low-energy effective theory on the D4-branes turns out to be the 4-dimensional $\SU
(N_c)$ YM theory. When the D6-branes are added to this system, the spectrum of an open string stretched between D4 and D6-branes contains $N_f$ Dirac fermions that belong to the fundamental representation of the $\SU(N_c)$ gauge group. These fermions are naturally interpreted as quarks. The $\U(N_f)$ gauge symmetry on D6-branes corresponds to the flavor symmetry. One of the advantages of this construction is that the masses of the quarks are understood geometrically as the position of the D6-branes in the $x^{8,9}$-plane. The quark mass is proportional to the distance between the D4 and D6-branes, and hence the massless QCD is realized when the D4 and D6-branes intersect with each other at $x^8=x^9=0$. In this case, the brane configuration is invariant under the SO(2) rotation acting on the $x^{8,9}$-plane. 
This rotational symmetry is interpreted as the axial $\U(1)_A$ symmetry in massless large $N_c$ QCD.\footnote{In QCD, $\U(1)_A$ is explicitly broken by axial anomaly. However, we ignore this anomaly in this section, since the effect of the anomaly is suppressed by $1/N_c$ in large $N_c$ QCD. The axial anomaly for this system was studied by \citet{Barbon:2004dq}. See section \ref{sec:U1A} for related discussion.}

As mentioned in section \ref{general}, the holographic description is obtained by replacing the D4-branes with the corresponding supergravity background. The D6-branes are extended along $x^\mu$ ($\mu=0,1,2,3$) and three other directions related to $x^{5,6,7}$ in the background. One notable fact is that the radial coordinate $u$ has a lower bound $\ukk$, and the radius of $S^4$ is non-zero at $u=\ukk$. In particular, the point corresponding to the origin $x^{5,6,7,8,9}=0$ in the 5-dimensional space transverse to the D4-branes is removed from the spacetime in this geometry. Therefore, the D6-branes cannot take a configuration that is invariant under the SO(2) rotation of the $x^{8,9}$-plane. This implies the spontaneous breaking of the $\U(1)_A$ symmetry, which is expected to occur in QCD by the chiral condensate $\langle\bar\psi\psi\rangle\ne 0$.
As a result, a massless mode interpreted as the Nambu-Goldstone mode associated with this spontaneous symmetry breaking appears in the spectrum of the fluctuation of the flavor D6-branes. This mode corresponds to the $\eta'$ meson. It is expected to be massless, when the quarks are all massless, in the large $N_c$ limit. As mentioned above, it is easy to make the quarks massive by placing D6-branes away from $x^8=x^9=0$. In this case, $\eta'$ becomes massive. It was shown by \citet{Kruczenski:2003uq} that the quark mass $m_q$ and the $\eta'$ meson mass $m_{\eta'}$ obey the Gell-Mann--Oakes--Renner relation: $m_{\eta'}^2= -m_q\langle\bar\psi\psi\rangle/f_\pi^2$, where $f_\pi$ is the pion decay constant, in agreement with what we expect in QCD.

In this way, this model succeeded in reproducing a part of the spontaneous chiral symmetry breaking. However, the full chiral symmetry of massless QCD is $\U(N_f)_L\times \U(N_f)_R$, and it is expected to be spontaneously broken down to the diagonal $\U(N_f)$ subgroup. The full chiral symmetry and its breaking are not manifestly realized in this model. A beautiful resolution of this issue was given by \citet{Sakai:2004cn}, which will be reviewed in the next subsection.

\subsubsection{D4-D8 (WSS) model}
\label{D4D8}
\citet{Sakai:2004cn} proposed a top-down holographic model of QCD based on D4 and D8-branes. The brane configuration of this model before moving to the holographic description is as follows.
The configuration of the color branes is the same as the previous model. We have $N_c$ D4-branes extended along $x^{0,1,2,3,4}$ and wrapped on an $S^1$ of radius $\mkk^{-1}$, parametrized by $\tau=x^4$, with the supersymmetry-breaking boundary condition. The flavor branes are chosen to be $N_f$ D8-\Dbar~pairs localized in the $\tau$-direction.\footnote{A D8-brane has two possible orientations. In order to distinguish the orientation, we refer to one of them as D8 and the other as \Dbar. The \Dbar-brane is obtained by flipping the D8-brane and reversing its orientation.} We place $N_f$ D8-branes at $\tau=0$ and $N_f$ \Dbar-branes at $\tau=\pi\mkk^{-1}$.\footnote{We may put \Dbar-branes at another point in the $\tau$ direction. As long as the separation between D8 and \Dbar~is larger than the string scale, possible tachyon modes created by open strings stretched between them become massive. We don't have to worry about possible instability due to closed string exchange between D8 and \Dbar-branes, as we take the decoupling limit. It is possible to show explicitly that there is no unstable mode in the holographic description. When the D8 and \Dbar-branes coincide, a tachyon field arises from the open strings stretching between them, which corresponds to the scalar field $X$ in the HW1 model discussed in section \ref{sec:bottomup}.} We refer to an open string stretched between D4 and D8 (or \Dbar) as 4-8 string (or 4-$\bar 8$ string). The massless mode in the spectrum of the 4-8 string turns out to be a left-handed Weyl fermion $\psi_L$ in 4 dimensions. Similarly, the 4-$\bar 8$ string produces a right-handed Weyl fermion $\psi_R$. These fermions are interpreted as quarks. Together with the gluons $A_\mu$ created by the open strings attached to the D4-branes, the massless degrees of freedom obtained from these strings are precisely those of massless QCD. (See Table \ref{tab:sym}.) In addition, there are closed strings, 8-8, $\bar8$-$\bar8$ and 8-$\bar8$ strings in this system. As we discussed in section \ref{AdS/CFT}, we take a decoupling limit $\ell_s\rightarrow 0$, 
while the QCD gauge interactions are kept finite, so that the closed strings are decoupled from the QCD sector. This argument also applies to the 8-8, $\bar8$-$\bar8$ and 8-$\bar8$ strings. One important consequence is that the gauge fields created by 8-8 and $\bar8$-$\bar8$ are decoupled, and hence the gauge symmetry on the D8 and \Dbar-branes, denoted as $\U(N_f)_L$ and $\U(N_f)_R$, respectively, is regarded as a global symmetry.
The symmetry of the system is summarized in Table \ref{tab:sym}.
\begin{table}[ht]
\centering
    \begin{tabular}{|c|c|c|c|c|}
    \hline
        string & field &$\overset{\rm D4}{\SU(N_c)}$& $\overset{\rm D8}{\U(N_f)_L}$ & $\overset{\overline{\rm D8}}{\U(N_f)_R}$ \\
        \hline
        4-4&$A_\mu$ & adj. &1&1\\
        4-8&$\psi_L$ & $\Box$ &  $\Box$&1\\
        4-$\bar 8$& $\psi_R$ & $\Box$ & 1& $\Box$\\
        \hline
    \end{tabular}
    \caption{Massless fields obtained from the open strings in the D4-D8-\Dbar~system}
    \label{tab:sym}
\end{table}
As is clear from this Table, $\U(N_f)_L$ and $\U(N_f)_R$ correspond to the chiral symmetry of massless QCD.
In this way, QCD with manifest chiral symmetry is realized within the framework of string theory. Since the chiral symmetry plays a crucial role in the low-energy properties of QCD, this is a significant advantage of this model.

In addition to the symmetries in QCD, this system has an $\SO(5)$ symmetry acting on the coordinates $x^{5,6,7,8,9}$ as a rotation. It is easy to see that the massless fields listed in Table \ref{tab:sym} (gluon and quark fields) are invariant under this $\SO(5)$ symmetry, while many of the KK artifacts (see section \ref{holoYM}) are not invariant. Therefore, any bound state in this system that is not invariant under the $\SO(5)$ symmetry cannot be interpreted as a QCD bound state, as it necessarily contains the KK artifacts among its constituents. This is one of the useful criteria for identifying QCD bound states in the spectrum.

\begin{figure}[h]
\begin{center}
\begin{picture}(300,90)(0,0)
\put(0,15){
\includegraphics[scale=0.30]{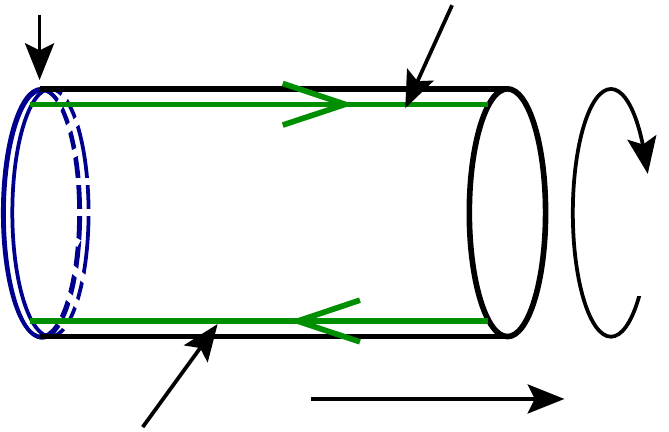}}
\put(9,82){\makebox(0,0){D4}}
\put(70,83){\makebox(0,0){D8}}
\put(20,9){\makebox(0,0){$\overline{\rm D8}$}}
\put(99,50){\makebox(0,0){$\tau$}}
\put(87,19){\makebox(0,0){$u$}}
\put(140,45){\makebox(0,0){{\Large $\Longrightarrow$}}}
\put(200,18){
\includegraphics[scale=0.30]{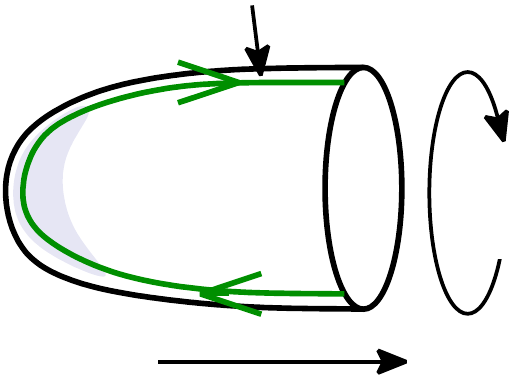}}
\put(237,78){\makebox(0,0){D8}}
\put(277,49){\makebox(0,0){$\tau$}}
\put(265,19){\makebox(0,0){$u$}}
\put(232,37){\makebox(0,0){\color{Green}{$Z$}}}
\put(180,32){\makebox(0,0){$u=\ukk$}}
\put(192,36){\vector(3,2){10}}
\end{picture}
\caption{The left panel depicts the brane configuration of the D4-D8-\AD~system that realizes QCD. The right panel is the brane configuration obtained after replacing the D4-brane with the corresponding supergravity background described in section \ref{holoYM}. In the right panel, the D8 and \AD-branes are connected, providing a geometric understanding of chiral symmetry breaking in QCD. 
\label{chiSB}}
\end{center}
\end{figure}

The holographic description is obtained in a similar way as explained in section \ref{D4D6}. We replace the D4-branes with the corresponding supergravity background, and the D8 and \Dbar-branes are treated as probe branes. One interesting consequence is that the D8-branes and \Dbar-branes have to be connected in this background. This is because the D8-branes and \Dbar-branes are placed at $\tau=0$ and $\tau=\pi\mkk^{-1}$, respectively, and extended along the $u$-direction. As it is shown in the right panel of Fig.~\ref{chiSB}, they meet at the tip of the cigar at $u=\ukk$. It is easy to see that the connection of D8 and \Dbar~is unavoidable as long as the asymptotic configuration at $u\rightarrow\infty$ approaches the original brane configuration discussed above. This phenomenon is interpreted as the spontaneous chiral symmetry breaking $\U(N_f)_L\times \U(N_f)_R\rightarrow \U(N_f)$ in QCD. Intuitively, this is understood as follows.
As we have seen, $\U(N_f)_L$ and $\U(N_f)_R$ correspond to the gauge symmetry of the D8 and \Dbar-branes, respectively, before replacing D4-branes with the supergravity background, which corresponds to the weakly coupled regime. In the holographic description, the D8-branes and \Dbar-branes are connected, and there is only one connected component of $N_f$ D8-branes, which gives one component of $\U(N_f)$ as the unbroken part of the symmetry.\footnote{This mechanism of chiral symmetry breaking also works in holographic QCD with gauge groups $\SO(N_c)$ and ${\rm USp}(N_c)$ \citep{Imoto:2009bf}.} More explicitly, this chiral symmetry breaking can be shown by the existence of the associated Nambu-Goldstone modes, which will be discussed in section \ref{pion}.

Since this holographic description is supposed to be dual to the D4-D8-\Dbar~system that describes QCD, it should contain all the hadrons in QCD. In fact, one can find natural candidates for the glueballs, mesons and baryons in the holographic description of QCD. As discussed in section \ref{holoYM}, the closed strings correspond to glueballs. Open strings attached on the probe D8-branes are interpreted as mesons. Baryons are obtained as D4-branes wrapped on $S^4$, which can also be described by solitons of the meson effective theory, as discussed in section \ref{baryon}.

It is possible to write down the low-energy effective action for the open strings on the D8-branes, which is interpreted as the meson effective action in this holographic model of QCD. Since the fields related to mesons in QCD have to be invariant under the $\SO(5)$ rotation of $S^4$
as discussed above, the $S^4$ directions can be reduced. Then, after reducing the $S^4$ directions, the meson effective action turns out to be a 5-dimensional $\U(N_f)$ Yang-Mills-Chern-Simons (YM-CS) theory. Let us just show the result without its detailed derivation:
\begin{equation}
    S_5=-\kappa\int d^4 x dZ\,{\rm tr}\left(
    \frac{1}{2}h(Z)F_{\mu\nu}F^{\mu\nu}+k(Z)\mkk^2 F_{\mu Z}F^{\mu Z}
    \right)+\frac{N_c}{24\pi^2}\int\omega_5(A)\ .
    \label{5dim_action}
\end{equation}
Here $\mu,\nu=0,1,2,3$ are 4-dimensional Lorentz indices, which are raised and lowered by the Minkowski metric, and $Z$ is the coordinate along the D8-brane in Fig.~\ref{chiSB}, related to $u$ by $u^3/\ukk^3=1+Z^2$. $A=A_\mu dx^\mu + A_Z dZ$ is the 5-dimensional $\U(N_f)$ gauge field.
$\kappa$ is a constant proportional to $N_c$ and the 't Hooft coupling $\lambda$ given by
\be
\kappa=\frac{\lambda N_c}{216\pi^3}\ .
\label{kappa}
\ee
The functions $h(Z)$ and $k(Z)$ are given by
\begin{equation}
    h(Z)=(1+Z^2)^{-1/3}\ ,~~~k(Z)=1+Z^2\ ,
\end{equation}
and
\begin{equation}\label{eq:omega5}
    \omega_5(A)={\rm tr}\left(A F^2-\frac{i}{2}A^3F-\frac{1}{10}A^5\right)
\end{equation}
is the Chern-Simons 5-form.

As we will see in section \ref{Bottom}, many of the bottom-up models of holographic QCD have similar 5-dimensional meson effective actions. This type of 5-dimensional gauge theory is the starting point of the analysis in the following sections.

Before closing this section, let us emphasize that, in the top-down approach, the chiral symmetry breaking, confinement and the existence of the hadrons as well as the effective action (\ref{5dim_action}) are all derived without making any phenomenological assumptions.
By comparing the predictions of holographic QCD with experimental data or other calculations in QCD, such as lattice QCD, we can make quantitative tests of holography and string theory as a theory of hadrons. This is very important not only for hadron physics but also for string theory, since it provides strong evidence that the idea of AdS/CFT correspondence (gauge/string duality) works even for the cases without supersymmetry and conformal symmetry.

\subsection{Bottom-up approach}
\label{Bottom}

Even before the above D4-D8 model was constructed,
some of its features were anticipated by 
dimensional deconstruction of QCD \citep{Son:2003et}, which we briefly review in
appendix \ref{appDD}, where an extra continuous dimension arises
from a continuum limit of a 4D open moose theory with a large number
of hidden symmetry groups. 

In the so-called bottom-up approach of holographic QCD,
one typically builds upon the insights
gained from top-down, string-theoretic constructions, and the
holographic dictionary established therein, to build models
that have more flexibility, in particular to
make contact with the high-energy behavior of QCD, which
cannot be described by a background geometry built upon D4 
color branes.

Real QCD is asymptotically free and has approximate conformal
symmetry at high energies, which can be implemented in
a (still strongly coupled) holographic model by a geometry 
which is asymptotically
anti-de Sitter (recall that pure anti-de Sitter space provides
the arena of the original AdS/CFT correspondence with
maximally supersymmetric and superconformal YM theory in
the limit of infinite 't Hooft coupling and infinite $N_c$).

\subsubsection{Minimal hard-wall and soft-wall models}\label{sec:bottomup}

The effective 5D action of the \WSS\ model \eqref{5dim_action} 
as well as the minimal
hard-wall (HW) and soft-wall (SW) models considered by \citet{Erlich:2005qh,DaRold:2005mxj,Hirn:2005nr,Karch:2006pv} 
involve a Yang-Mills action of the form
\be\label{S5YM}
S_{5,\mathrm{YM}} = -\frac{1}{4g_5^2} \int d^4x \int_0^{z_0} dz
\sqrt{g}\, g^{PR}g^{QS} e^{-\Phi(z)}
\,\text{tr}\left(\mathcal{F}^\mathrm{L}_{PQ}\mathcal{F}^\mathrm{L}_{RS}
+\mathcal{F}^\mathrm{R}_{PQ}\mathcal{F}^\mathrm{R}_{RS}\right),
\ee
where $P,Q,R,S=0,\dots,3,z$ and $\mathcal{F}_{MN}=\partial_M \mathcal{B}_N-\partial_N \mathcal{B}_M+i[\mathcal{B}_M,\mathcal{B}_N]$.
The $\U(N_f)_L\times\U(N_f)_R$ gauge fields $\mathcal{B}_\mu^{L,R}$ 
(also written as $L_\mu,R_\mu$ in the following)
are dual to
the operators $ (\bar\psi\gamma_\mu\psi)^{L,R}$.
Following the \WSS model, one adds a Chern-Simons action 
\be\label{SCS}
S_{5,\rm CS}=\frac{N_c}{24\pi^2}\int \left( \omega_5(L) - \omega_5(R) \right)
\ee
with $\omega_5$ defined in \eqref{eq:omega5},
up to potential boundary terms at $z_0$.

The top-down \WSS model described in Sec.~\ref{D4D8} can be brought
into the form \eqref{S5YM} with holographic UV boundary at $z=0$
by setting $u=1/z$, 
involving a nontrivial dilaton and 
a metric that does not
asymptote to anti-de Sitter. It therefore does not reproduce
the conformally symmetric UV limit of four-dimensional QCD.

In the models \citep{Erlich:2005qh,DaRold:2005mxj,Hirn:2005nr,Karch:2006pv}, 
the metric is simply chosen as pure $\AdS_5$,
\be\label{ds2AdS}
ds^2=z^{-2}(\eta_{\mu\nu}dx^\mu dx^\nu - dz^2),
\ee
with conformal symmetry broken either by a sharp HW cut-off \citep{Polchinski:2001tt}
at $z_0$ with trivial $\Phi=0$, or, in the standard SW model, by
a nontrivial dilaton $\Phi(z)=c^2 z^2$ with $z_0\to\infty$.
The energy scale $z_0^{-1}$, or $c$ in the standard SW model,
represents the QCD scale $\Lambda_\mathrm{QCD}$ and directly determines
glueball and vector-meson masses. The motivation for the original SW model with quadratic dilaton \citep{Karch:2006pv}
is to have linear Regge trajectories; 
an early attempt towards an improved bottom-up model is given by the
semi-hard-wall model of \citet{Kwee:2007nq,Kwee:2007dd}.

The 5D gauge coupling is fixed by matching to the large-$N_c$ QCD two-point function of
vector currents at short distances (OPE at $Q^2=-q^2\to\infty$, cf.\ Eq.~\eqref{PiVas} and Sec.\ \ref{sec:OPE}) \citep{Erlich:2005qh}:\footnote{Note that with Eq.~\eqref{ds2AdS} the metric $g_{MN}$ has
dimension of inverse length squared. This convention,
where the AdS radius is set to unity, leads to a dimensionless 5D coupling $g_5$.}
\begin{equation}
  g_5^2 = \frac{12\pi^2}{N_c}\,.
  \label{eq:g5}
\end{equation}

In the original HW \citep{Erlich:2005qh} and SW models \citep{Karch:2006pv,Kwee:2007dd},
chiral symmetry is broken by an additional bifundamental scalar $X(z,x)$ dual to the chiral
condensate and sourced by dimension-3 quark mass terms
with action
\be
S_{5,X}=\int d^4x \int_0^{z_0} dz\,\sqrt{g}\,\text{tr}\left(
|DX|^2-M_X^2|X|^2 \right),\label{Sscalar}
\ee
where $D_M X=\partial_M X-i \mathcal{B}_M^\mathrm{L} X+iX\mathcal{B}_M^\mathrm{R}$
and $M_X^2=-3$. In a pure $\AdS_5$ background, one obtains
vacuum solutions
\be
X(z)=\frac12 v(z)=\frac12(M_q z+\Sigma\, z^3),\label{ScalarVEV}
\ee
where $M_q$ and $\Sigma$ are
model parameters related to the quark mass matrix and the quark condensate,\footnote{According to \citep{DaRold:2005vr,Cherman:2008eh,Gorsky:2009ma,Abidin:2009aj} both
$M_q$ and $\Sigma$
need to be rescaled
by $C=\sqrt{N_c}/(2\pi)$ to be interpreted as quark masses and
condensates in QCD: $m_q=M_q/C$ and $\langle q \bar q\rangle=C\Sigma$.\label{footnote:rescaling}}
which can be freely adjusted by postulating a suitable potential localized
on the IR brane \citep{DaRold:2005vr}.
The tachyonic mass value $M_X^2=-3$ is determined by the scaling dimension of the quark
mass term; generalizations away from this value have been
considered by \citet{Domenech:2010aq} and \citep{Leutgeb:2021mpu}
(the Breitenlohner-Freedman bound \citep{Breitenlohner:1982bm}
permits tachyonic values $-4 \le M_X^2 < 0$).

This bifundamental field can be viewed as the tachyon of open strings
stretching between left and right handed flavor branes.
In the original top-down \WSS model, the tachyon $X$ is omitted.\footnote{It has been considered, however, by \citet{Bergman:2007pm,Aharony:2008an,Hashimoto:2008sr,Dhar:2008um,McNees:2008km,Niarchos:2010ki}}
There, chiral symmetry breaking occurs as a direct consequence of the
geometry which forces left and right flavor branes to connect either
at the minimal value $u=\ukk$ or, with non-maximal separation, at some $u_0>\ukk$.
This leads to different IR boundary conditions for vector and axial-vector
modes $\mathcal{V}=\frac12(\mathcal{B}_L+\mathcal{B}_R)$
and $\mathcal{A}=\frac12(\mathcal{B}_L-\mathcal{B}_R)$.
The same mechanism is used in the HW model of 
\citet{Hirn:2005nr},
which also omits $S_{5,X}$. \citet{Domenech:2010aq} have adopted these
symmetry-breaking boundary conditions on the bulk Yang-Mills fields also in a HW model with bifundamental scalar $X$, because it naturally
avoids infrared
boundary contributions from the CS action.
Following the terminology of \citet{Cappiello:2010uy,Leutgeb:2019zpq,Leutgeb:2019gbz,Leutgeb:2021mpu},
the original HW model of Erlich et al. \citep{Erlich:2005qh}, the Hirn-Sanz model
\citep{Hirn:2005nr}, and the latter combination of chiral symmetry breaking
possibilities \citep{Domenech:2010aq} will be referred to as HW1, HW2, and HW3, respectively.

\begin{figure}[htbp]
 \centering \includegraphics[width=.75\linewidth]{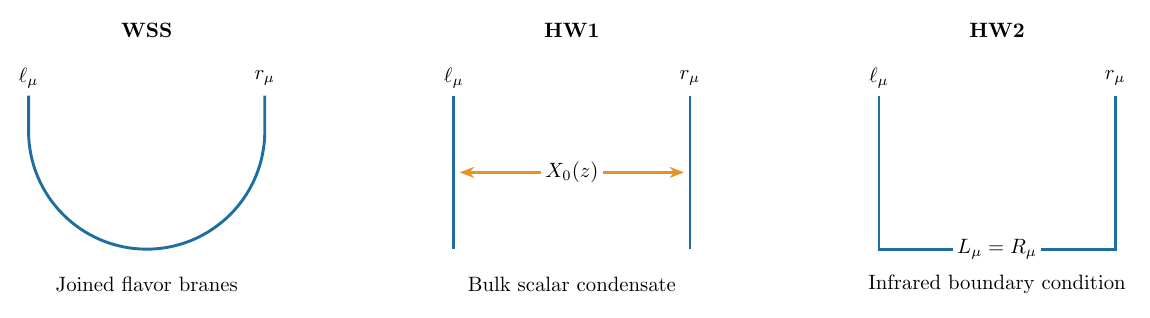}
\caption{Three mechanisms implementing the same non-singlet chiral symmetry-breaking pattern. The drawing is schematic: the HW1 model with scalar background $X_0$ couples the two gauge sectors throughout the bulk, while the gauge-only models identify their infrared transformations. The lower-case $\ell_\mu$ and
$r_\mu$ are the sources, i.e., 
boundary values of the flavor gauge fields $L_\mu$, $R_\mu$.}
\label{fig:mechanisms}
\end{figure}

The three holographic mechanisms of chiral symmetry breaking are sketched in Figure~\ref{fig:mechanisms}.


\subsubsection{Improved hQCD models}

In the minimal SW model the quadratic dilaton and quadratic scalar potential impose, in the axial
sector, an overly restrictive relation between the quark mass and the condensate: infrared
regularity selects a solution whose condensate is proportional to $m_q$, excluding an independent
nonzero condensate in the chiral limit. Extensions therefore modify the scalar potential, the
background profiles, or both.

The dynamical soft-wall program replaces the prescribed background by solutions of 5D equations
of motion. An early example is the Einstein-frame gravity-dilaton-tachyon model
of \citet{Batell:2008zm}, which gives a dynamical soft wall with linear radial trajectories. Its UV
profiles need modification for a QCD interpretation: if $T$ is identified with the quark bilinear,
$T\propto z$ near the boundary describes explicit, not spontaneous, breaking.
\citet{Gherghetta:2009ac} instead keep a fixed AdS metric and reconstruct a modified dilaton
from a scalar profile with independently adjustable mass and condensate and a quartic potential.
The dilaton is linear in the infrared, giving a common radial slope and a nonzero asymptotic
vector-axial mass splitting, although the construction is not a fully backreacted solution. The
coupled pseudoscalar and longitudinal axial system was studied by \citet{Kelley:2010mu}, recovering
the Gell-Mann-Oakes-Renner relation and a reasonable radial pion spectrum, with a nonzero chiral
condensate at $m_q=0$
(see also \citep{Li:2012ay,Chen:2022goa}).

A related approach uses the open-string tachyon on flavor branes and
antibranes \citep{Bigazzi:2005md,Casero:2007ae,Iatrakis:2010zf,Iatrakis:2010jb}
(concretely, space-filling D4-$\overline{\mathrm{D4}}$ branes). 
The tachyon $T$ is a bifundamental of
$\mathrm{U}(N_f)_L\times\mathrm{U}(N_f)_R$, whose source and normalizable mode encode the quark mass
and condensate; a condensate proportional to the identity preserves the vector subgroup. Its
dynamics follows from a nonlinear Dirac-Born-Infeld action motivated by brane-antibrane
systems due to 
\citet{Sen:2003tm},
\begin{equation}
S_{\rm DBI}=-\!\int d^5x\,V(|T|)
\left[\sqrt{-\det\mathcal A_L}+\sqrt{-\det\mathcal A_R}\right],
\label{eq:Sen_DBI}
\end{equation}
\begin{equation}
(\mathcal A_i)_{MN}
=g_{MN}+c_FF^{(i)}_{MN}
+\frac{c_T}{2}\left[(D_MT)^*D_NT+(D_NT)^*D_MT\right],
\qquad i=L,R,
\label{eq:Sen_matrix}
\end{equation}
with $D_MT=(\partial_M+iA_M^{(L)}-iA_M^{(R)})T$, $g$ the induced metric, and $c_F$, $c_T$ the gauge
and tachyon normalizations. In the model studied by \citet{Iatrakis:2010zf,Iatrakis:2010jb} the flavor fields propagate
on the induced geometry of a 6D AdS soliton with constant dilaton, and the Gaussian potential
$V(|T|)=V_0\exp(-a_T|T|^2)$ is a phenomenological ansatz. Tachyon condensation then supplies the
infrared suppression giving linear radial trajectories without a prescribed quadratic dilaton, and
the nonlinear tachyon equation with its infrared condition determines the condensate for a given
mass, generally numerically. Chiral anomalies and the singlet pseudoscalar mass require additional
Wess-Zumino couplings and a Ramond-Ramond kinetic sector \citep{Casero:2007ae}. Combining this flavor
sector with improved gluon dynamics and its leading backreaction leads to V-QCD.

\subsubsection{The Veneziano limit and V-QCD models}

A further aim of improved hQCD is to encode the renormalization-group flow, whose infrared end may
be a mass gap, massless Goldstone bosons, or an interacting conformal theory. Veneziano's
limit \citep{Veneziano:1976wm},
\begin{equation}
N_c\to\infty,\qquad N_f\to\infty,\qquad x_f=\frac{N_f}{N_c}\ \text{fixed},
\label{eq:veneziano}
\end{equation}
with $g^2N_c$ fixed, differs from the 't~Hooft limit \citep{tHooft:1973alw}, where sea-quark effects
are suppressed by $N_f/N_c$: at fixed $x_f$ planar quark loops contribute at the same order as planar
gluon diagrams, so flavor contributions to the running and the backreaction of the chiral sector
must be kept at leading order.

For $\mathrm{SU}(N_c)$ with $N_f$ fundamental flavors, the two-loop beta
function \citep{Caswell:1974gg,Banks:1981nn}, with one-loop coefficient $\tfrac{11}{3}N_c-\tfrac23N_f$,
has for $x_f$ just below $11/2$ a weakly coupled Caswell-Banks-Zaks infrared fixed point. The lower
edge of the conformal window is beyond perturbation theory; a walking regime with Miransky scaling
at the transition \citep{Miransky:1996pd,Kaplan:2009kr} is one proposed scenario, realized in V-QCD.
Treating $x_f$ as continuous is useful for critical scaling, although walking can also be studied at
finite $N_c$ and $N_f$ \citep{Cacciapaglia:2020kgq}. 

V-QCD combines improved holographic Yang-Mills dynamics \citep{Gursoy:2007cb,Gursoy:2007er} with a
tachyonic flavor sector \citep{Casero:2007ae,Jarvinen:2011qe}. Probe flavor branes, as in the
standard WSS implementation \citep{Sakai:2004cn}, have an action of order $N_cN_f$, compared with
$N_c^2$ for glue; at fixed $x_f$ both sectors are leading, and the metric, dilaton and tachyon
equations must be solved together. The action $S=S_g+S_f+S_a$ contains the Einstein-Hilbert term
with dilaton kinetic and potential terms ($S_g$), the matrix DBI action for the complex tachyon
coupled to the left and right flavor gauge fields ($S_f$), and an axion dual to the topological
density ($S_a$).

Hadronic applications include meson and glueball spectra, scalar mixing, and vector and
axial-vector correlators, connecting UV normalizations to infrared observables and testing the
$V-A$ difference and the $S$-parameter \citep{Arean:2012mq}; real-time spectral functions and
transport have also been studied \citep{Iatrakis:2014txa}. Matching the beta function and the
quark-mass anomalous dimension does not by itself fix every Wilson coefficient or power correction
of the QCD OPE, which must be checked separately for each correlator. At finite temperature and
density V-QCD has been used for the phase diagram and equations of state by \citet{Alho:2012mh,Alho:2013hsa}, and,
combining a modeled baryonic sector with low-density nuclear input, for hybrid neutron-star
equations of state by \citet{Ishii:2019gta,Jokela:2020piw}, with results depending on the baryonic
approximation and matching prescription.

\subsubsection{Light-front holographic QCD}
Light-front (or light-cone) holographic QCD (LFhQCD) is a particular realization of the broader holographic approach to QCD. Like conventional AdS/QCD, it uses higher-dimensional geometry and infrared modifications to describe confinement, hadron spectra, and form factors. However, while  bottom-up hQCD emphasizes the correspondence between 5D bulk fields and local QCD operators, providing a framework for current correlators, chiral symmetry breaking, and, in improved models, renormalization-group flow, LFhQCD introduces a more explicit partonic interpretation: it identifies the holographic coordinate \(z\) with a boost-invariant transverse variable  and maps the bulk equations onto an effective light-front bound-state equation. Its solutions describe the transverse structure of valence light-front wave functions within a semiclassical approximation \citep{Brodsky:2006uqa,Brodsky:2008pf}. 

The theory is quantized on the null plane $x^+=0$,
with a Hamiltonian eigenvalue equation $H_\mathrm{LF}|\psi\rangle = M^2|\psi\rangle$
that governs the hadronic mass spectrum.
In the valence (lowest Fock-state) sector, the dynamics of a $q\bar{q}$ meson
can be reduced to an effective one-dimensional Schr\"odinger-like equation in
the boost-invariant transverse separation $\zeta$.
The identification $\zeta \leftrightarrow z$ (the AdS$_5$ radial coordinate)
is the central holographic mapping: the 
LFhQCD Schr\"odinger-like equation is
identical in form to the wave equation for a field propagating in a
(deformed) AdS$_5$ background.

The choice of a quadratic dilaton profile,
as for the SW model, translates  into a harmonic
confinement potential, and linear Regge
trajectories for both mesons and baryons with a single universal scale
$c \simeq 0.5\,\text{GeV}$ \citep{Brodsky:2008pf,Brodsky:2014yha}. It must be noticed, however,  that their quadratic dilaton profile has opposite sign to that of \citet{Karch:2006pv} and, as discussed by \citet{Karch:2010eg}, such a choice would lead to an unphysical massless pole in the vector two-point  function in the SW model (see also \citep{Zuo:2009dz,Gutsche:2011vb,Afonin:2021cwo,Leutgeb:2025jmv}).

An underlying superconformal quantum mechanics \citep{deAlfaro:1976je} that interconnects  LFhQCD
equations for mesons, baryons, and tetraquarks, organizing them into
hadronic supermultiplets with degenerate
trajectories
was shown to be an  emergent property
of the LFhQCD Hamiltonian. The confining potential---and with it the dilaton profile---is entirely determined by the underlying symmetry, down to a constant term that turns out to be essential. Indeed,  the constant term fixed by the superconformal algebra is exactly what renders the ground-state pion massless in the chiral limit \citep{Brodsky:2013ar,Brodsky:2016yod}.

A comprehensive review of the LFhQCD framework and its phenomenological applications
is \citet{Brodsky:2014yha}.

\section{Mesons}

\subsection{Spectrum of (axial) vector mesons and pseudoscalars}
\label{pion}

Holographic QCD models embed the vector and axial-vector sectors 
together with the pseudoscalar Goldstone bosons
in a single 5D gauge system.
In the case of the top-down \WSS model and the bottom-up HW2 model
by just a 5-dimensional YM+CS action
(Eqs.~\eqref{5dim_action}, \eqref{eq:omega5} or
Eqs.~\eqref{S5YM}, \eqref{SCS}, respectively) 
and
supplemented by a bifundamental bulk scalar in the case of the HW1 model. 

In this subsection, we compare the results for the spectrum 
and decay constants of
vector and axial-vector mesons as well as of the (pseudo-)Goldstone pseudoscalars 
obtained in the top-down \WSS model and in the bottom-up models with
AdS geometry.

\subsubsection{\WSS model}
\label{specWSS}


The 5-dimensional gauge field in \eqref{5dim_action} can be
expanded in terms of even and odd eigenmodes in the (dimensionless)
coordinate $Z=-\infty\ldots+\infty$ running along the joint D8-\AD\ branes,
\be
A_\mu(x,Z)=\sum_{n=1}^\infty \psi_n(Z) v^{(n)}_\mu(x)\ ,~~~
A_Z(x,Z)=\sum_{n=0}^\infty \phi_n(Z) \varphi^{(n)}(x)\ .
\label{expansion}
\ee

The 4-dimensional modes $v^{(n)}$ can be interpreted as
vector and axial-vector mesons for $n$ odd and even, respectively,
with masses $\sqrt{\lambda_n}\MKK$ determined by
the eigenvalue equation
\be\label{psinSS}
-(1+Z^2)^{1/3}\6_Z\left[ (1+Z^2)\6_Z \psi_n \right]=\lambda_n \psi_n,\quad \psi_n(\pm\infty)=0.
\ee
Identifying the $n=1$ mode with the $\rho$ meson with mass of $\approx 776$ MeV,
the numerical result $\lambda_1=0.669314\ldots$
fixes the Kaluza-Klein mass of the \WSS model at $\MKK\approx 949$~MeV.

In the expansion of $A_Z$, one can choose $\phi_n(Z) \propto \6_Z \psi_n(Z)$ for $n\ge1$
and
absorb all the $\varphi^{(n)}(x)$ in $v^{(n)}_\mu(x)$ by a 4D gauge transformation, except
for $\varphi^{(0)}(x)$ with $\phi_0(Z)\propto 1/k(Z)=1/(1+Z^2)$.

Integrating over $Z$, the term involving $F_{\mu Z}F^{\mu Z}$ in \eqref{5dim_action} thus
provides a Proca mass term $m_n^2 v_\mu^{(n)}v^{(n)\mu}$ for the
vector and axial-vector mesons, and also a kinetic term for
the massless field $\varphi^{(0)}(x)$, which represents the pion
(or an entire Goldstone boson multiplet when $N_f>1$).

In order to show that $\varphi^{(0)}(x)$ corresponds to the Nambu-Goldstone mode associated with the chiral symmetry breaking, we define a $\U(N_f)$- valued field:
\be\label{pionU}
U(x):=\mathrm{P}\exp \left(i\int A_Z(x,Z)\, dZ\right)\ .
\ee
The gauge transformation of the 5-dimensional gauge field $A=A_\mu dx^\mu+ A_Z dZ$:
\be
A\ra gAg^{-1}+ig dg^{-1}
\ee
with $\U(N_f)$-valued function $g(x,Z)$ induces
a transformation
\be
U(x)\rightarrow  g_L U(x) g_R^{-1}\ ,
\label{Utransf}
\ee
where
\be
g_L:=\lim_{Z\ra+\infty} g\ ,~~g_R:=\lim_{Z\ra-\infty} g\ .
\label{gLgR}
\ee
Recall that the $\U(N_f)$ gauge transformation on the D8-brane at $Z\ra \pm\infty$ corresponds to the gauge transformation on the D8 and \AD-branes in the D4-D8-\AD \ system. (See section \ref{D4D8}.) From Table \ref{tab:sym}, the transformation (\ref{Utransf}) is interpreted as a chiral symmetry transformation with parameters $(g_L,g_R)\in \U(N_f)_L\times \U(N_f)_R$. Then,  (\ref{Utransf}) is nothing but the transformation of the $\U(N_f)$-valued pion field under the chiral symmetry, and hence the fluctuation of the $\U(N_f)$-valued field (\ref{pionU}) is the Nambu-Goldstone modes associated with the chiral symmetry breaking. When $A_Z$ is small, (\ref{pionU}) is expanded as
\be
U(x)\simeq 1+i\int A_Z(x,Z)dZ+\cdots\simeq 1+i\varphi^{(0)}(x)\int dZ\phi_0(Z)+\cdots\ ,
\label{Uexp}
\ee
where the contributions from $\varphi^{(n)}(x)$ with $n\ge1$ have been dropped because $\phi_n(Z)$ with $n\ge 1$ is a total derivative. From the relation (\ref{Uexp}), we find that the massless field $\varphi^{(0)}(x)$ corresponds to the Nambu-Goldstone mode associated with the spontaneous chiral symmetry breaking.

Alternatively, one can set $A_Z=0$ by a large gauge transformation
involving the (non-normalizable) zero-mode of (\ref{psinSS}) in the axial-vector sector, $\psi_0 = \frac2\pi \arctan(Z)$,
which moves the pion field from the holonomy (\ref{pionU})
to nontrivial boundary conditions on $A_\mu$ at $Z=\pm\infty$
\citep{Sakai:2004cn}.\footnote{See \citet{Lau:2016dxk} for an expression of the pion field that can be used for any gauge choice.}

Either way, with $\kappa={\lambda N_c}/(216\pi^3)$ and $\lambda=g_{\rm YM}^2 N_c$,
the pion decay constant, which follows from the kinetic term
of the pions, turns out to be given by \citep{Sakai:2004cn}
\begin{equation}
 f_\pi^2=\kappa M_{\mathrm{KK}}^2\int\mathrm{d} Z\,k(Z) (\psi_0')^2
 =\frac{4\kappa M_{\mathrm{KK}}^2}{\pi}=\frac{\lambda N_c}{54\pi^4}M_{\mathrm{KK}}^2 .
 \label{eq:fpiwss}
\end{equation}
Setting $f_\pi=92.4$ MeV corresponds to $\kappa=0.00745$ or $\lambda\approx 16.63$ when extrapolated to $N_c=3$.

In the \WSS model, the pions are massless; a mass term 
requires an additional construction, e.g.\ an open Wilson line \citep{Aharony:2008an} or world-sheet instantons \citep{Hashimoto:2008sr}.%
\footnote{Such an extension, together with the Witten-Veneziano mechanism
for the anomalous mass of the $\eta'$ meson (see Sec.~\ref{sec:U1A}),
permits, however, a remarkably good fit \citep{Brunner:2015oga}
of all the pseudoscalar masses
for $N_f=N_c=3$.}

\subsubsection{AdS/QCD models}

In the bottom-up models the conformal boundary is at $z=0$, and $z$
is cut off at some $z_0 \sim \Lambda^{-1}_\mathrm{QCD}$ in the HW case, while SW models
have exponential damping factors from a nontrivial dilaton
as $z\to\infty$.

\paragraph{Vector sector}

For the vector fields 
$\mathcal{V}=\frac12(\mathcal{B}_L+\mathcal{B}_R)$,
the mode equation does not couple to $S_{5,X}$.
With AdS metric it reads
\begin{equation}\label{psinSW}
\partial_z \left( \frac{e^{-\Phi(z)}}{z} \partial_z \psi_n \right) + \frac{e^{-\Phi(z)}}{z} M_n^2 \psi_n = 0.
\end{equation}
In the HW models, where $\Phi(z)\equiv 0$,
this is solved by 
$\psi_n(z)\propto zJ_1(M_n z)$ with $M_n$ determined
by the boundary conditions $\psi_n(0)=\psi'_n(z_0)=0$ through the zeros of the Bessel function $J_0$. Identifying again
the lowest mode with the $\rho$ meson,
$z_0$ gets fixed to $z_0=\gamma_{0,1}/M_\rho$, where $\gamma_{0,1}= 2.40483\ldots$
denotes the first zero of $J_0$. For $M_\rho=775$ MeV, this gives
$z_0=3.103 \, \text{GeV}^{-1}$.

Neither the \WSS model nor the HW models show the expected
behavior $m^2_n \propto n$ from linear confinement. This
is instead realized, by construction, in the SW models with $\Phi=c^2 z^2$ \citep{Karch:2006pv}.
Defining a new function
\begin{equation}
\psi_n(z) = e^{{\Phi(z)}/{2}} z^{1/2} \tilde\psi_n(z),
\end{equation}
the above mode equation can be recast into a Schr\"odinger-like form:
\begin{equation}
- \tilde\psi_n''(z) + V(z)\, \tilde\psi_n(z) = M_n^2 \, \tilde\psi_n(z),
\end{equation}
with an effective potential
\begin{equation}
V(z) = c^4 z^2 + \frac{3}{4z^2}.
\end{equation}
This leads to strictly linear eigenvalues $M_n^2=4c^2 n$, $n=1,2,3,\ldots$, such that $c= M_\rho/2$.

Table \ref{tab:vmasses} compares the vector-meson masses obtained in
the \WSS model and in the AdS/QCD models with experimental values
of the ground-state $\rho(770)$ meson and excited $\rho$ vector mesons,
grouped in radial Regge trajectories as proposed by \citet{Masjuan:2012gc}. 

In the \WSS model and the HW model, the mass of the
first excited $\rho$ meson is significantly higher than that of $\rho(1450)$,
and could perhaps be identified with the lowest state on the second radial
trajectory, which according to \citep{Masjuan:2012gc} would be the $\rho(1700)$.
In fact, in the D4/D8-brane construction underlying the \WSS model, one can
also consider massive open string excitations, which are not only responsible
for mesons with spin greater than 1, but also provide additional lower-spin
mesons, in particular those appearing as radial excitations (see Sec.~\ref{ISS}). \citet{Imoto:2010ef}
indeed have suggested interpreting the $\rho(1450)$ as the first massive string
excitation, forming a multiplet with the states $a_2(1320)$, $b_1(1235)$, $a_0(1450)$,
and $\pi(1300)$.\footnote{Since VMD in the \WSS model only involves the states
described by the flavor gauge fields, this would suggest that the $\rho(1700)$
and not so much the $\rho(1450)$ is participating in electromagnetic interactions
described by VMD. This is not supported by experimental data \citep{BaBar:2012bdw,Achasov:2018ujw,BESIII:2020xmw,SND:2023gan}; however,
see ch.~63, \textit{Spectroscopy of Light Meson Resonances}, of the PDG \citep{ParticleDataGroup:2026mpi} for the
confusing status of $\rho$ meson resonances, which clearly cannot be well
described in a narrow-width approximation.
}

The SW model, on the other hand, aims at describing the leading radial Regge trajectory of the vector mesons by the modes of the 5-dimensional flavor gauge field. However, when fitted to the mass of the lowest $\rho$ meson, the strict proportionality $M_n^2\propto n$ underpredicts
the excited states. 
This can be corrected by modifying the dilaton such that it becomes quadratic in $z$ only
at large $z$ as done in the simple semi-hard-wall (SHW) model \citep{Kwee:2007dd}.
Another possibility that has appeared recently is to use deep learning
to determine the geometry of hQCD models \citep{Akutagawa:2020yeo,ThomasArun:2025uyi}.

A central quantity describing correlation functions involving
vector currents is
the vector bulk-to-boundary propagator. It is obtained by replacing $M_n^2$
by $q^2=-Q^2$ and the boundary conditions by
$\mathcal{J}(Q,0)=1$ and $\partial_z \mathcal{J}(Q,z_0)=0$. In the
HW models this is given in closed form by modified Bessel functions $K$ and $I$,
\be\label{HWVF}
\mathcal{J}(Q,z)=
Qz \left[ K_1(Qz)+\frac{K_0(Q z_0)}{I_0(Q z_0)}I_1(Q z) \right]=
\sum_{n=1}^\infty \frac{g_5 F_n^V \psi_n(z)}{Q^2+M_n^2},
\ee
where $F_n^V=\lim_{z\to0}|\psi_n'(z)/(g_5 z)|$ is the decay constant of
the $n$th vector meson,
\be\label{FVdef}
\langle 0 | J_\mu^a(0) | v_n^b(q,\lambda) \rangle = F_n^V \varepsilon_\mu(q,\lambda) \delta^{ab}.
\ee
While no closed-form expression seems to be known in the \WSS model,
the bulk-to-boundary propagator of the SW model
can be expressed in terms of the confluent hypergeometric Tricomi function 
$U(a,b,z)$ \citep{Grigoryan:2007my}:
\be\label{SWVF}
\mathcal{J}^{\rm SW}(Q,z)=\Gamma(1+{\frac{Q^2}{4c^2}})\, U\left({\frac{Q^2}{4c^2}},0,(c z)^2\right).
\ee

Requiring that the vector current two-point function matches
the leading-order pQCD result at large $Q^2$ \citep{Erlich:2005qh},
\be\label{PiVas}
\Pi_V(Q^2)=-\frac1{g_5^2 Q^2} \left( \frac1z \partial_z \mathcal{J}(Q,z) \right)\Big|_{z\to0} 
\sim -\frac{N_c}{24\pi^2}\ln Q^2,
\ee
determines $g_5$ as given in Eq.~\eqref{eq:g5} for both HW and SW models,
because their bulk-to-boundary propagators have the same asymptotic behavior, $\J(Q\to\infty,z)\to (Qz) K_1(Qz)$. The \WSS model, however,
cannot be matched like this.

As also shown in Table \ref{tab:vmasses}, the HW models
come closest to the experimental value for $F_\rho$. The latter
can be matched by reducing
$g_5^2\to 0.894(2\pi)^2$ for $N_c=3$, which
in fact turns out to lead to better phenomenological
predictions in the case of hadronic light-by-light scattering
as well as hadronic vacuum polarization \citep{Leutgeb:2022cvg}
(see Sec.\ \ref{sec:hlbl}).

When the mass of the strange quark is included, 
either by following \citet{Aharony:2008an,Hashimoto:2008sr} in the case
of the \WSS or HW2 models or by the non-normalizable mode of the
bifundamental scalar $X$ in the HW1 model \citep{Abidin:2009aj},
a shortcoming of 
the holographic models is that the
$\phi$ meson remains degenerate with the $\rho$ and $\omega$ mesons.

\begin{table}
\centering
\begin{tabular}{lccccc}
\toprule
 & $F_V^{1/2}$ & $M_V$ & $M_{V^*}$ & $M_{V^{**}}$ & $M_{V^{***}}$ \\
\colrule
\WSS \citep{Sakai:2004cn} & 405 & \textit{775} &  1606 & 2432 & 3254 \\
HW \citep{Erlich:2005qh}  & 329 &  \textit{775} & 1779 & 2789 & 3800 \\ 
SW  \citep{Karch:2006pv} & 260 & \textit{775} & 1096 & 1342 & 1550 \\
SHW \citep{Kwee:2007dd,Leutgeb:2022cvg} & 314 & \textit{775} & \textit{1465} & 1903 & 2230 \\
\colrule
$F_\rho$, $M_{\rho}$ (exp.) & 348(1) & 775.26(23) & 1465(25) &  1900 & 2150 \\ 
\quad 2nd radial traj. &&& 1720(20) & 2000$^\dagger$ & 2270$^\dagger$ \\
\botrule
\end{tabular}
\caption{
Values of the masses of the four
lowest vector mesons obtained in the \WSS model, the minimal HW and SW models,
and the interpolating semi-hard-wall (SHW) model \citep{Kwee:2007dd} 
in comparison with the masses of the first and second radial Regge trajectories
as identified by \citet{Masjuan:2012gc}, with
PDG values and errors for the listed
$\rho$ 
mesons in MeV if available \citep{ParticleDataGroup:2026mpi} (a dagger marks states not listed).
All holographic models are with $M_V=775$ MeV as input;
the extra parameter of the SHW model \citep{Kwee:2007dd} was used
to also match $M_{V^*}$ with the mass of $\rho(1450)$ (matched values indicated by italic fonts).
}
\label{tab:vmasses}
\end{table}

\paragraph{Axial sector}

As concerns axial-vector mesons, the spectrum depends on the details of how
chiral symmetry breaking is implemented.

The Hirn-Sanz (HW2) model, like the \WSS model\footnote{In the following, the index $n$ enumerates vector and axial-vector modes separately. In the \WSS model Eq.~\eqref{psinSS} yields $M_n\equiv M^V_n=\sqrt{\lambda_{2n-1}}\mkk$ and $M^A_n=\sqrt{\lambda_{2n}}\mkk$.}, does not introduce
a bifundamental scalar for chiral symmetry breaking.
In the \WSS model, chiral symmetry breaking is a consequence of the joining of the
D8-\AD\ branes on which gauge modes with odd parity in $Z$
correspond to axial-vector mesons; in the HW2 model,
this corresponds to the boundary conditions
$\psi^A_n(0)=\psi^A_n(z_0)=0$ and $\psi_n^A{}'(z_0)\not=0$
for the same equation as for vector mesons, \eqref{psinSW} with $\Phi\equiv 0$,
leading to
\be
\psi^A_n(z) \propto z J_1(M^A_n z),\quad M^A_n=\gamma_{1,n}/z_0,
\ee
where $\gamma_{1,n}$ is the $n$th zero of $J_1$.
Since $z_0$ is already fixed by $M_\rho$, this leads to the prediction
$M^A_1=(\gamma_{1,1}/\gamma_{0,1})  M_\rho = 1235\, \mathrm{MeV}$, 
while the \WSS model yields\footnote{This is for $M_\rho=775$ MeV to be in line with the other models; for the traditional choice of $\MKK=949$ MeV this is slightly higher at 1189 MeV.} 1186.5 MeV,
which both agree
remarkably well with the experimental value $M_{a_1(1260)}=1230(40)$ MeV.

In the models with bifundamental scalar, the mode equation for
axial-vector mesons $\mathcal{A}_\mu=(\mathcal{B}^L_\mu
-\mathcal{B}^R_\mu)/2$ have an extra potential contribution
from the background $X(z)=v(z)/2$, effectively replacing
$M_n^2$ by $(M^A_n)^2-(g_5v(z)/z)^2$ in \eqref{psinSW}.
This raises their masses
compared to the vector-meson masses when the same boundary
conditions are used as for the vector mesons (HW1), and even more
with the
boundary condition $\psi_n^{A}(z_0)_\mathrm{HW3}=0$.
See Table \ref{tab:avmasses} for a comparison of the predictions of
$a_1$ and (ideally mixed) $f_1$ meson masses in the chiral \WSS and HW2 models and
in the HW1, HW3, and SW models with quark masses and chiral
condensate matched to reproduce $m_\pi$ and $f_\pi$.
Note that in this case the SW model as
set up by \citet{Colangelo:2024xfh} has linear radial
Regge trajectories only in the vector sector; in contrast to the
latter, it now strongly overestimates the masses of the known axial-vector mesons.

\begin{table}
\centering
\begin{tabular}{lcccc}
\toprule
 & $M_A$ & $M_{A^*}$ & $M_{A^{**}}$ & $M_{A^{***}}$ \\
\colrule
\WSS (chiral) \citep{Sakai:2004cn} & 1186 & 2020 & 2843 & 3664 \\
HW2 (chiral) \citep{Hirn:2005nr}  &  1235 & 2261 & 3279 & 4294 \\ 
HW1  ($m_\pi=135$) \citep{Leutgeb:2021mpu} & 1367 & 2141 & 2987 & 3934 \\
HW3 ($m_\pi=135$) \citep{Leutgeb:2021mpu} & 1431 & 2421 &3398 &4387 \\
SW ($m_\pi=135$) \citep{Colangelo:2024xfh,Leutgeb:2025jmv} & 1679 & 2674 & 3503 & 4241 \\
\colrule
$M_{a_1}$ (exp.) & 1230(40) & 1655(16) & 2095$^\dagger$ &   \\ 
\quad 2nd radial traj. && 1930$^\dagger$ & 2270$^\dagger$ \\
\botrule
\end{tabular}
\caption{
Values of the masses of the four
lowest axial-vector mesons obtained in the \WSS model, the chiral Hirn-Sanz (HW2) model and models with bifundamental scalar (the latter with fitted $m_\pi$ and $f_\pi$)
in comparison with the masses of the first and second radial Regge trajectories
as identified by \citet{Masjuan:2012gc}, with
PDG values and errors for the listed
$a_1$ mesons 
in MeV \citep{ParticleDataGroup:2026mpi} (a dagger denotes not listed).
}
\label{tab:avmasses}
\end{table}

\paragraph{Pseudoscalars and GOR (without U(1)$_A$)}

In the HW2 case, like in the \WSS model, the pion is massless
unless a mass term is added by hand, and it is still given
by the holonomy \eqref{pionU}, only with a simpler holographic wave function
\be\label{HW2alpha}
\alpha(z)=1-z^2/z_0^2,
\ee
yielding
$f_\pi^2=2/({g_5^2z_0^2})$.

However, the extra scalar field $X$ in the HW1-type models
also produces a nontrivial coupling between the longitudinal component of the axial-vector field and the pion field(s). Defining 4D Fourier modes $A_{\mu \,\parallel}^a(q,z)= -iq_\mu \varphi(q,z)$ and $U(x,z)=\exp{\left [2i t^a \pi^a(x,z)/f_\pi\right]}$ (for $\SU(2)$, $t^a=\sigma^a/2$, with $\sigma^a$ being the Pauli matrices) one gets the system of coupled equations \citep{Erlich:2005qh,Kwee:2007dd,Abidin:2009aj}:
\begin{eqnarray}
\partial_z\left(\frac{e^{-\Phi}}{z} \, \partial_z \varphi^a \right)
+\frac{g_5^2 \, v(z)^2 e^{-\Phi}}{z^3} (\pi^a-\varphi^a)&=&0~,\label{HW-AL}\\
-q^2\partial_z\varphi^a+\frac{g_5^2 \, v(z)^2}{z^2} \, \partial_z \pi^a &=&0~.\label{HW-Az}
\end{eqnarray}
In the HW case, one has two options for consistent boundary conditions
at $z_0$ corresponding to HW1 and HW2/HW3 defined above, where the former
requires one to include a subtraction of an IR boundary term in $S_\mathrm{CS}$
which vanishes for HW2/HW3 \citep{Leutgeb:2021mpu}; in the SW case
the condition is regularity as $z\to\infty$.

There are now normalizable modes with $\varphi(q,0)=0=\pi(q,0)$
for an infinite tower of pseudoscalars with $q^2=m_n^2$, and
the nonnormalizable solution with $\varphi_S(q,0)=0$, $\pi_S(q,0)=-1$
can be represented as
\citep{Abidin:2009aj}
\begin{equation}
\varphi_{S}(q,z) = -g_5 \sum_{n=1}^\infty \frac{f_{\pi_n} m_n^2 \varphi_n(z)}{q^2-m_n^2},\quad
\pi_{S}(q,z) = -g_5 \sum_{n=1}^\infty \frac{f_{\pi_n} m_n^2 \pi_n(z)}{q^2-m_n^2},
\label{phipisumrule}
\end{equation}
where $f_{\pi_n}=-\6_z\varphi_n(\epsilon)/(g_5\epsilon)$
are the decay constants defined by 
\be\label{deffpi}
\langle 0 | J^{A,a}_{\mu}(0)|\pi_n^b(q)\rangle
=i f_{\pi_n} q_\mu \delta^{ab}.
\ee

In the chiral limit $M_q\to0$ but $\Sigma$ finite,
$m_1$ approaches zero, with $\pi_1(z)\to -1/(g_5 f_{\pi_1})$ except
very close to the boundary at $z=0$, whereas the modes $n\ge2$ remain
massive ($m_2=1.9$ GeV in HW1, 1.7 GeV in HW3, 2.1 GeV in the SW model,
all significantly above the mass of the physical $\pi(1300)$; HW2 has no excited pseudoscalars at all).

For sufficiently small quark mass matrix $M_q=m_q\mathbf1$ and
finite $\Sigma=\sigma\mathbf1$ one can show that \citep{Erlich:2005qh,Abidin:2009aj,Leutgeb:2021mpu}
$
f_{\pi_n}^2 m_n^2 \to 2 m_q \sigma.
$
For the massive excited pseudoscalars, i.e., for $n>1$, this implies that $f_{\pi_n}\to 0$, so they decouple from the axial current (while still having interactions with other mesons, see Sec.\ \ref{sec:hlbl}); and the lightest pseudoscalars with $f_{\pi_1}=f_\pi$ obey the Gell-Mann--Oakes--Renner (GOR) relation
\be
f_\pi^2 m_\pi^2=2 m_q \sigma.
\ee

\subsection{Chiral Lagrangian and interactions with (axial) vector mesons}

This section summarizes the main predictions of the WSS and hard-wall (HW) models for the leading chiral low-energy constants, pion form factors and resonance sum rules. 
Consistently with the large-$N_c$ limit, meson loops and resonance widths are not included. The $L_i$ are the three-flavor Gasser-Leutwyler (GL) \ $\mathcal O(p^4)$ coefficients at the scale $\mu\simeq  M_\rho$ \citep{Gasser:1983yg,Gasser:1984gg}.

The chiral Lagrangian is written in terms of $U(x)=u(x)^2=\exp(2i\pi^a(x)t^a/f_\pi)$, where the $\pi^a$ is the octet of Goldstone bosons (hereafter referred to as pions) of the spontaneous chiral symmetry breaking. 
We take $U$ to transform as $U\to g_RUg_L^\dagger$, following the
convention of \citet{Gasser:1984gg,Bijnens:1999sh} customary in
chiral perturbation theory, which is also used in Sec.~\ref{sec:weak} below.
This differs from the convention of Eq.~\eqref{Utransf} 
and most of the holographic literature by the exchange
$g_L\leftrightarrow g_R$.

The Lagrangian is  an expansion in the number of derivatives:
\begin{align}
 \mathcal L_2={}&\frac{f_\pi^2} {4}\mathrm{Tr}\left( u_\mu u^\mu+\chi_+\right) ,\label{chiral2}\\
 \mathcal L_4={}&L_1\mathrm{Tr}\left( u_\mu u^\mu\right) ^2
 +L_2\mathrm{Tr}\left( u_\mu u_\nu\right)\mathrm{Tr}\left( u^\mu u^\nu\right)
 +L_3\mathrm{Tr}\left( u_\mu u^\mu u_\nu u^\nu\right)\notag\\
 &\qquad\qquad\qquad\qquad-iL_9\mathrm{Tr}\left( f_{+\mu\nu}u^\mu u^\nu\right) 
 +\frac{L_{10}}4\mathrm{Tr}\left( f_{+\mu\nu}f_+^{\mu\nu}-f_{-\mu\nu}f_-^{\mu\nu}\right) +\ldots,
 \label{eq:chiral4}
\end{align}
with
\begin{align}
u_\mu=i\{u^\dagger(\partial_\mu-ir_\mu)u-u(\partial_\mu-i\ell_\mu)u^\dagger\},\quad f_{\pm\mu\nu}=uF_{L\mu\nu}u^\dagger\pm u^\dagger F_{R\mu\nu}u,\quad \chi_\pm=u^\dagger \chi u^\dagger \pm u\,\chi^\dagger u,
\quad \chi=2B_0\mathcal{M},
\end{align}
where $r_\mu$ and $\ell_\mu$ are external vector sources and $F_{\{R,L\}\mu\nu}$ their field strength,  $\mathcal M$ is the quark-mass matrix, $B_0$ is the chiral condensate parameter, and the dots stand for the mass terms $L_4,\ldots,L_8$ and the contact terms from source-only.


In hQCD, the pion field in the chiral Lagrangian is written as (\ref{pionU}). This pion field can be decomposed as
\be
U(x)=\xi_R^{-1}(x)\xi_L(x)
\label{Uxixi}
\ee
with
\be
\xi_R^{-1}(x)=\mathrm{P}\exp\left(i\int_0^{\infty} A_Z(x,Z)\,dZ\right)\ ,~~~
\xi_L(x)=\mathrm{P}\exp\left(i\int^0_{-\infty} A_Z(x,Z)\,dZ\right)\ .
\ee
These fields transform under the 5-dimensional gauge transformation as
\be
\xi_R(x)\ra h(x)\,\xi_R(x)\, g_R^{-1}\ ,~~~\xi_L(x)\ra h(x)\,\xi_L(x)\, g_L^{-1} ,
\label{xi}
\ee
where $g_R$ and $g_L$ are as given in (\ref{gLgR})\footnote{Note that $g_L$ and $g_R$ in this section are exchanged from those in (\ref{gLgR}).}, and $h(x)=g(x,Z=0)$. The decomposition (\ref{Uxixi}) and the transformation (\ref{xi}) are the same as those of
the pion fields in the hidden local symmetry approach \citep{Bando:1984ej}\footnote{See \citet{Bando:1987br} and \citet{Harada:2003jx} for reviews.} to meson Lagrangians. In this approach, the gauge transformation associated with $h(x)$ in (\ref{xi}) is called the hidden local symmetry, and the $\rho$-meson is introduced as its gauge field.
Thus, hQCD realizes hidden local symmetry as the gauge symmetry at $Z=0$.\footnote{See Appendix \ref{appDD} for a bottom-up holographic model of QCD based on hidden local symmetry \citep{Son:2003et}.} Since the action of hQCD, such as (\ref{5dim_action}), is invariant under the 5-dimensional gauge transformation, it is automatically invariant under hidden local symmetry, as well as chiral symmetry. In this formalism, using the hidden local symmetry, it is possible to choose a gauge such that $\xi_R^{-1}(x)=\xi_L(x)=:u(x)$, with which (\ref{Uxixi}) is written as $U(x)=u(x)^2$ as above. In this gauge, the chiral symmetry transformation (\ref{xi}) can be realized by choosing $h(x)$ such that $g_R u(x) h^{-1}(x)=h(x)u(x) g_L^{-1}$. In order to emphasize that $h(x)$ satisfying this relation depends on $g_L$, $g_R$ and $u(x)$, we write it as $h(g,u)$. Then, the low energy effective action of hQCD written with $u(x)$ as the pion field is invariant under the transformation
\be
u(x)\ra g_R\, u(x)\, h^{-1}(g,u)=h(g,u)\,u(x)\,g_L^{-1}\ .
\ee
This is precisely the symmetry imposed on the Lagrangians (\ref{chiral2}) and (\ref{eq:chiral4}).

In  hQCD models, the contributions to the GL coefficients due  to vectors  and axial vectors come from the YM part of the 5D Lagrangian. The $\mathrm{Tr}(u_\mu u^\mu)$ term is generated by the $\mathrm {Tr}(F_{\mu z})^2$ term, whereas a single quartic term in the chiral field $\propto \mathrm {Tr}[u_\mu,u_\nu]^2$ (a Skyrme term), is generated by the $\mathrm {Tr}(F_{\mu\nu})^2$ term.  As a consequence, the models predict relationships among the GL parameters that would hold only in the large-$N_c$ limit of QCD, such as, for instance,  the relation  $L_9=-L_{10}$, which binds the pion charge radius to $\Pi_V(0)-\Pi_A(0)$. Since $L_9+L_{10}$ is measurable in $\pi\to e\nu\gamma$, additional bulk operators would be needed to reproduce radiative pion decays.  

In the WSS model, the GL coefficients depend on two overlap integrals of the axial-vector (pion) zero mode: $I_1=\int\mathrm{d} Z\,k(Z)^{-1/3}(1-\psi_0^2(Z))=4.686$, $I_2=\int\mathrm{d} Z\,k(Z)^{-1/3}(1-\psi_0^2(Z))^2=2.505$, with
\begin{equation}
L_1=\frac{\kappa I_2}{32},\qquad L_2=2L_1,\qquad L_3=-6L_1,\qquad L_9=-L_{10}=\frac{\kappa I_1}{4}.
 \label{eq:wsslec}
\end{equation}
Numerical values, for $\kappa=0.00745$, having set $f_\pi=92.4$ MeV  (see Eq.\eqref{eq:fpiwss}),  are shown in Table \ref{tab:GLWSSHS}.\footnote{Corrections, producing 
a $4.5\%$ shift of $L_{1,2}$, scaling as $N_c/\lambda$ rather than $N_c\lambda$, preserving  $L_2=2L_1$ but breaking $L_3=-6L_1$, have been discussed by \citet{Hoyos:2022ptd,Lublinsky:2025hks}.}

In the HW2 model, similarly to the WSS model, the GL coefficients follow from overlaps of the axial-vector zero mode (pion wave function)  $\alpha(z)=1-z^2/z_0^2$:
\begin{equation}
 L_1=\frac{1}{32\,g_5^2}\int_0^{z_0} \frac{dz}{z}\left(1-\alpha^2(z)\right)^2=\frac{11}{768g_5^2},\qquad 
 L_9=-L_{10}=\frac{1}{4\,g_5^2}\int_0^{z_0}\frac{dz}{z}\,\left(1-\alpha^2(z)\right)=\frac{3}{16\,g_5^2}, 
 \label{eq:hslec}
\end{equation}
with again $L_2=2L_1,\;L_3=-6L_1.$

As Table~\ref{tab:GLWSSHS} shows, the UV-fixed coupling $ g_5^2=4\pi^2$ cannot also reproduce the physical $f_\pi$; the fit with $f_\pi=87\,\mathrm{MeV}$ (in the chiral limit) corresponds to $12\pi^2/g_5^2\simeq4.30$ rather than $N_c=3$, so a deformation of the model is needed to satisfy both \citep{Hirn:2005nr}.

The HW1 model adds a bifundamental scalar $X$ dual to $\bar q q$. The extra parameter $\sigma$ removes the HW2 tension and allows  fixing $z_0^{-1}=323\,\mathrm{MeV}$, $m_q=2.29\,\mathrm{MeV}$, $\sigma=(327\,\mathrm{MeV})^3$ from $M_\rho,m_\pi,f_\pi$.

The scalar sector also provides the mass-dependent coefficients $L_4,\ldots,L_8$, which the HW2 models cannot predict. Scalar exchange modifies $L_3=-6L_1$, while $L_2=2L_1$, $L_4=L_6=0$ and, for equal vector and axial kinetic functions, $L_9=-L_{10}$ survive \citep{DaRold:2005vr,Domenech:2010aq}.

\begin{table}[h]
\centering
\begin{tabular}{lrrrrr}
\toprule
 & WSS & HW1: & HW2: & HW2: & \\
Quantity & benchmark & UV matching & UV matching & low-energy fit & Experiment \\
\midrule
$M_\rho$ (MeV) & $\simeq 776$ (input) & 776 (input) & 776 (input) & 776 (input) & 775.26(23) 
\\
$f_\pi$ (MeV) & 92.4 (input) & 92.4 (input) & 72.63 & 87 (input) & 92.2(1) 
\\
$g_5^2$ & --- & $4\pi^2 = 39.478$ & $4\pi^2 = 39.478$ & 27.514 & --- \\
$10^3 L_1$ & 0.583 & 0.498 & 0.363 & 0.521 & 0.53(6) 
\\
$10^3 L_2$ & 1.166 & 0.996 & 0.726 & 1.041 & 0.81(4) 
\\
$10^3 L_3$ & $-3.499$ & $-2.989$ & $-2.177$ & $-3.123$ & $-3.07(20)$ 
\\
$10^3 L_9 = -10^3 L_{10}$ & 8.727 & 6.005 & 4.749 & 6.815 & 5.93(43) 
\,/\,4.06(39) 
\\
$\langle r_\pi^2 \rangle$ (fm$^2$) & 0.477 & 0.329 & 0.421 & 0.421 & 0.434(5) 
\\
\bottomrule
\end{tabular}
\caption{Predictions with explicitly different input choices, compared with experiment. All holographic $L_i$ values are tree-level results of the 5D gauge sector, obtained by integrating out the complete vector and axial-vector towers in closed form; scalar-meson exchange is not included, which in HW1 would shift $L_3$ by roughly $+0.2\times 10^{-3}$ and is the sole tree-level source of $L_4$, $L_5$, $L_6$ and $L_8$, not listed here. WSS inputs and conventions are those of \citet{Sakai:2005yt} with $M_{KK}=0.949$\,GeV, $\lambda=16.63$ and $N_c=3$. The HW2 low-energy fit uses the chiral-limit normalization 87\,MeV of the original comparison by \citet{Hirn:2005nr}; it does not simultaneously satisfy the $N_c=3$ UV matching $g_5^2=12\pi^2/N_c$. Radii are tree-level quantities, with no pion-loop contribution.  Experimental column: $M_\rho$ is the $\rho(770)$ mass, $f_\pi=F_\pi/\sqrt{2}$ with $F_\pi=130.2(8)$\,MeV from $\pi\to\mu\nu$, and $\langle r_\pi^2\rangle$ the charged-pion charge radius, all from the PDG \citep{ParticleDataGroup:2024cfk}.  The experimental $L_i$ values are renormalized couplings $L_i^r(\mu=M_\rho)$ from $O(p^6)$ analyses: $L_{1,2,3}$ from the fit of \citet{Bijnens:2014lea}, $L_9$ from the pion vector form factor \citep{Bijnens:2002hp}, and $-L_{10}$ from hadronic $\tau$ decays \citep{Gonzalez-Alonso:2008sva}. Physical $f_\pi$ and scale-dependent $L_i^r$ are not strictly the chiral-limit, large-$N_c$ quantities predicted at tree level, so the comparison is indicative at the level of $O(1/N_c)$ and chiral corrections.}
\label{tab:GLWSSHS}
\end{table}

In the holographic model,  resonances in the vector and axial vector towers and the pion are characterized by their decay constants defined in \eqref{FVdef} (with an analogous formula for the axial vectors) and in \eqref{deffpi} for the pion, which depend on the values of derivatives of the normalizable modes  in the UV-limit $z\to0$
(see Sec.~\ref{pion} for more details).

The vector--pion couplings are defined by the cubic vertex $\mathcal{L}\supset g_n\,\epsilon^{abc}\,V^a_{n\mu}\,\pi^b\partial^\mu\pi^c$, with dimensionless $g_n\equiv g_{V_n\pi\pi}$ also obtained as overlap integrals 
 of the normalizable modes. For example, in the WSS model, one obtains  $|g_{\rho\pi\pi}| \approx 34.0/\sqrt{\lambda N_c} \approx
 4.81$  \citep{Sakai:2005yt},  with a  resulting width $\Gamma_{\rho\to\pi\pi}=g_{\rho\pi\pi}^2M_\rho(1-4m_\pi^2/M_\rho^2)^{3/2}/(48\pi)$ of $119\,\mathrm{MeV}$ (massless pions) or $97\,\mathrm{MeV}$ (physical phase space) 
 for $\lambda\approx 16.63$, below the observed $\simeq150\,\mathrm{MeV}$ \citep{Brunner:2015oqa,Hechenberger:2023ljn};
 a similar deficit is found in the HW1 model \citep{Erlich:2005qh}. The $a_1\rho\pi$ coupling can be calculated analogously, yielding $|g_{a_1\rho\pi}|\approx 34.43M_{\mathrm{KK}}/\sqrt{\lambda N_c}$ 
 in the case of the \WSS model 
 \citep{Hechenberger:2023ljn}.

\subsection{Interactions with photons: VMD}
The action of QCD with external fields is
\begin{equation}
\mathcal{L}=\mathcal{L}_{\rm QCD}+\bar q_L\,\gamma^\mu\ell_\mu q_L+\bar q_R\,\gamma^\mu r_\mu q_R.
\end{equation}
 The photon is a vector source of the form
\begin{equation}
\ell_\mu=r_\mu=e\,Q_{\rm em}\,A^\gamma_\mu,\qquad Q_{\rm em}={\rm diag}\left(\tfrac23,-\tfrac13,-\tfrac13\right).
\end{equation}
In hQCD, external  sources are the UV boundary values of the 5D gauge fields $L_M$ and $R_M$. 
Thus $A^\gamma_\mu$ fixes a non-normalizable mode and is never itself a bulk state. In terms of Euclidean momentum (virtuality $Q=\sqrt{-q^2}$), the solution of the 5D EOM factorizes as $V_\mu^\perp(Q,z)=e\,Q_{\rm em}A^{\gamma\perp}_\mu(Q)\mathcal{J}(Q,z)$ (in the gauge $V_z=0$), where $\mathcal{J}$ is the vector bulk-to-boundary propagator, whose expansion in vector eigenfunctions is given in Eq.~\eqref{HWVF}.

Thus, any electromagnetic vertex is a 5D overlap of $\mathcal{J}$ with hadronic wavefunctions, so it is a pure sum of $\rho_n$ poles with no contact term. 

As an  important example, the electromagnetic form factor of the pion can be written as an overlap of the vector bulk-to-boundary propagator $\mathcal{J}$ with the pion charge density $\rho_\pi(z)$,
\begin{equation}
 F_\pi(Q^2)=\int_0^{z_0}dz\,\rho_\pi(z)\,\mathcal{J}(Q,z),\qquad
 g_n=g_5\int_0^{z_0}dz\,\rho_\pi(z)\,\psi^V_n(z),\qquad
 \int_0^{z_0}dz\,\rho_\pi(z)=1 .
 \label{eq:gndef}
\end{equation}
For the HW2 model $\rho_\pi=(\partial_z\alpha)^2/(g_5^2f_\pi^2\,z)$ with $\alpha$ given in \eqref{HW2alpha}, so that
\begin{equation}
 g_n=\frac{1}{g_5f_\pi^2}\int_0^{z_0}\frac{dz}{z}\,\psi^V_n(z)\,\big[\partial_z\alpha(z)\big]^2=\frac{2g_5}{z_0^2}\int_0^{z_0}dz\,z\,\psi^V_n(z).
 \label{eq:gnHW}
\end{equation}
Then, using \eqref{HWVF},  the pion form factor can be written: 
\begin{equation}
 F_\pi(Q^2)=\sum_n\frac{F^V_n\,g_n}{Q^2+M_{V_n}^2},\qquad\mathrm{with}\quad 
F_\pi(0)=\sum_n\frac{F^V_n\,g_n}{M_{V_n}^2}=1.
 \label{eq:Fpi}
\end{equation}
The sum rule follows from $\mathcal{J}(0,z)=1$ and expresses charge normalization.
In the HW2 model, the terms in the series of $F_\pi(0)$ are pure numbers, independent of the parameters of the model, the first terms being 
$1.11$,$-0.14$, $0.05$, $-0.02$,$\ldots$ with the $\rho$   overshooting the charge by $11\%$ and heavier states restoring its value. 

These numbers are, however, model dependent. In the \WSS model the successive contributions are $1.31$, $-0.34$, $0.05$, $-0.01$, \ldots, with the $\rho$ overshooting the total by $31\%$, whereas in HW1 (with $f_\pi=92.4$ MeV) they follow a different pattern, $0.80$, $0.29$, $-0.09$, $-0.01$, \ldots, with the $\rho$ accounting for only $80\%$. These numbers are largely determined by the form of pion wave function and hence, ultimately, by how chiral symmetry is broken in each model.

This shows that hQCD realizes Sakurai's idea   of vector meson dominance (VMD) \citep{Sakurai:1960ju}, albeit in a generalized form involving not only the $\rho$ meson but also its entire tower of radial excitations.

Similarly, the couplings $g_n\equiv g_{V_n\pi\pi}$ satisfy
\begin{equation}
 \sum_n\frac{g_n^2}{M_{V_n}^2}=\frac1{3f_\pi^2},
 \label{eq:pipisumrules}
\end{equation}
which, when saturated by the $\rho$ alone, gives $M_\rho^2\simeq3g_{\rho\pi\pi}^2f_\pi^2$ (the WSS model yields $3.05$ in place of $3$), rather than the KSRF value $2$ \citep{Kawarabayashi:1966kd,Riazuddin:1966sw}. The factor $1/3$ is, however, not universal: it changes once scalar exchange or higher-derivative terms are included. Finally, when only a finite set of resonances is retained, the heavier ones must be integrated out covariantly, as in hidden local symmetry, rather than simply discarded, if these sum rules are to be preserved \citep{Harada:2010cn}.

A comparison of the shape of $F_\pi(Q^2)$ in HW and SW models with experimental
data has been carried out by \citet{Kwee:2007dd,Kwee:2007nq}, with somewhat better agreement
for HW models.

\subsubsection{VMD and the Gasser-Leutwyler coefficients}

In  chiral resonance Lagrangians \citep{Ecker:1988te}\footnote{As discussed by \citet{Ecker:1989yg}, there are two different ways of introducing spin-1 resonances into the chiral Lagrangian: one describes vector and axial-vector mesons by Proca fields, the other by antisymmetric tensor fields.}, coupling the lightest vector nonet to the two $O(p^2)$ chiral building blocks (one coupling the vector to $f_+^{\mu\nu}$, governing the $\rho$--$\gamma$ transition, the other to $u^\mu u^\nu$, governing $\rho\pi\pi$) and integrating the vector out, at lowest order in $1/M_V^2$, generates local $O(p^4)$ terms. 

Only $L_1$, $L_2$, $L_3$, $L_9$ and $L_{10}$ receive vector contributions; $L_{4,5,6,7,8}$ are generated by scalar and $\eta'$ exchange. The relations $L_2 = 2L_1 ,\; L_3 = -6L_1$ 
are a pure consequence of the vector-exchange tensor structure.

Using also the  KSRF relation determines the vector exchange  contributions to the $L_i$  in terms $\propto f_\pi^2/M_{V}^2$ with $M_V=M_\rho$.  The axial vector exchange adds  a positive contribution $\propto f_\pi^2/M_{A}^2$ with $M_A=M_{a_1}$  to $L_{10}$ only. The agreement with the values derived from experimental fits is quite good. 
The one place where naive VMD visibly overshoots is $L_3$,  reflecting a scalar-octet contribution of opposite sign, so that for $L_3$ the vector and the scalar exchange partially cancel. For $L_9$ and $L_{10}$ the vector (and axial) saturation is essentially exact.

In holographic QCD, the same formulas hold summed over the whole tower.
In the HW2 model, integrating out the vector and axial towers gives rise to infinite sums for the GL coefficients \citep{Hirn:2005nr,Sakai:2005yt}
\begin{equation}
 L_1=\frac{L_2}{2}=-\frac{L_3}{6}=\frac{f_\pi^4}{8}\sum_n\frac{g_n^2}{M_{V_n}^4},\qquad
 L_9=\frac{f_\pi^2}{2}\sum_n\frac{F^V_n\,g_n}{M_{V_n}^4},\qquad
 L_{10}=-\frac14\sum_n\frac{(F^V_n)^2}{M_{V_n}^4}+\frac14\sum_n\frac{(F^A_n)^2}{M_{A_n}^4},
 \label{eq:rcht}
\end{equation}
 once again trading the exchange of the single lowest vector and axial-vector resonances for that of the full vector and axial vector  towers. 
 
The vector and axial sums in $L_{10}$ diverge separately, reflecting the UV divergence of the contact term $H_1$, depending only on the external sources; only their difference is finite. 

Using the completeness of the 5D resonance wave functions, the expressions \eqref{eq:rcht} can be shown to be equivalent to \eqref{eq:hslec}. This is yet another example of the remarkable ability of  hQCD models to provide closed analytic expressions for infinite sums of resonance contributions to 4D physical quantities. 

\subsection{OPE and short distance matching}
\label{sec:OPE}

Beyond the leading perturbative logarithm, the vector and axial-vector two-point functions
constrain the short-distance behavior of holographic QCD and relate its ultraviolet description to
the nonperturbative structure of the QCD vacuum. For flavor-nonsinglet currents
$J^a_{V\mu}=\bar q\gamma_\mu t^a q$ and $J^a_{A\mu}=\bar q\gamma_\mu\gamma_5t^a q$ in the chiral
limit,
\begin{equation}
 i\!\int\!\mathrm{d}^4x\,e^{iq\cdot x}
 \langle0|\mathrm{T}\{J^a_{J\mu}(x)J^b_{J\nu}(0)\}|0\rangle
 =\delta^{ab}(q_\mu q_\nu-q^2g_{\mu\nu})\Pi_J(-q^2),
 \qquad J=V,A,
 \label{eq:ope_current_convention}
\end{equation}
where $\Pi_A$ includes the pion contribution; the anomalous singlet axial current is not considered.

For the vector channel, a convenient quantity is the Euclidean Adler function \citep{Adler:1974gd},
\begin{equation}
 D(Q^2)=-12\pi^2Q^2\frac{\mathrm{d}\Pi_V(Q^2)}{\mathrm{d}Q^2},
 \qquad Q^2\equiv-q^2>0,
 \label{eq:ope_adler}
\end{equation}
which for the electromagnetic current is related to the inclusive ratio $R(s)$ by
\begin{equation}
 D_{\mathrm{em}}(Q^2)=Q^2\int_{s_{\mathrm{th}}}^{\infty}
 \frac{R(s)}{(s+Q^2)^2}\,\mathrm{d}s,
 \label{eq:ope_dispersion_em}
\end{equation}
connecting the Euclidean correlator to $e^+e^-\to$ hadrons data and to lattice
determinations \citep{Eidelman:1998vc,Davier:2023hhn}. Non-strange vector and axial-vector spectral functions are
also measured in hadronic $\tau$ decays \citep{Davier:2013sfa}.

At large Euclidean momentum, the QCD operator product expansion (OPE) reads \citep{Shifman:1978bx,Shifman:1978by}
\begin{equation}
 \Pi_{V,A}(Q^2)\sim\Pi_{\mathrm{pert}}(Q^2;\mu)
 +\frac{\mathcal C_4(Q^2,\mu)}{Q^4}
 +\frac{\mathcal C_6^{V,A}(Q^2,\mu)}{Q^6}
 +\mathcal O(Q^{-8}),
 \label{eq:ope_expansion}
\end{equation}
where $\mathcal C_d$ collects Wilson coefficients times vacuum matrix elements of dimension-$d$
operators. The leading perturbative term, common to both channels in the chiral limit, is
\begin{equation}
 \Pi_{\mathrm{pert}}^{\mathrm{LO}}(Q^2;\mu)
 \simeq-\frac{N_c}{24\pi^2}\log\frac{Q^2}{\mu^2}.
 \label{eq:ope_partonic}
\end{equation}
There is no $1/Q^2$ term in the standard chiral-limit OPE (see however the end of this section);
the first power correction is the gluon condensate, and dimension six contains four-quark
operators. With large-$N_c$ factorization,
\begin{align}
 \Pi_{V,A}(Q^2)
 &=-\frac{N_c}{24\pi^2}
 \left[\log\frac{Q^2}{\mu^2}+C(\mu)+\mathcal O(\alpha_s)\right]
 +\frac{1}{24Q^4}
 \left\langle\frac{\alpha_s}{\pi}G^a_{\mu\nu}G^{a\mu\nu}\right\rangle
 \nonumber\\
 &\quad+\epsilon_{V,A}\frac{14}{9}
 \frac{\pi\alpha_s\langle\bar q q\rangle^2}{Q^6}
 +\mathcal O(Q^{-8}),
 \qquad \epsilon_V=-1,\quad \epsilon_A=\frac{11}{7},
 \label{eq:ope_explicit_largeNc}
\end{align}
with $\langle\bar qq\rangle$ the condensate per light flavor, all quantities at a common scale
$\mu$, and $C$ a subtraction constant.

In an asymptotically AdS model, matching the leading logarithm fixes $g_5^2=12\pi^2/N_c$ (AdS radius
set to one) \citep{Erlich:2005qh}; the condensates supply further matching conditions, and HW2 and
HW1 behave differently. In HW2 the only corrections involve ratios of modified Bessel functions of
$x=Qz_0$, and since $K_\nu(x)/I_\nu(x)\sim\pi e^{-2x}$ the IR wall effects are exponentially
suppressed, beyond every inverse power of $Q$.

In HW1, chiral symmetry breaking is due to a bifundamental scalar, with action  \eqref{Sscalar} and   vacuum profile, containing a term proportional to the quark condensate,  given in \eqref{ScalarVEV}. 
At large $Q$ it corrects only the axial channel,
\begin{equation}
 \Pi_A^{\mathrm{HW1}}(Q^2)
 =-\frac{1}{2g_5^2}\log\frac{Q^2}{\mu^2}+c(\mu)
 +\frac{8\sigma^2}{5Q^6}+\cdots .
 \label{eq:ope_hw1_axial}
\end{equation}
Neither channel has a $1/Q^4$ gluon-condensate term, and the vector channel lacks the four-quark
$1/Q^6$ term: HW1 gives the expected power and sign of the chiral difference, but not the separate
OPEs of Eq.~\eqref{eq:ope_explicit_largeNc}.

In the minimal SW model with $\Phi(z)=c^2z^2$ the renormalized vector correlator
is \citep{Karch:2006pv,Zuo:2008re}
\begin{equation}
 \Pi_V^{\mathrm{SW}}(Q^2)
 =-\frac{1}{2g_5^2}
 \psi\!\left(1+\frac{Q^2}{4c^2}\right)+C,
 \label{eq:ope_sw_digamma}
\end{equation}
whose poles give the linear spectrum $m_n^2=4c^2 n$. The asymptotic expansion of the
digamma function yields
\begin{equation}
 \Pi_V^{\mathrm{SW}}(Q^2)
 =-\frac{1}{2g_5^2}\left[
 \log\frac{Q^2}{\mu^2}+\frac{2c^2}{Q^2}
 -\frac{4c^4}{3Q^4}+\mathcal O(Q^{-8})
 \right]+c(\mu),
\label{eq:ope_sw_expansion}
\end{equation}
with a vanishing $Q^{-6}$ coefficient but a surviving $Q^{-2}$ term. Identifying the $Q^{-4}$ term
with the gluon condensate in Eq.~\eqref{eq:ope_explicit_largeNc} gives
\begin{equation}
 \left\langle\frac{\alpha_s}{\pi} G^a_{\mu\nu}G^{a\mu\nu}\right\rangle_{\mathrm{SW}}
 =\frac{16c^4}{g_5^2}
 =\frac{4N_c}{3\pi^2}c^4.
 \label{eq:ope_sw_gluon}
\end{equation}

The missing condensates can be introduced through additional UV structure. \footnote{Condensates have been studied using 5D string worldsheets  ending on 4D Wilson loops by \citet{Andreev:2007vn,Andreev:2010zm}.}
 A power deformation of
the AdS warp factor in conformal coordinates,
\begin{equation}
 w_J(z)=\frac1z\left[1+c_4z^4+c_6^Jz^6+\mathcal{O}(z^8)\right],
 \label{warpdef}
\end{equation}
gives
\begin{equation}
 \Pi_J(Q^2)=-\frac{1}{2g_5^2}\log\frac{Q^2}{\mu^2}+c(\mu)
 +\frac{8c_4}{3g_5^2Q^4}
 +\frac{96c_6^J}{5g_5^2Q^6}
 +\mathcal O(Q^{-8}),
 \label{eq:ope_warp_46}
\end{equation}
a $z^2$ term in \eqref{warpdef}, omitted here, would generate a $Q^{-2}$ term. If all condensate
effects are encoded in these weights, matching the leading power coefficients (neglecting their
running) requires
\begin{equation}
 c_4=\frac{3g_5^2}{8}\mathcal C_4
 =\frac{g_5^2}{64}
 \left\langle\frac{\alpha_s}{\pi}G^a_{\mu\nu}G^{a\mu\nu}\right\rangle,
 \qquad
 c_6^{V,A}=\frac{5g_5^2}{96}\mathcal C_6^{V,A}.
 \label{eq:ope_warp_matching}
\end{equation}
A single metric with identical vector and axial couplings gives $c_6^V=c_6^A$; distinct effective
warp factors $w_V\neq w_A$ require chirality-sensitive interactions, such as the left-right
field-strength coupling of \citep{Hirn:2005vk}. In HW1 the scalar contribution at dimension six adds
to that of the metric deformation.

Warp-factor deformations also modify the normalizable wave functions, and hence pole positions and
residues of $\Pi_V$ and $\Pi_A$: masses, decay constants, $f_\pi$, meson couplings and the
Gasser-Leutwyler coefficients \citep{Gasser:1983yg} must therefore be re-evaluated consistently. This has been done
for HW1 by \citet{DaRold:2005mxj} and for HW2 by \citet{Hirn:2005vk,Cappiello:2010egv}.

The $1/Q^2$ term unavoidable in the minimal SW model has prompted debate about a possible QCD origin
(see \citet{Jugeau:2013zza} and references therein). Fits to data bound its coefficient \citep{Narison:1992ru}.
A short-distance linear term in the static potential motivated an effective tachyonic gluon
mass \citep{Chetyrkin:1998yr}, also explored in AdS/CFT \citep{Andreev:2006vy,Andreev:2006ct}. From
the quark-hadron duality side, large-$N_c$ Regge-like spectra generically produce a $1/Q^2$ term
unless their parameters are tuned to cancel it \citep{Shifman:2000jv,Golterman:2001nk,Afonin:2004yb}.

These two-point constraints probe only the vacuum and quadratic source sector. Transition form
factors and multi-current amplitudes, including hadronic light-by-light scattering, depend on
additional vertices, and their short-distance constraints must be checked independently.

\subsection{Chiral anomalies, the axial anomaly and massive pions}

As discussed by \citet{Sakai:2004cn} for the \WSS model, 
the CS term of the probe D8 branes
correctly reproduces the chiral anomaly of QCD and the Wess-Zumino-Witten (WZW) term in the chiral Lagrangian.

This can be analyzed by including background gauge fields for the chiral
$\U(N_f)_L\times\U(N_f)_R$ symmetries in the form of
non-normalizable modes for the 5D gauge potential.
Including a background electromagnetic field, this gives rise to the
Adler-Bell-Jackiw anomaly responsible for the decay of $\pi^0\to\gamma\gamma$.
However, because the CS term involves all the (axial) vector mesons
described by the flavor gauge fields it determines all their
anomalous interactions. Some of those 
(such as $\omega\to\pi^0\gamma$ and $\omega\to\pi^0\pi^+\pi^-$) have been worked out
for the \WSS model already by \citet{Sakai:2005yt}; a more complete
analysis has been carried out by \citet{Hechenberger:2023ljn, Hechenberger:2024piy}, including
also anomalous interactions with glueballs.

In the bottom-up models, the necessary CS term is added
following the example of the \WSS model, where it
is derived from first principles. 
However, in HW models, this requires the subtraction
of a boundary term at $z=z_0$, unless HW2/HW3 boundary conditions
are chosen. For the chiral limit of the HW1 model,
this has been specified by \citet{Grigoryan:2007wn}; with
finite quark masses the required generalization has been
obtained by \citet{Leutgeb:2021mpu}. 

For the transition form factor which describes the decay
of (massive) pions into two virtual photons, this leads to
\be\label{TFFpi}
F_{\pi_n\gamma^*\gamma^*}(Q_1^2,Q_2^2)=
\frac{N_c}{12\pi^2}g_5 K_n(Q_1^2,Q_2^2),
\ee
with
\be
K_n(Q_{1}^{2},Q_{2}^{2})= -\int_{0}^{z_{0}}dz\,\mathcal{J}(Q_{1},z)\mathcal{J}(Q_{2},z)\partial_{z}\phi_n(z)
+\mathcal{J}(Q_{1},z_{0})\mathcal{J}(Q_{2},z_{0})\left[\phi_n(z_0)-\pi_n(z_0)\right].
\ee
This implies (equally in the HW1 and HW3 models)
$K_n(0,0)=-\pi_n(z_0)$,
since $\mathcal{J}(0,z)\equiv 1$.
In the chiral limit where $\pi_1(z)\equiv -1/(g_5 f_{\pi})$ one obtains
the well-known result
\be
F_{\pi_1 \gamma \gamma}\equiv 
F_{\pi_1 \gamma^* \gamma^*}(0,0) = \frac{N_c}{12 \pi^2 f_{\pi}}.
\ee
With nonzero quark masses, $K_n(0,0)$ has to be evaluated numerically.
Using \eqref{phipisumrule} one can, however, derive the sum rule \citep{Leutgeb:2021mpu}
\be
\sum_{n=1}^\infty f_{\pi_n} F_{\pi_n\gamma\gamma}=\frac{N_c}{12\pi^2},
\ee
showing that all massive pions contribute to the anomaly.

While in the chiral limit all $f_{\pi_n}$ with $n\ge2$ vanish and
thereby decouple from the anomaly,
the excited (massive) pions retain
a finite $F_{\pi_n\gamma\gamma}$ and thus do not decouple from the photons
(see also Sec.\ \ref{sec:hlbl}, where additionally
the anomalous transition form factor of axial-vector mesons is discussed).

\subsection{Theta term, U(1)$_A$ anomaly, and $\eta'$ mass}\label{sec:U1A}

Although subleading in the 't Hooft limit with fixed $N_f$ but large $N_c$,
the effects of the gluonic U(1)$_A$ anomaly oftentimes cannot be neglected
when hQCD results are extrapolated to realistic values of $N_c=3$ and
$N_f=3$, because it reduces the multiplet of Goldstone bosons from
a nonet to an octet. In fact, this anomaly is naturally realized in the
top-down approach and provides an anomalous mass for the $\eta'$ 
meson \citep{Armoni:2004dc,Barbon:2004dq,Sakai:2004cn}; 
this is also the case for the $\theta$ term of QCD \citep{Witten:1998uka}.

\subsubsection{\WSS model}

The CP-breaking topological $\theta$ term of QCD, 
\be\frac{\theta}{8\pi^2}\text{Tr}\int_{\mathbb{R}^4}\FYM\wedge \FYM,
\ee
is a natural
possibility in the Witten model of low-energy QCD introduced in section \ref{holoYM}, because it can be
provided through the type-IIA RR 1-form gauge field $C_1$ \citep{Witten:1998uka}.
It can be identified from 
the D4 brane action
\be
S^{\rm D4}=-T_4 \mathrm{Tr}\int d^4x d\tau e^{-\Phi}\sqrt{|\det (g_{(5)}+2\pi\alpha' \FYM)|}+ T_4\tfrac12(2\pi\alpha')^2\mathrm{Tr} \int C_1 \wedge\FYM\wedge\FYM + \ldots,
\ee
with $T_4=(2\pi)^{-4}l_s^{-5}$ and $\alpha'=l_s^2$.
This gives
\be
\theta + 2\pi k = \frac{1}{l_s}\int_{S_\tau} C_1
\ee
with $k$ an integer.
Thus, with a nonzero $C_1$ background field, one can obtain
a nonzero $\theta$ parameter. For small $\theta$ parameter such that the backreaction to
the original Witten background can be neglected\footnote{The
fully backreacted case has been studied by \citet{Bigazzi:2015bna}},
one can readily find a regular solution of the field equations of 
$C_1$ with a constant value at $u\to\infty$
and use this to derive the topological susceptibility from the 
kinetic terms of the $C_1$ field with the result
\citep{Hashimoto:1998if,Barbon:2004dq,Sakai:2004cn,Bartolini:2016dbk,Leutgeb:2019lqu}
\be\label{chig}
S_{C_1}=-\frac1{4\kappa_{10}^2}\int d^{10}x \sqrt{|g|}\,\left|dC_1\right|^2+\ldots
\to-\frac{\chi_g}{2} \int d^4x\, \theta^2, \quad \chi_g=\frac{\lambda^3 \MKK^4}{4(3\pi)^6}.
\ee

In the presence of flavor branes, their CS terms also encode the
U(1)$_A$ anomaly. In fact, the U(1)$_A$ transformation, which corresponds to the 5-dimensional gauge transformation with $g_L=g_R^{-1}=e^{i\alpha}{\bf 1}_{N_f}$ with $g_L$ and $g_R$ given in (\ref{gLgR}), induces the transformation of the $\theta$ parameter \citep{Sakai:2004cn}:
\be
\theta\ra\theta+2N_f\alpha\ .
\ee
This shows that the U(1)$_A$ symmetry is broken to $\mathbf{Z}_{2N_f}$ by anomaly.\footnote{There is a mixed anomaly between the discrete axial symmetry and the flavor symmetry in QCD \citep{Tanizaki:2018wtg}. A derivation of this anomaly in the \WSS model was presented by \citet{Akhond:2025wwn}.}

Because the U(1)$_A$ symmetry is broken by the anomaly, the associated would-be Nambu-Goldstone mode $\eta_0$ acquires a mass. This mechanism can be demonstrated explicitly in the \WSS model.
Indeed, the isoscalar would-be Goldstone mode $\eta_0$ 
turns out to mix with the non-normalizable mode of $C_\tau$,
which induces a mass term proportional to the topological susceptibility
\be
m_0^2=\frac{2N_f}{f_\pi^2}\chi_g
\ee
in accordance with the Witten-Veneziano formula
\citep{Witten:1979vv,Veneziano:1979ec,Armoni:2004dc,Barbon:2004dq}.

\subsubsection{Katz-Schwartz model}\label{sec:KS}

The first implementation of the U(1)$_A$ anomaly in bottom-up
AdS/QCD models due to Katz and Schwartz \citep{Katz:2007tf,Schafer:2007qy} did not attempt to follow the mechanism obtained
in the \WSS model, but added
a massless complex field $Y$ accounting for the gluon field strength squared $\alpha_s G_{\mu\nu}^2$ by its modulus and $\alpha_s G\tilde G$ by its phase, with
an interaction with the bifundamental scalar $X$ of the HW1 model of the form
\be
\mathcal{L}_{X,Y}^{5D}/\sqrt{g}=
\text{tr}\left[ |DX|^2+3|X|^2\right]+\frac1{2(\ln z\Lambda)^2}|DY|^2+\frac{\kappa_0}{4}\left[\bar{Y}^{N_f}
\det(X)+\text{h.c.}\right],\label{LXY}
\ee
where  the logarithm in front of $|DY|^2$ accounts for
the fact that its dual operators approach scaling dimension 4 only
asymptotically. 

Without the logarithm in \eqref{LXY}, the field equations for $Y$
in AdS$_5$ would give a background $\langle Y\rangle = C + \Xi z^4$, where
$\Xi$ represents a gluon condensate.
The axial anomaly relation and an OPE fit of the flavor-singlet axial vector correlation function lead to \citep{Katz:2007tf}
$
C=\frac{\alpha_s}{2\pi^2}\sqrt{2N_f}$.
To make contact with the (one-loop) running of 
$\alpha_s$ in QCD, $\alpha_s$ is made dependent on $\ln(z\Lambda)$
and $\Lambda$ identified with $z_0^{-1}$.
The extra coupling constant $\kappa_0$ is chosen as some
large parameter, because it turns out that the results such
as the mass of the $\eta'$ would-be Goldstone boson become
insensitive to it in this limit.

The model has a nonzero topological susceptibility and a pseudoscalar glueball field through fluctuations
of the phase of $Y$ which mixes with the singlet $\eta_0$.
By tuning the gluon condensate parameter $\Xi$, \citet{Leutgeb:2022lqw}
found that the masses of $\eta$ and $\eta'$ as well as their
two-photon widths could be matched with experimental values at
the percent level. 

\subsubsection{Superconnections}

A formulation which follows more closely a string-theoretic construction
and which moreover leads to a generalization of the
Chern-Simons terms has been put forward in
holographic QCD models from tachyon condensation \citep{Iatrakis:2010zf,Iatrakis:2010jb}
(and their extensions to the Veneziano limit \citep{Jarvinen:2011qe,Arean:2016hcs,Jarvinen:2022mys}),
where the open-string tachyon $T$ is identified up to a prefactor with the
bifundamental scalar $X$ of the bottom-up models.
With the latter, one employs a generalized Chern-Simons term involving a so-called 
(Quillen \citep{Quillen:1985vya}) superconnection
    \begin{align}
    \label{eq:CSterm}
    S_\mathrm{CS}&=T_p \int_{\Sigma_{p+1}}C \wedge \text{Str} \exp[{i2\pi \alpha' \mathcal{F}} ]    
    = T_p \int_{\Sigma_{p+1}}C \wedge  i\,d\Omega
\end{align}
with
\begin{align}
     i\mathcal{F}&=\begin{pmatrix}
iF_L-T^{\dagger}T & DT^{\dagger} \\
DT & iF_R -TT^{\dagger} \end{pmatrix},
\end{align}
where
$C$ is a formal sum of Ramond-Ramond (RR) $n$-form potentials $C_n$
and the active integrand is only the $(p+1)$-form part of the entire expression.

The background geometry has
RR flux proportional to $N_c$, which gets picked up by the 5-form part $\Omega_5$.
There is also a coupling of the $C_3$ RR potential and the fields contained in $\Omega_1$, and this is responsible for the implementation of the $U(1)_A$ anomaly. 

Dualizing the $C_3$ RR potential in 5 dimensions one obtains \citep{Leutgeb:2024rfs}
\begin{align}\label{eq:Sa}
   S_{a}= 
   \int \frac{1}{2} B(z)\, \left(da-T_p \Omega_1\right)
   \wedge * \left(da-T_p \Omega_1\right),
\end{align}
where $a$ corresponds to the pseudoscalar glueball field, which
eventually mixes with the pseudoscalar quarkonia, and, to leading order in pseudoscalar fluctuations in $T=e^{-i \eta}T_0 e^{-i\eta}$,
\begin{align}
  \Omega_1= 4\pi \alpha' \text{tr}\left[\eta \,d \exp(-2 \pi \alpha' T_0^2) +A\exp(-2 \pi \alpha' T_0^2) \right],
  \qquad\text{with}\quad A=(A_L-A_R)/2.
\end{align}
In \eqref{eq:Sa}, the function $B(z)$ is a phenomenological function that
allows one, as in \ref{sec:KS}, to introduce a logarithmic running of $\alpha_s$.

Explicitly, the Lagrangian can be written as
\begin{align}
\mathcal{L}_a \propto \sqrt{g} g^{MN}(\partial_M a- \text{tr}(V_t A_M+\eta \partial_M V_t)(\partial_N a- \text{tr}(V_t A_N+\eta \partial_N V_t))
\end{align}
with $V_t=\exp(-2 \pi \alpha' T_0^2)$. Expanded at small $z$ it
has the form
\begin{align}
    \mathcal{L}_a&\propto \sqrt{g} g^{MN}(\partial_M \tilde{a}-A^0_M)(\partial_N \tilde{a}-A^0_N)\qquad\mbox{with}\quad\tilde{a}=a \sqrt{{2}/{N_f}},
\end{align}
which reproduces, asymptotically, the form of the Lagrangian of the
Katz-Schwartz model.
The axial anomaly is thus implemented through a St\"uckelberg coupling;
see also \citep{Csaki:2026arv}.

\subsection{Scalar mesons}

Light scalar mesons are broad, overlapping resonances near thresholds, possibly with molecular or
multiquark components. Their masses and widths are reliably extracted from amplitude poles on
unphysical sheets, using unitarity, analyticity, chiral symmetry and coupled channels; dispersive
analyses and lattice QCD have established a light scalar nonet, and the quark-mass dependence of
its poles sheds light on its nature. 

Large $N_c$, and hence hQCD, gives a complementary picture in terms of narrow resonance towers. In the
HW1 model, normalizable modes  of the scalar field give rise to a tower of resonances with $m_n\sim n$. Such  a behavior may be modified  once additional
potentials or boundary terms are added. In the SW case, the scalars  lie on a linear Regge trajectory,
$m_{S,n}^2=4c^2(n+\tfrac12)$, $n=1,2,\ldots$. The analysis of \citet{Colangelo:2008us} accommodates $a_0(980)$
and $f_0(980)$ but finds too small scalar-pseudoscalar couplings; a modified dilaton and a quartic
scalar term \citep{Gherghetta:2009ac} separate explicit from spontaneous chiral symmetry breaking and
improve the spectra; see also \citet{Ghoroku:2005vt,Branz:2010ub}.
 In the \WSS model, scalar mesons, except for the light pseudoscalar meson nonet, are described as excited open string states \citep{Imoto:2010ef}.\footnote{\citet{Sakai:2004cn} interpreted the fluctuations of the scalar field corresponding to the position of D8-branes in the transverse direction as scalar mesons. However, it was shown by \citet{Imoto:2010ef} that they cannot be identified with QCD mesons, because they are odd under a $\mathbf{Z}_2$ symmetry, called $\tau$-parity, that is absent in QCD.} They lie on a Regge trajectory, $m_N^2\sim N/\tilde\ell_s^2$ at the leading order for large $\lambda$, where $N$ is the excitation level. Candidates for $\pi(1300)$, $a_0(1450)$ and $f_0(1370)$ were found among the $N=1$ states, whereas no candidates for $a_0(980)$, $f_0(500)$ or $f_0(980)$ were found in the spectrum. This absence may support interpretations of the latter mesons as four-quark states.\footnote{See the review ``64. Scalar Mesons below  1  GeV" by the PDG  \citep{ParticleDataGroup:2026mpi} for possible interpretations of these states.} See section \ref{ISS} for more discussion on this point.

The gluonic $0^{++}$ sector differs: a massless probe scalar dual to $\mathrm{tr}\,G^2$ gives
$m_{G,n}^2=4c^2(n+1)$, $n=1,2,\ldots$, in the soft wall \citep{Colangelo:2007pt}, while in improved hQCD
glueballs are fluctuations of the coupled gravity-dilaton background \citep{Gursoy:2007cb,Gursoy:2007er}.
In the \WSS model, 
radiative \citep{Hechenberger:2023ljn} glueball decays have been computed. Neither the
identification of glueballs with physical isoscalars nor quarkonium-glueball mixing follows from
these minimal spectra.

The two-photon transition form factor 
of a neutral scalar enters both its decays and the
scalar-exchange contribution to HLbL 
(see Sec.~\ref{sec:hlbl})
in the muon $g-2$. It has two independent gauge-invariant
structures \citep{Knecht:2018sci,Hoferichter:2020lap},
\begin{equation}
 \Gamma_S^{\mu\nu}=T_1^{\mu\nu}F_1^S+T_2^{\mu\nu}F_2^S,\qquad
 T_1^{\mu\nu}=(q_1\cdot q_2)g^{\mu\nu}-q_2^\mu q_1^\nu,
 \label{eq:scalar_vertex}
\end{equation}
with $T_2^{\mu\nu}$ of higher order in the photon momenta. The minimal HW1 action has no $SVV$
vertex; the model of \citet{Cappiello:2021vzi} adds two chirally invariant operators quadratic in $X$
and in the field strengths, producing a vertex $\propto X_0SF_V^2$, so that $F_{1,2}^S$ are overlap
integrals of the scalar mode, $X_0$ and the photon bulk-to-boundary propagators. The superconnection
model of \citet{Leutgeb:2024rfs}, motivated by tachyon condensation in brane-antibrane
systems \citep{Casero:2007ae} (as in the Veneziano-limit models of \citet{Jarvinen:2011qe}), obtains
the vertex from the DBI expansion, which also adds an $X_0^2$-dependent mass term to the scalar
fluctuations.

At large Euclidean virtualities, the overlaps are dominated by $z\sim1/Q$. In the chiral limit,
$X_0\sim\sigma z^3$ gives $F_1^S\sim Q^{-6}$, $F_2^S\sim Q^{-8}$ \citep{Cappiello:2021vzi}, while for
$m_q\ne0$ the term $X_0\sim m_qz$ dominates, giving $F_1^S\sim m_qQ^{-4}$,
$F_2^S\sim m_qQ^{-6}$ \citep{Leutgeb:2024rfs}. The leading-twist light-cone analysis instead gives
$F_1^S\sim Q^{-2}$, $F_2^S\sim Q^{-4}$ \citep{Hoferichter:2020lap}, with a different dependence on
the asymmetry $w=(Q_1^2-Q_2^2)/(Q_1^2+Q_2^2)$ than in the models of \citet{Cappiello:2021vzi,Leutgeb:2024rfs}. The
discrepancy remains unresolved, although violating these short-distance constraints has mild
consequences for low-energy phenomenology.

\subsection{Tensor mesons}\label{sec:tensormesons}

In top-down holography, the energy-momentum tensor of the boundary theory is
dual to the metric field in the bulk. Fluctuations in the latter correspond
to tensor glueballs (see Sec.~\ref{sec:GB}). Tensor quarkonia which come
in flavor multiplets are instead described by massive open-string modes
on the flavor branes \citep{Imoto:2010ef}.

In bottom-up models, which do not have D-brane constructions,
tensor quarkonia are nevertheless usually described by metric
fluctuations, because the corresponding operators $\bar q\, t^a \gamma^{(\mu} \overset{\leftrightarrow}{D}\,^{\nu)}q$
have dimension 4 like the energy-momentum operator. The dual field
therefore should be a massless tensor field like the metric field.

In hard-wall AdS/QCD, 
traceless-transverse metric fluctuations 
$h_{\mu\nu}=\epsilon_{\mu\nu}^T(q)h_n(z)$
with boundary conditions $h_n(0)=0=h_n'(z_0)$ are given by
\be
    h_n(z) \propto
   z^2 J_2(\gamma_{1,n}z/z_0),
\ee
yielding a tower of tensor mesons
with masses fixed by
$
m_n^T/ M_\rho=\gamma_{1,n}/\gamma_{0,1}=
1.593,\, 2.917,\, 4.230,\, \ldots 
$
The lowest tensor resonance mass is thus $m^T_1=1.235$ GeV, differing  by only about $3\%$ from the physical value $M_{f_2}=1.2754$ GeV of the lightest
tensor meson $f_2(1270)$.
(In the soft-wall model, one would instead have $M_n^T/M_\rho=\sqrt{n+1}$ and thus
$m^T_1=1.096$ GeV.)

\citet{Katz:2005ir} proposed
to fix the gravitational constant in front of the Einstein-Hilbert action
such that 
the two-point function of the 4D energy-momentum tensor can be identified with the leading term of the OPE of the energy-momentum tensor correlator in QCD \citep{Novikov:1981xi}
\be
\langle T^{\mu\nu}T^{\rho\sigma} \rangle_\mathrm{QCD}
=-P^{\mu\nu\rho\sigma} \left(\frac{N_c N_f}{160\pi^2}+\frac{N_c^2-1}{80\pi^2}\right) Q^4 \ln Q^2+\ldots,
\ee
with the transverse-traceless projector
$P_{\mu\nu\rho\sigma}=\frac{1}{2}(P_{\mu\rho}P_{\nu\sigma}+P_{\mu\sigma}P_{\nu\rho}-\frac{2}{3}P_{\mu\nu}P_{\rho\sigma})$
where $P_{\mu\nu}=\eta_{\mu\nu}-q_\mu q_\nu/q^2$.
This then determines the magnitude of all tensor couplings.
While this strongly underestimates the amplitude of $f_2\to\pi\pi$,
it turns out to give a very good match with the two-photon decay width of
$f_2(1270)$;\footnote{Because the result for the $f_2\to\gamma\gamma$ does not depend on the details
of chiral symmetry breaking, \citep{Katz:2005ir} argued that this result
is more robust than the one for $f_2\to\pi\pi$.}
see also \citep{Mamedov:2023sns}.

This model was therefore adopted in the study of hadronic light-by-light
scattering by \citet{Mager:2025pvz,Cappiello:2025fyf} (see Sec.~\ref{sec:hlbl}),
with a modified normalization such that the tensor meson tower
saturates the longitudinal short-distance constraint of
the light-by-light amplitude. 
With the original normalization of
\citep{Katz:2005ir}, the tensor coupling has a large-$N_c$
counting typical of
tensor glueballs rather than tensor quarkonia, which is
rectified by the rescaling of \citet{Mager:2025pvz,Cappiello:2025fyf}.
Since the metric field is a single flavor-neutral field,
it is simply assumed that it describes the multiplet of tensor meson towers
collectively. Dividing the two-photon width of the ground-state mode
over $f_2(1270)$, $a_2(1320)$, and $f_2'(1525)$ 
with the assumption of ideal mixing in fact
gives a remarkably good match of the experimental values.

\subsection{Low-energy weak interactions}\label{sec:weak}

Below the electroweak scale, nonleptonic $\Delta S=1$ transitions are described by an effective
Hamiltonian of four-quark operators. In the large-$N_c$ limit the dominant term is the
current--current operator, $\mathcal H_{\rm eff}=-\kappa_-\,Q_-+{\rm h.c.}$, with
$\kappa_-=-\frac{G_F}{\sqrt2}V_{ud}V_{us}^*C_-(\mu)$ and
\begin{equation*}
 Q_-=4\,(\bar s_L\gamma^\mu u_L)(\bar u_L\gamma_\mu d_L)-4\,(\bar s_L\gamma^\mu d_L)(\bar u_L\gamma_\mu u_L),
\end{equation*}
where color indices are contracted within each bracket.

At low energies these interactions are described by the weak chiral Lagrangian, the most general
Lagrangian in the Goldstone fields $U=u^2$ that transforms like the quark operators under
$\SU(3)_L\times \SU(3)_R$. With $L_\mu=iU^\dagger D_\mu U$, $u_\mu=iu^\dagger D_\mu U u^\dagger$,
$f_\pm^{\mu\nu}=uF_L^{\mu\nu}u^\dagger\pm u^\dagger F_R^{\mu\nu}u$, $\lambda=(\lambda_6-i\lambda_7)/2$
projecting onto $s\to d$, and $\Delta=u\lambda u^\dagger$, the leading term is
\begin{equation*}
 \mathcal L_2^{\Delta S=1}=G_8F^4\mathrm{Tr}\left(\lambda L_\mu L^\mu\right)+G_{27}F^4\left(L_{\mu23}L^\mu_{11}+\tfrac23L_{\mu21}L^\mu_{13}\right)+{\rm h.c.},
 \qquad \mathrm{Tr}\left(\lambda L_\mu L^\mu\right)=\mathrm{Tr}\left(\Delta u_\mu u^\mu\right) ,
\end{equation*}
where the octet coupling $G_8$ and the 27-plet coupling $G_{27}$ are fixed by $K\to\pi\pi$; their
ratio, $|G_{27}/G_8|\approx1/18$, is the $\Delta I=1/2$ rule. At $\mathcal O(p^4)$ the octet
Lagrangian is $\mathcal L_4^{\Delta S=1}=G_8F^2\sum_{i=1}^{37}N_iW_i$, where the $W_i$ are the 37
independent $(8_L,1_R)$ operators of order $p^4$ built from $\Delta$, $u_\mu$, $f_\pm^{\mu\nu}$ and the
quark-mass spurions, and the $N_i$ are dimensionless couplings \citep{Ecker:1992de}. Those relevant for
radiative decays are
\begin{equation*}
 W_{14}=i\mathrm{Tr}\left(\Delta\{f_+^{\mu\nu},u_\mu u_\nu\}\right),\quad
 W_{15}=i\mathrm{Tr}\left(\Delta u_\mu f_+^{\mu\nu}u_\nu\right),\quad
 W_{16}=i\mathrm{Tr}\left(\Delta\{f_-^{\mu\nu},u_\mu u_\nu\}\right),\quad
 W_{17}=i\mathrm{Tr}\left(\Delta u_\mu f_-^{\mu\nu}u_\nu\right),
\end{equation*}
\begin{equation*}
 W_{18}=\mathrm{Tr}\left(\Delta(f_{+\mu\nu}f_+^{\mu\nu}-f_{-\mu\nu}f_-^{\mu\nu})\right),
\end{equation*}
together with the odd-intrinsic-parity operators $W_{28},\ldots,W_{31}$, e.g.\
$W_{30}=\varepsilon_{\mu\nu\rho\sigma}\mathrm{Tr}\left(\Delta u^\mu\right)\mathrm{Tr}\left( f_+^{\rho\sigma}u^\nu\right)$.
The $N_i$ cannot be fixed by symmetry alone, yet they control radiative and rare kaon decays such
as $K\to\pi\ell^+\ell^-$, $K\to\pi\gamma\gamma$ and $K\to\pi\pi\gamma$. They are usually estimated with
factorization and resonance models. In the Weak Deformation Model (WDM)
\citep{Ecker:1990in,Ecker:1992de}, the weak operator is built from the full hadronic left-handed current,
including resonance exchange, so that the weak couplings are fixed by the strong ones.

Holographic QCD offers a framework in which such estimates follow from a single five-dimensional
action. Early work computed four-point current correlators in a hard-wall AdS/QCD model, including
the self-interactions of the bulk gauge fields, to address the $\Delta I=1/2$ rule and the $B_K$
parameter \citep{Hambye:2005up,Hambye:2006av}. The correlators interpolate between chiral perturbation theory
at low momenta and perturbative QCD at high momenta, and the results agree with data at the 25\%
level.

The idea of implementing four-fermion weak operators, which are products of single-trace currents, as double-trace deformations of holographic QCD was proposed by \citet{Gazit:2008gz}, who applied it to $\pi\to\mu\nu$, neutron $\beta$ decay and parity-violating pion--nucleon coupling using the WSS, HW and SW models and extended to
a systematic construction of the nonleptonic $\Delta S=1$ weak chiral Lagrangian  by \citet{Cappiello:2011re} based on the HW2 model, in which the pion is
directly represented in terms of the chiral field $U$. 

The key observation is that, at large-$N_c$, $Q_-$ factorizes into a product of left-handed
currents and is therefore a double-trace operator. Its insertion amounts to a deformation of the
ultraviolet boundary conditions of the bulk gauge fields \citep{Witten:2001ua}. Each current is the
holographic response to its source,
$J^a_{L\mu}(x)=\delta S_5/\delta\ell^{a\mu}(x)\propto-\frac{1}{g_5^2}\big[z^{-1}\partial_zL^a_\mu(x,z)\big]_{z\to0}$,
and the weak Lagrangian follows as
$\mathcal L_W\propto\kappa_-\,\mathrm{Tr}\left(\lambda\,J_{L\mu}J_L^\mu\right)$, evaluated on the solutions of
the bulk equations.

Because the holographic current is computed from the full bulk solution, it automatically contains
the whole tower of vector and axial-vector resonances. The construction is thus a 5D realization of
the WDM. It reproduces the weak chiral Lagrangian in both the normal and the anomalous (odd
intrinsic parity) sectors, and it predicts the $\mathcal O(p^4)$ weak couplings $N_i$ as overlap
integrals of the bulk wave functions, with no parameters
beyond those of the strong sector and the overall normalization $G_8$. Applied to $K\to3\pi$, it
shows that the cancellation of resonance contributions found in earlier models is model dependent,
and it gives agreement with experiment \citep{Cappiello:2011re}.

Two limitations should be kept in mind. First, factorization captures the leading large-$N_c$ dynamics
 but not the enhancement responsible for the $\Delta I=1/2$ rule, so $G_8$ is taken from
data rather than predicted. Second, penguin operators such as $Q_6$, which factorize into products
of scalar and pseudoscalar densities, would require the corresponding bulk scalar field and its
double-trace deformation. Within these limits, the same bulk dynamics fixes the couplings of the
weak vertices to photons, which makes the framework an interesting setting for radiative and rare kaon
decays.

\section{Baryons}
\label{baryon}

Around 1960, even before the quark model appeared, Skyrme proposed to describe baryons as solitons \citep{Skyrme:1961vq}, called Skyrmions, in an effective theory of pions (Skyrme model). Although this looks completely different from baryons in QCD, where a baryon is considered as a bound state of three quarks, the predictions of the Skyrme model turned out to be roughly consistent with the observed properties of baryons \citep{Adkins:1983ya}.
The description of baryons in holographic QCD is based on this idea. As we have seen in the previous section, the meson effective theory is given as a five-dimensional gauge theory. Thus, the baryons are obtained as solitons in this five-dimensional gauge theory, which will be reviewed in section \ref{baryon_as_soliton}.

In the top-down approach, there is another (physically equivalent) way to construct a baryon. That is, the baryons are described as D-branes in string theory. This is a natural extension of the idea of Skyrme in the sense that the fundamental degrees of freedom in string theory are strings, which correspond to mesons, and typical solitonic objects in string theory are D-branes, which describe baryons.
As we will see in section \ref{baryon_as_D-brane}, the D-brane describing a baryon must have $N_c$ open strings attached to it \citep{Witten:1998xy,Gross:1998gk}.
These open strings correspond to quarks, and this description is intuitively related to the picture of a baryon as a bound state of $N_c$ quarks in QCD.

\begin{figure}[ht]
\begin{center}
\begin{picture}(330,70)(0,0)
\put(0,23){
\includegraphics[scale=0.15]{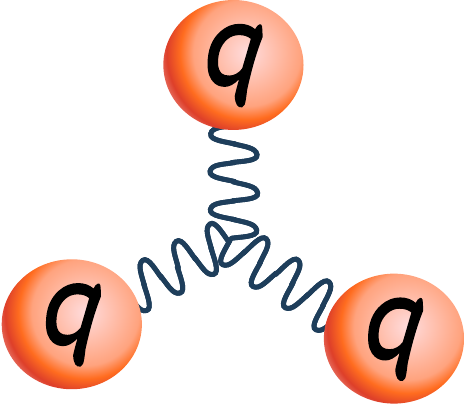}}
\put(13,67){\makebox(0,0){bound state of $N_c$ quarks}}
\put(63,37){\makebox(0,0){$\Longleftrightarrow$}}
\put(20,10){\makebox(0,0){$n_B=\frac{\sharp~\mbox{quarks}}{N_c}$}}
\put(95,23){
\includegraphics[scale=0.20]{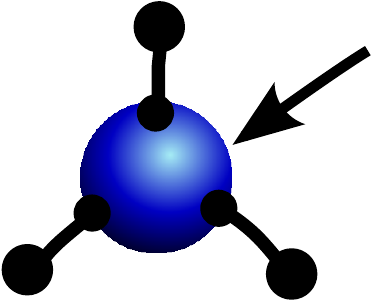}}
\put(110,67){\makebox(0,0){D4 with $N_c$ strings}}
\put(150,49){\makebox(0,0){D4-brane}}
\put(163,37){\makebox(0,0){$\Longleftrightarrow$}}
\put(115,10){\makebox(0,0){$n_B=\sharp \,{\rm D}4$}}
\put(190,20){
\includegraphics[scale=0.15]{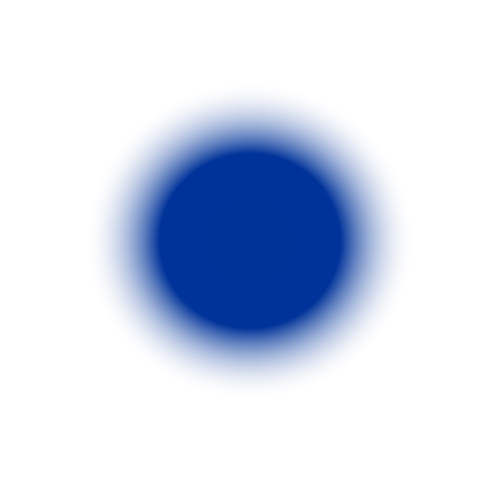}}
\put(210,67){\makebox(0,0){soliton in 5 dim}}
\put(212,39){\line(2,1){10}}
\put(235,45){\makebox(0,0){
{$F\ne0$}
}}
\put(210,10){\makebox(0,0){
{$n_B=\frac{1}{8\pi^2}\int{\rm tr}(F\wedge F)$}
}}
\put(263,37){\makebox(0,0){$\Longleftrightarrow$}}
\put(290,20){
\includegraphics[scale=0.15]{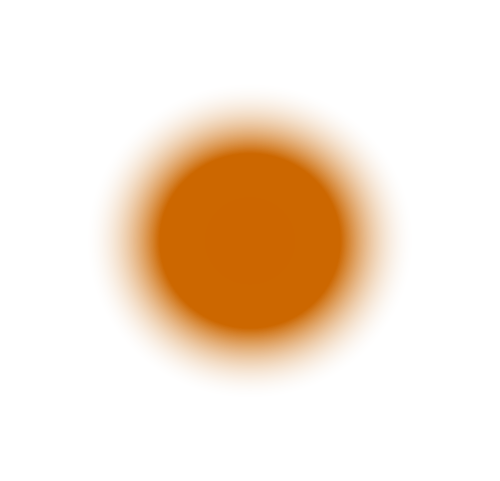}}
\put(313,39){\line(2,1){10}}
\put(345,46){\makebox(0,0){
{$U^{-1}dU\ne0$}
}}
\put(310,67){\makebox(0,0){Skyrmion}}
\put(320,10){\makebox(0,0){
{$n_B=\frac{1}{24\pi^2}\int{\rm tr}\left[(U^{-1}dU)^3\right]$}}}
\end{picture}
\caption{Four descriptions of a baryon. They are connected in string theory.
}
\label{hadron}
\end{center}
\end{figure}

Thus, there are at least four seemingly different descriptions of a baryon: a Skyrmion, a soliton in five-dimensional gauge theory, a D-brane, and a bound state of $N_c$ quarks. Remarkably, all these descriptions appear in the framework of the top-down holographic QCD based on string theory, introduced in section \ref{D4D8}, and they are argued to be equivalent. (See Fig.~\ref{hadron}.)

\subsection{Baryon as D-brane}
\label{baryon_as_D-brane}

We first consider a baryon realized as a D4-brane in the \WSS model.
Recall that the supergravity background (\ref{D4metric2})--(\ref{D4dilaton2}) used in the top-down approach of holographic QCD described in section \ref{topdown} contains a non-contractible $S^4$. Let us consider a D4-brane wrapped on this $S^4$.\footnote{This D4-brane should not be confused with the $N_c$ D4-branes introduced to realize the color degrees of freedom in section \ref{holoYM}.} It is easy to see that the minimal energy
configuration is given by  $r=0$ ($u=\ukk$) and $Z=0$. Therefore, it behaves as a point particle in the four-dimensional spacetime parametrized by $x^\mu$ ($\mu=0,1,2,3$). As we will soon show, the wrapping number of the D4-brane corresponds to the baryon number, and hence this particle is interpreted as a baryon.

The D4-brane is coupled to the RR-fields via the CS-term, which induces
\bea
S_{\rm CS}^{\rm D4}=\frac{1}{(2\pi\ell_s)^3}\int dC_3\wedge a_1=N_c\int a_1\ ,
\label{CSD4}
\eea
where $a_1$ is the $\U(1)$ gauge field on the D4-brane, in the background with (\ref{D4dilaton-RR}). This term provides a source for $N_c$ units of electric charge. However, because $S^4$ is a compact space, the net electric charge on it must be canceled.\footnote{
The equation of motion in electrodynamics is $d*F=*J$, where $F=da_1$ is the field strength and $J$ is the current. The integration of this equation over a compact space implies $\int *J=0$. This argument holds even if one uses the DBI action for the $\U(1)$ gauge field.}
To cancel the contribution from (\ref{CSD4}), $N_c$ open strings have to be attached to the D4-brane. Each of these open strings has one endpoint on the D4-brane and the other endpoint on the flavor D8-brane.
Therefore, the D4-brane wrapped on $S^4$ with $N_c$ open strings carries $N_c$ flavor indices, which is naturally interpreted as a bound state of $N_c$ quarks in QCD.

As argued above, because $N_c$ open strings attached to the D4-brane have endpoints on the D8-brane, it behaves as a particle with $N_c$ units of electric charge on the D8-brane. This electric charge induces a unit baryon number charge. To see this, consider the variation of the action (\ref{5dim_action}) and pick up the boundary terms:\footnote{Here, we work in units with $\mkk=1$.}
\be
\delta S|_{\rm boundary}=\kappa\int d^4x\,2{\rm tr}
\left[\delta A^\mu k(Z) F_{\mu Z}
\right]^{Z=+\infty}_{Z=-\infty}\ ,
\label{varS}
\ee
from which the currents associated with $\U(N_f)_L\times\U(N_f)_R$ are obtained as
\be
J_{L\mu}=-\kappa\left(k(Z)F_{\mu Z}\right)\Big|_{Z=+\infty}\ ,~~~
J_{R\mu}=+\kappa\left(k(Z)F_{\mu Z}\right)\Big|_{Z=-\infty}\ .
\label{JLJR}
\ee
Then, the baryon number current is 
\be
J_{B\mu}=-\frac{2}{N_c}\kappa\left[
k(Z)\hat F_{\mu Z}
\right]^{Z=+\infty}_{Z=-\infty}\ ,
\label{JB}
\ee
where $\hat F_{MN}$ denotes the $\U(1)$ part of the field strength $F_{MN}$.
From (\ref{JB}), we find that the baryon number charge is proportional to the electric flux at the boundary, which implies that $N_c$ units of electric charge on the D8-brane induce a unit baryon number charge.

The analysis of the baryon spectrum can be done by quantizing the fields created by the open strings attached to the D4-brane. The lightest modes in the open string spectrum correspond to the degrees of freedom of the instanton moduli space in the solitonic approach, which will be discussed in section \ref{baryon_as_soliton}.
For the $N_f=2$ case, by quantizing these lightest modes, protons, neutrons, $\Delta$, and other excited baryon states with $I=J$, where $I$ and $J$ are isospin and spin, respectively, are obtained \citep{Hashimoto:2010je}. (See section \ref{baryon_as_soliton} for more details.)
Contributions from excited open string states were analyzed by \citet{Hayashi:2020ipd}, and many excited baryon states with $I\ne J$ were found in the spectrum.\footnote{States with $I\ne J$ were also discussed by \citet{Hashimoto:2019wmg} in a different approach.} Although there are some ambiguities both in the analysis and in the interpretation, the results are reasonably close to the baryon spectrum found in the experiments.

This approach can be extended to multi-baryon systems. For example, $k$ baryons are described by $k$ D4-branes wrapped on $S^4$. As we saw in section \ref{briefintro}, the fields created by the open strings connecting D4-branes become $k\times k$ matrices. Based on this idea, a matrix model describing multi-baryon systems was proposed by \citet{Hashimoto:2010je} and further developed to describe three-body nuclear forces
\citep{Hashimoto:2010ue} and various nuclei \citep{Hashimoto:2011nm,Hashimoto:2019wmg}.\footnote{See also \citet{Aoki:2012th} for a review and \citet{Hashimoto:2008jq} for an earlier related work.}

\subsection{Baryon as soliton}
\label{baryon_as_soliton}

Another way to describe baryons is to consider solitonic gauge configurations with non-trivial instanton number in the 5-dimensional gauge theory, that is the meson effective theory of holographic QCD. In this description, the baryon number $n_B$ is given by
\be
n_B=\frac{1}{8\pi^2}\int{\rm tr}(F\wedge F)\ ,
\label{nB}
\ee
where the integration in the right-hand side is over the 4-dimensional space parametrized by $x^{1\sim 3}$ and $Z$ \citep{Son:2003et,Sakai:2004cn}. To see this, note that the CS-term in the action (\ref{5dim_action}) contains a term of the form $N_c\int\hat A\wedge{\rm tr}(F\wedge F)$, where $\hat A$ is the $\U(1)$ part of the $\U(N_f)$ gauge field. Then, using the expression for the baryon number current (\ref{JB}) and the equations of motion for $\hat A$, we obtain\footnote{See also
\citep{Hashimoto:2008zw,Lau:2016dxk} for the derivation of this formula.}
\be
n_B=\int d^3x\, J_B^0=\frac{2}{N_c}\kappa\int d^3x\int_{-\infty}^\infty dZ\,\partial_Z\left(k(Z)\widehat F^{0Z}\right)
=\frac{1}{8\pi^2}\int{\rm tr}(F\wedge F)\ .
\ee
The right-hand side of (\ref{nB}) is called the instanton number, which is known to be integer-valued. Therefore, a gauge configuration with instanton number $n_B$ corresponds to a system with $n_B$ baryons.
Since this description does not need to introduce D-branes or strings, it can be used in both top-down and bottom-up approaches of holographic QCD.
In the \WSS model, it can be argued that the gauge configuration with instanton number $n_B$ is equivalent to $n_B$ D4-branes wrapped on $S^4$ \citep{Douglas:1995bn}, and hence the two descriptions of baryons are the same.\footnote{In fact, it is known that $k$ D$(p-4)$-branes embedded in the world-volume of a stack of D$p$-branes are equivalent to a gauge configuration on the D$p$-branes with instanton number $k$ \citep{Douglas:1995bn}.}


Interestingly, the idea of describing a baryon as an instanton configuration was already proposed by Atiyah and Manton in the late 1980's in connection with the Skyrmions \citep{Atiyah:1989dq}. They introduced the Euclidean ``time" coordinate $Z$ and showed that an instanton configuration in the 4-dimensional space parametrized by $(x^1,x^2,x^3,Z)$
with non-zero instanton number $n_B$ given in (\ref{nB}) induces a Skyrmion configuration with the winding number $n_B$
\be
n_B=\frac{1}{24\pi^2}\int{\rm tr}\left[(U^{-1}dU)^3\right]\ ,
\ee
through the relation (\ref{pionU}). Thus, holographic QCD provides physical interpretations of $Z$ as the holographic coordinate and of $U(x^\mu)$ defined by (\ref{pionU}) as the pion field in the Skyrme model, with the correct transformation law under chiral symmetry. See \citet{Sutcliffe:2015sta} for more discussion on the relation between Skyrmions and instantons.

Therefore, the description of baryons in this approach is in many respects very similar to that of Skyrmions, and various techniques developed for analyzing Skyrmions can be applied to the analysis of baryons in holographic QCD. For example, baryon spectrum and other static properties have been analyzed for the Skyrmions by quantizing the fluctuations around the classical solution \citep{Adkins:1983ya}. Similar analysis can be done for baryons described as solitons in holographic QCD. Compared to the Skyrme model, holographic QCD has an advantage in including the contributions from vector/axial vector mesons in a systematic way, and hence we expect to have some improvements in the predictions of physical quantities.\footnote{
See \citet{Nawa:2006gv,Nawa:2007gh,Nawa:2009su,Bartolini:2017sxi,Bartolini:2023oxs} for the analysis of Skyrmions using 4-dimensional effective meson theories derived from the \WSS model.}

Several approaches have been developed to analyze the solitons that describe baryons in holographic QCD. One approach is to apply the idea of \citet{Adkins:1983ya} directly to the 5-dimensional theory and study the properties of baryons by quantizing fluctuations around a soliton configuration satisfying (\ref{nB}). This approach was employed by \citet{Hata:2007mb,Hata:2008xc,Hashimoto:2008zw,Seki:2008mu} for the WSS model and  by \citet{Pomarol:2007kr,Pomarol:2008aa,Panico:2008it,Domenech:2010aq} for the HW models.
Another approach is to introduce 5-dimensional baryon fields representing 5-dimensional particles associated with these solitons and study baryon properties using an effective field theory that includes these baryon fields. This approach was developed by \citet{Hong:2007kx,Hong:2007ay,Hong:2007dq,Park:2008sp} for the WSS model.
See also \citep{deTeramond:2005su,Hong:2006ta,Abidin:2009hr,Ahn:2009px,Maru:2009ux,Ballon-Bayona:2024yuz} for bottom-up models with 5-dimensional Dirac or Rarita-Schwinger fields representing baryons. 

Below, we show some of the results in the \WSS model with $N_f=2$, mainly following \citet{Hata:2007mb} and \citet{Hashimoto:2008zw}.\footnote{See \citet{Sugimoto:2010srw} and \citet{Manton:2022fcb} for reviews.} In general, it is difficult to obtain exact analytic solutions of the equations of motion for the 5-dimensional YM-CS theory in (\ref{5dim_action}) with $n_B\ne 0$, although it is possible to get numerical solutions.\footnote{See, e.g., \citep{Pomarol:2007kr,Panico:2008it,Bolognesi:2013nja,Rozali:2013fna,Gorsky:2015pra,Suganuma:2020jng,Hori:2023fxq,Sugimoto:2025btn} for numerical studies.} One way to analyze the system with a baryon is to start with an exact solution in the flat spacetime and then include the effect of curvature perturbatively. When the 't Hooft coupling $\lambda$ is very large, the effect of the warp factors $h(Z)$, $k(Z)$, and the CS-term in the action (\ref{5dim_action}) are negligible. In this case, the solution can be approximated by the BPST instanton solution \citep{Belavin:1975fg} for flat 4-dimensional space.\footnote{When $\lambda$ is large, the size of the instanton core turns out to be small and it fits entirely within the region with $h(Z)\approx k(Z)\approx 1$, where the spacetime is approximately flat. However, the flat space approximation of the instanton solution is valid only near the center of the instanton, and its long-distance behavior cannot be trusted. See \citet{Cherman:2009gb,Cherman:2011ve,Colangelo:2013pxk,Bolognesi:2013nja} for the issue of model-independent long-distance properties of baryons.}
This solution depends on eight parameters, called instanton moduli, corresponding to the position $(\vec x_0,Z_0)\in {\bf R}^4$ in the 4-dimensional space, size $\rho\in {\bf R}_{>0}$ and $\SU(2)$ orientation $\boldsymbol{a}\in \SU(2)$ of the instanton. Taking into account the effect of the warp factors and CS-term perturbatively,
it can be shown that the size $\rho$ is stabilized at a finite value of order $\lambda^{-1/2}$ \citep{Hong:2007kx,Hong:2007ay,Hata:2007mb}. (See also \citep{Pomarol:2007kr} for a similar observation in a bottom-up model.)
Intuitively, the gravitational force due to the warp factors tends to make the size smaller, while the repulsive force due to the $\U(1)$ charge induced from the CS-term makes the size larger.\footnote{The latter effect corresponds to the role played by $\omega$-meson to stabilize the Skyrmion \citep{Adkins:1983nw}.} Therefore, the size $\rho$ is no longer an exact modulus, but has a potential with a minimum at around $\rho\sim\lambda^{-1/2}$. Similarly, the energy of the instanton configuration depends on the position of the instanton in the $Z$-direction, denoted as $Z_0$, and the minimal energy configuration is at $Z_0=0$. Therefore, the moduli space of the one-baryon solution consists of $\vec x_0\in {\bf R}^3$ and $\boldsymbol{a}\in \SU(2)$, which is the same as that of the Skyrmion.

Among the fluctuations around the classical solution, light modes are expected to be more important than heavier modes at low energies.\footnote{Here, light (heavy) modes refer to fluctuations with small (large) characteristic excitation energies, as determined by the quadratic action around the classical solution.} A natural first step in quantization of fluctuations is to quantize the degrees of freedom corresponding to the moduli space $\{(\vec x_0,\boldsymbol{a})\}$.
The idea is to promote the moduli parameters to slowly varying time-dependent variables $(\vec x_0(t),\boldsymbol{a}(t))$ and substitute the instanton solution parametrized by these variables into the action (\ref{5dim_action}) to obtain an effective quantum mechanical action for the moduli.\footnote{This procedure is called ``collective coordinate quantization" or ``moduli space quantization". See \citet{Gervais:1974dc} for the original work, \citet{Tomboulis:1975qt,Manton:1981mp,Gibbons:1986df} for the application to monopoles, which is closely related to the instantons, and \citet{Adkins:1983ya} for the application to the Skyrmions.}
The physical states of the resulting quantum mechanics are identified with various one-baryon states.

\begin{figure}
\begin{center}
\begin{picture}(100,120)(0,-20)
\put(-70,13){
\includegraphics[scale=0.20]{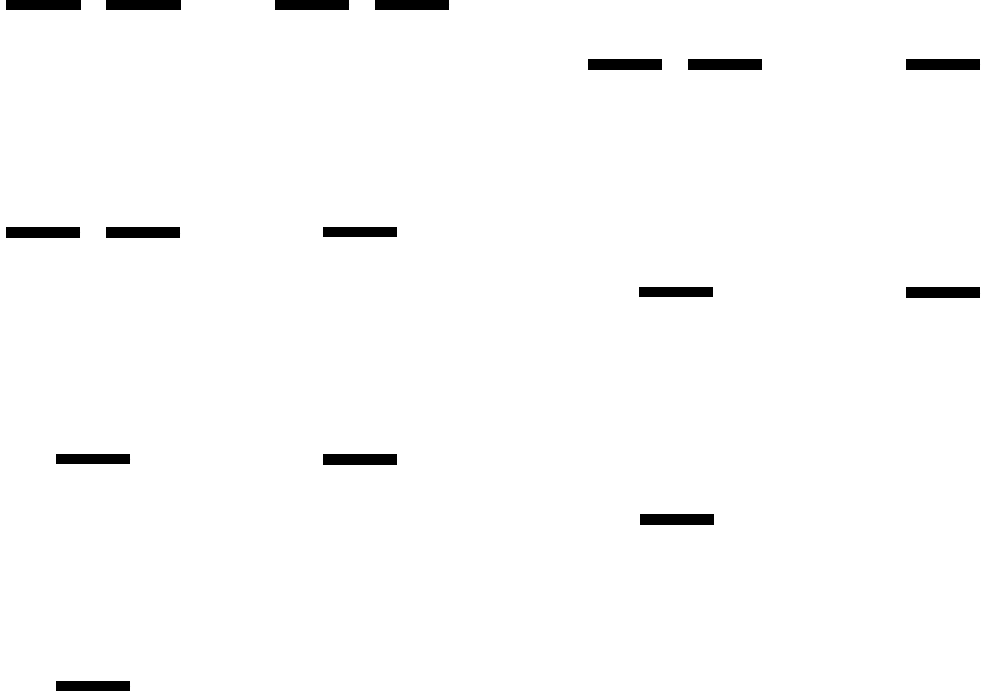}}
\put(-80,-10){\makebox(0,0){$J^P$}}
\put(-54,-10){\makebox(0,0){$\frac{1}{2}^+$}}
\put(-29,-10){\makebox(0,0){$\frac{1}{2}^-$}}
\put(1,-10){\makebox(0,0){$\frac{3}{2}^+$}}
\put(26,-10){\makebox(0,0){$\frac{3}{2}^-$}}
\put(-80,6){\vector(0,1){80}}
\put(-85,92){\makebox(0,0){mass}}
\put(-85,15){\vector(0,1){20}}
\put(-85,15){\vector(0,-1){0}}
\put(-85,37){\vector(0,1){20}}
\put(-85,37){\vector(0,-1){0}}
\put(-85,59){\vector(0,1){20}}
\put(-85,59){\vector(0,-1){0}}
\put(-105,25){\makebox(0,0){$\sqrt{\frac{2}{3}}\mkk$}}
\put(50,-28){
\includegraphics[scale=0.18]{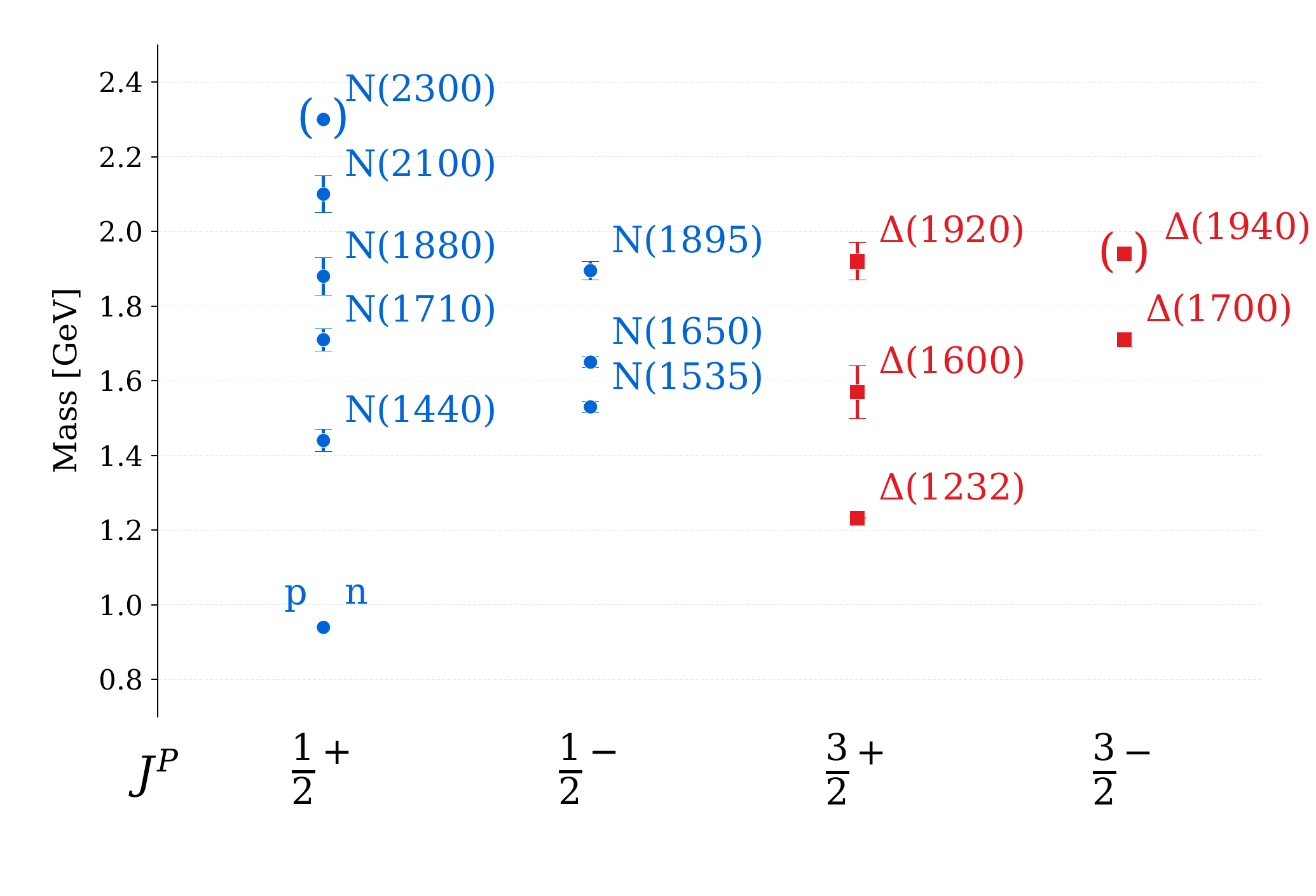}}
\end{picture}
\caption{
The left panel is the baryon spectrum obtained in holographic QCD \citep{Hata:2007mb},
while the right panel is the spectrum of the
baryons with $I=J$ taken from PDG particle listings \citep{ParticleDataGroup:2026mpi}. $N(2300)$ and $\Delta(1940)$ are not regarded as well-established resonances in PDG particle listings.
}
\label{spec}
\end{center}
\end{figure}

Since the moduli space of the one-baryon solution is identical to that of the Skyrmion, the analysis for the Skyrmion by \citet{Adkins:1983ya} can be directly applied to our case. As a result, we find the proton, neutron, and $\Delta(1232)$ states in the spectrum. In addition to the moduli $(\vec x_0,\boldsymbol{a})$, it is reasonable to include the contributions from the collective coordinates $\rho$ and $Z_0$, which can be shown to be parametrically light in the large-$\lambda$ limit. 
Quantizing $\rho$ and $Z_0$ gives rise to excited baryon states that can be identified with $N(1440)$ and $N(1535)$. The former is known as the Roper resonance, a nucleon resonance with $I=1/2$ and $J^P=\frac{1}{2}^+$, while the latter is the lightest negative-parity nucleon resonance with $J^P=\frac{1}{2}^-$. The negative-parity states arise from the excitation of the $Z_0$ mode, because $Z_0$ is odd under the parity transformation.
These two states are predicted to have nearly degenerate masses. The decay properties of these resonances were studied by \citet{Fujii:2021tsw,Iwanaka:2022uje}. See also \citep{Hori:2023fxq} for a numerical analysis of the Roper resonance.
The spectrum of heavier states also has a similar structure. The prediction of holographic QCD \citep{Hata:2007mb} and the experimental data \citep{ParticleDataGroup:2026mpi} are shown in Fig. \ref{spec}. Note that in this approach only states with $I=J$ are obtained,\footnote{See section \ref{baryon_as_D-brane} for a few comments on the states with $I\ne J$.} 
and hence we only show the baryons with $I=J$ in the right panel of Fig. \ref{spec}. Note also that the identification of the excited baryon states obtained in this analysis (left panel) and those found in the experiments (right panel) is not entirely clear, because more states with $I=J=1/2,3/2$ are predicted by \citet{Hayashi:2020ipd}, in which the contributions from excited open string states are also considered.

Once the baryon states are constructed by this procedure, the matrix elements $\langle B| J_{L/R}^\mu(x)|B'\rangle$ of the current operators given in (\ref{JLJR}) with baryon states $|B\rangle$ and $|B'\rangle$, can be calculated. Using these matrix elements, various properties of baryons, such as magnetic moments, charge radii, axial coupling, meson-baryon-baryon couplings, form factors, etc., can be extracted. For example, the magnetic moments for the proton and neutron measured in units of the nuclear magneton were estimated by \citet{Hashimoto:2008zw} as\footnote{Similar results were also obtained by \citet{Hong:2007kx,Hong:2007ay,Hata:2008xc,Kim:2008pw,Pomarol:2008aa}.}
\be
\mu_p\simeq 2.18\ ,~~~\mu_n\simeq -1.34\ ,
\ee
which can be compared with the experimental values:
$\mu_p|_{\rm exp}\simeq 2.79$, $\mu_n|_{\rm exp}\simeq -1.91$.
Although the predicted values are considerably smaller in magnitude than the experimental values\footnote{It was argued by \citet{Hong:2007kx,Hong:2007ay} that one could get better estimates by the shift $N_c\ra N_c+2$, which was suggested by the $N_c$ dependence in the constituent quark model, for these quantities.}, they can be regarded as an improvement over the predictions for the Skyrme model given by \citet{Adkins:1983ya}:
$\mu_p|_{\rm ANW}\simeq 1.87$, $\mu_n|_{\rm ANW}\simeq -1.31$. Another interesting observation is that the predicted proton electric and magnetic form factors closely follow the dipole behavior suggested by the experimental data at low momentum transfer, even though they can be expressed as sums of contributions from the vector meson exchange diagrams as expected from vector meson dominance \citep{Hong:2007dq,Hashimoto:2008zw,Panico:2008it}. This result appears highly non-trivial and warrants further investigation.

The analysis shown here has been extended in various directions, including the extension to $N_f=3$ and study of hyperons
\citep{Hashimoto:2009st,Ishii:2010ib,Liu:2017xzo,Liu:2017frj,Druks:2018hif,Liu:2022urb}\footnote{A subtle issue about a constraint in the baryon spectrum for $N_f=3$ was pointed out by \citet{Hata:2007tn}, but was resolved by \citet{Lau:2016dxk}.}, H-dibaryon \citep{Suganuma:2016lmp},
nucleon-nucleon potential
\citep{Kim:2008iy,Hashimoto:2009ys,Kim:2009sr,Kaplunovsky:2010eh,Baldino:2017mqq,Baldino:2021uie},
 electric dipole moments \citep{Hong:2007tf,Bartolini:2016jxq,Bartolini:2019shp}, 
N-body nuclear forces \citep{Hashimoto:2009as}, electromagnetic transition form factors between nucleons and excited baryons
\citep{Grigoryan:2009pp,Bayona:2011xj}, gravitational form factors \citep{Fujita:2022jus,Sugimoto:2025btn} (see section \ref{GFF}),
the electric and magnetic polarizabilities
\citep{Castellani:2024dxr}.
The phase structure at finite baryon density has also been extensively studied, as will be reviewed in section \ref{density}.

\section{Glueballs}\label{sec:GB}

\subsection{Spectrum}

The calculation of glueball spectra from type-IIA supergravity was one of the first
applications of holographic QCD, namely the Witten model based on type-IIA supergravity 
(see Sec.~\ref{holoYM})
\citep{Gross:1998gk,Csaki:1998qr,
Hashimoto:1998if,Csaki:1999vb,Constable:1999gb,Brower:2000rp}. (Glueballs have 
subsequently been studied further in bottom-up holographic models
in, e.g., \citep{BoschiFilho:2002ta,Colangelo:2007pt,Forkel:2007ru}.)

\begin{figure}[t]
\centerline{\includegraphics[width=0.5\textwidth
]{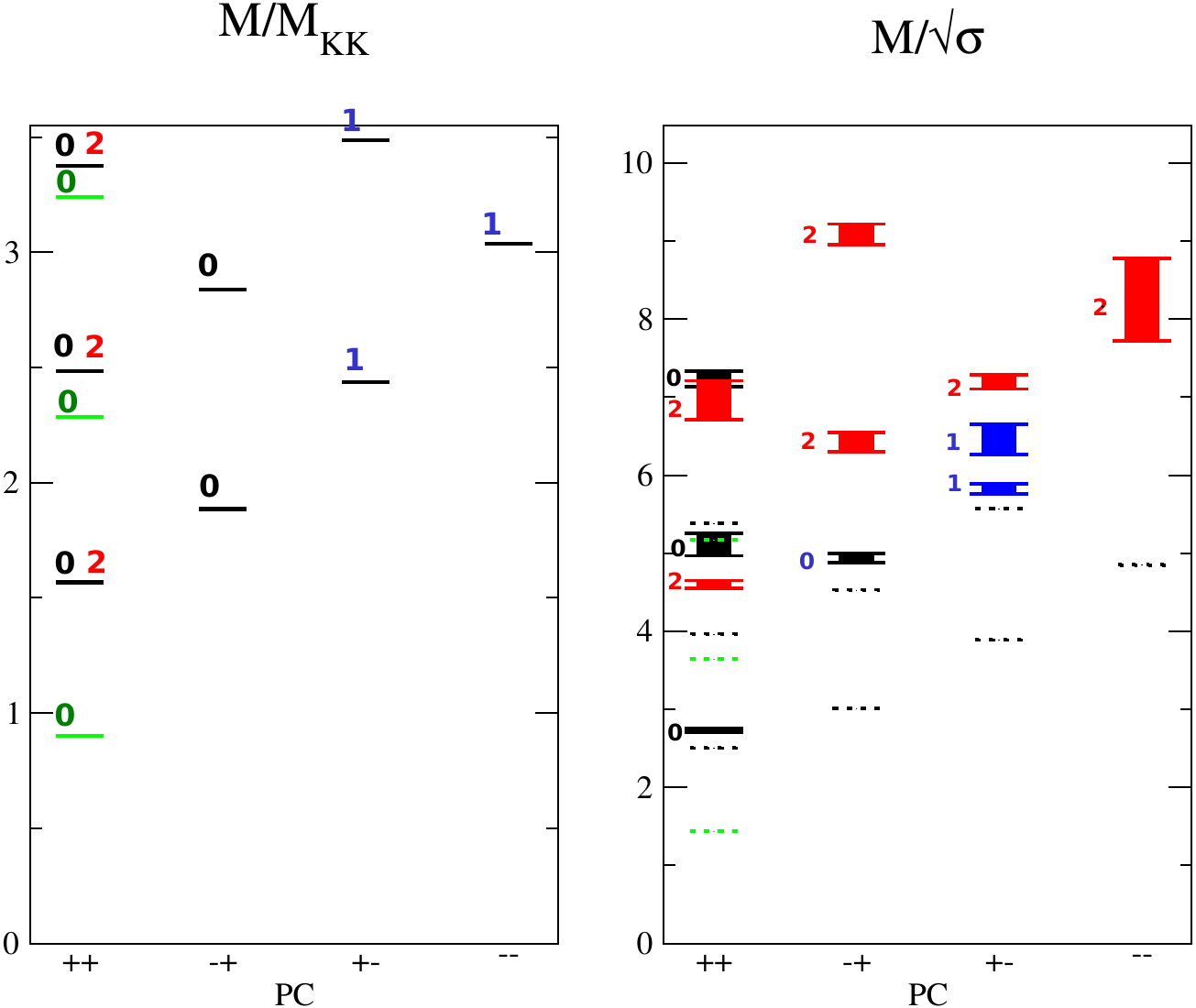}} 
\centerline{\small \hfil (a) \hfil\hfil (b) \hfil}
\medskip
\caption{
The glueball spectrum of the Witten model (a) in units of $\MKK$, (``exotic'' scalar
modes in green), compared to the spectrum obtained in the large-$N_c$ lattice
calculations of \citet{Lucini:2010nv} (b) in
units of the square root of the string tension
$\sqrt{\sigma}$, with the vertical axis scaled such that the lowest tensor modes in
both plots are matched. The dotted lines in figure (b) give the glueball spectrum
of the Witten model when expressed in terms of the string tension
of the Witten model with the usual choice $\lambda\approx 16.63$
for the Sakai-Sugimoto model. (Figure taken from \citep{Brunner:2015oqa})}
\label{figGB}       
\end{figure}

In 5-dimensional bottom-up models, tensor glueballs described by traceless-transverse
metric fluctuations are typically degenerate in mass with the scalar glueballs
described by dilaton fluctuations, whereas lattice QCD 
\citep{Morningstar:1999rf,Chen:2005mg,Gregory:2012hu,Athenodorou:2020ani}
finds that the lightest
glueball is a $J^{PC}=0^{++}$ scalar with mass $\sim 1.7$ GeV,
well below the $2^{++}$ tensor glueball with mass $\sim 2.4$ GeV,
followed by the $0^{-+}$ pseudoscalar glueball with $\sim 2.6$ GeV.
As shown in Fig.~\ref{figGB}(a), this pattern is qualitatively reproduced by
the spectrum of glueballs in the \WSS model.
In the latter, the lightest scalar glueball has mass \citep{Constable:1999gb,Brower:2000rp,Brunner:2015oqa} $0.901 \MKK$,
that is only 855 MeV when $\MKK$ is fixed by the $\rho$ meson mass;
the tensor mode, which is degenerate with a (predominantly) dilatonic
mode has $M_T=1.567\MKK=1487$ MeV.

A comparison that does not refer to the mass of the $\rho$ meson
can be made by relating the glueball masses
to the string tension. Fig.~\ref{figGB}(b), where the
string tension is denoted as $\sigma$, displays
the results of the large-$N_c$ lattice
study of \citet{Lucini:2010nv} as thick error bars.
In the \WSS model, the string tension is given by \eqref{tension}
and the resulting values, which only depend on
$\lambda$, are plotted in Fig.~\ref{figGB}(b)
by dotted lines when the value of $\lambda\approx 16.63$
is used.
Evidently, the lightest scalar glueball of the \WSS model is
again at only half the lattice value;
a significantly lower value of $\lambda\sim 4.6$ would be needed for
it to match.

On the other hand, there is a certain proliferation of $0^{++}$ states 
in the Witten background due to the existence of 
modes involving $\delta g_{\tau\tau}$, which have been termed ``exotic'' by \citet{Constable:1999gb}, where
they were first considered, and which gives rise to the lightest scalar glueball.
Independently of the above numerical discrepancy, \citet{Constable:1999gb} suspected that only one of the towers of scalar
states may survive in the limit $\MKK\to\infty$, where the Witten model would turn into an exact
string-gauge dual of large-$N_c$ Yang-Mills theory.
Another argument for discarding the ``exotic'' scalar glueball mode of the Witten model
has come up in the study of glueball decay patterns, see Sec.~\ref{sec:GBdecay}.

In fact, the spectrum displayed in Fig.~\ref{figGB} discards
various modes of the fields in the Witten model as having no
counterpart in 4-dimensional Yang-Mills theory by requiring
positive parity in the $\tau$-direction \citep{Brower:2000rp}.
Furthermore, the (rather massive) excitations involving the $S^4$
have been ignored.

In the \WSS model, the pseudoscalar glueball is provided by 1-form
field component $C_\tau$ and its mass is higher than the tensor
glueball, in agreement with lattice findings, although its mass
is again too low, $M_P=1.885\MKK=1789$ MeV.

Vector ($1^{--}$) and pseudovector  ($1^{+-}$) glueballs arise from joint
fluctuations of the $B_2$ and $C_3$ form fields \citep{Brower:2008cy,Hechenberger:2024piy} (originally
obtained by \citet{Brower:2000rp} as modes in
the 11D gauge field $A_3$ from which they can be obtained
by dimensional reduction).
While tensor glueballs are related to the leading Pomeron trajectory \citep{Brower:2006ea},
vector glueballs can be associated with the $C=-1$ Odderon
\citep{Brower:2008cy}.
(See \citet{Domokos:2009hm} and \citet{Anderson:2014jia} for applications
of the \WSS model to scattering and central exclusive production
processes in the Regge regime.)

Lattice gauge theory studies have actually identified further
glueballs, with different $J^{PC}$ quantum numbers for $J=0$ and $J=2$,
as well as glueballs with $J>2$. In the \WSS model, those
can be described by massive (closed) string modes.

\subsection{Decay pattern}
\label{sec:GBdecay}

In the \WSS model, the decay pattern of the lightest scalar glueball
mode with model mass of 855 MeV has been studied by \citet{Hashimoto:2007ze}
and compared with the glueball candidate \citep{Amsler:1995td}
$f_0(1500)$.
These calculations have been revisited by \citet{Brunner:2015oqa} 
with the finding that the decay width is too large, while the
heavier, predominantly dilatonic glueball with model mass of 1487 MeV
is a significantly narrower state. Since it appears unnatural that
the first excited scalar glueball should have a smaller width
than the ground state scalar glueball, and because the width
of the dilatonic glueball is more compatible with the most discussed
glueball candidates $f_0(1500)$ and $f_0(1710)$, \citet{Brunner:2015oqa} 
suggested that the ``exotic'' scalar glueball tower be regarded as
an artefact of the \WSS model.

In a simple extension of the \WSS model that includes quark masses
through world-sheet instantons, \citet{Brunner:2015oga,Brunner:2015yha}
derived a decay pattern that would fit best the one of
the glueball candidate $f_0(1710)$
(or $f_0(1770)$ when this is split off from a tetraquark $f_0(1710)$ \citep{Sarantsev:2021ein}), in line with 
the extended linear sigma model study of \citet{Janowski:2014ppa}.

The decay pattern of the less ambiguous $2^{++}$ tensor glueball was
also determined by \citet{Brunner:2015oqa}, with the conclusion that
it is predicted to be rather a broad resonance when the mass is extrapolated
to the one obtained in lattice simulations. The dominant decay modes, when
kinematically allowed, are decays in two vector mesons, which by VMD
also entails a larger two-photon decay rate 
than usually considered (with the exception of \citet{Cotanch:2005ja}).
Radiative decays of glueballs and quarkonia have been studied
in detail by \citet{Hechenberger:2023ljn} within the \WSS model,
finding also a two-photon width of the scalar glueball above 1 keV,
exceeding even the one obtained for $f_0(1710)$ by the Belle collaboration \citep{Belle:2013eck}, which was taken as evidence against its glueball nature
assuming that glueballs should at most have rates in the eV range.

Two-photon widths in the keV range have been obtained by \citet{Hechenberger:2023ljn}
also for the pseudoscalar glueball. The coupling to $F\tilde F$ of the latter arises from a mixing with the $\eta$ and $\eta'$ mesons as a consequence of the $\U(1)_A$ anomaly, which in the \WSS model
was worked out to leading order in $\sqrt{N_f/N_c}$
(treated as a formally small parameter in the probe approximation) by \citet{Leutgeb:2019lqu}. Since this mixing involves a numerically small
coefficient, the
corresponding first-order result may be 
meaningful even for the $N_f=N_c=3$ case of
physical interest. In the 
model of \citet{Katz:2007tf}, 
the mixing of a pseudoscalar
glueball mode with $\eta,\eta'$ and the coupling to two photons
was obtained by \citet{Leutgeb:2022lqw}; the mass of the 
pseudoscalar glueball in this model turns out to be 
comparable to the one obtained in the \WSS model and thus
smaller than that of quenched lattice QCD (the unquenched results
of \citet{Gregory:2012hu} do not include the pseudoscalar glueball).

\section{Further developments}

\subsection{Finite temperature}

Quantum field theory in thermal equilibrium at a nonzero temperature $T$ is
described by compactifying the Wick-rotated imaginary time on a circle with
circumference $\beta=1/T$. In the AdS/CFT correspondence for $\mathcal{N}=4$
SYM in flat Minkowski space,\footnote{If space is not flat, but has a finite curvature radius, there are two different phases: for small temperatures with respect to
the inverse radius, the dual geometry is still pure AdS space apart from
the compactified imaginary time; at higher temperatures, 
there is a first-order (Hawking-Page) transition
to AdS with a Schwarzschild black hole \citep{Hawking:1982dh}.} the dual geometry is then replaced by
that of a black three-brane with Hawking temperature $T$ and
a horizon at a nonzero value of the radial holographic coordinate.

Because bosons and fermions have different periodic boundary conditions 
in imaginary time, supersymmetry is broken, which makes SYM theory at
finite temperature an interesting model for (nonsupersymmetric) 
QCD at finite temperatures not too far
above its deconfinement transition where it is still strongly coupled.
Indeed, heavy-ion experiments have shown that many properties of the
quark-gluon plasma produced there are very different from a weakly coupled plasma.
In particular, its hydrodynamic behavior is rather close to the extremely small
universal viscosity/entropy ratio $\eta/s=\hbar/(4\pi)$
of strongly coupled field theories with a dual Einstein geometry
\citep{Son:2007vk,Casalderrey-Solana:2011dxg}.
Moreover, it was found that the extremely fast apparent thermalization
and hydrodynamic behavior can be understood and described by
holographic shock wave collisions \citep{Chesler:2008hg,Chesler:2015lsa}.
By introducing flavor probe branes, this setup was also used to
study
mesons and their melting
above the deconfinement transition \citep{Mateos:2006nu,Erdmenger:2007cm}.

In the Witten model of low-energy Yang-Mills theory, which involves
an extra dimension compactified on a circle with radius $\MKK^{-1}$,
one obtains a first-order Hawking-Page phase transition
at the temperature where $\beta=2\pi\MKK^{-1}$ \citep{Witten:1998zw}.
With the standard choice of $\MKK=949$ MeV, one finds $T_c=\MKK/(2\pi)\approx
151$ MeV, which is close to the (crossover) temperature observed in
real QCD, but it is actually much too small when compared to
the first-order transition temperature of pure YM theory (or quenched QCD).
Indeed, it has been pointed out by \citet{Mandal:2011ws} that the
$Z_{\Nc}\times Z_{\Nc}$ center symmetry for the two circle compactification
involved matches the phases of 4D YM theory only in the low-temperature
configuration, but not in the high-temperature black 4-brane geometry.
It was proposed that a more adequate description would be through
a Gregory-Laflamme transition where the D4 branes are replaced by
an inhomogeneous distribution of D3 branes, which in fact
leads to an estimate of the transition temperature closer to that of
quenched QCD \citep{Isono:2015uda}. 
The deconfined geometry corresponding to the black 4-brane background
has nevertheless found interesting applications as a model
for the chiral symmetry restoration, because the latter can
be separated from the deconfinement transition if one modifies
the original setup of the \WSS model to non-antipodal D8-\AD\ branes \citep{Aharony:2006da,Parnachev:2006dn,Horigome:2006xu,Bergman:2007wp}.

In the limit of asymptotic brane separation $L\ll\MKK^{-1}$,
the dynamics of the flavor brane approaches that of the decompactified
case, which can be viewed as the holographic version of an NJL
model \citep{Antonyan:2006vw,Davis:2007ka,Preis:2012fh}, where the large-$N_c$ gluon dynamics
becomes suppressed and which might therefore come closer to
the phase structure of real QCD.

For finite-temperature improved holographic QCD and V-QCD see \citet{Gursoy:2010fj} and \citet{Alho:2012mh,Alho:2013hsa}.

\subsection{Finite baryon density}
\label{density}

Because finite baryon density makes QCD inaccessible by standard lattice QCD
due to a severe sign problem and perturbation theory can only be
trusted at extreme densities beyond direct physical relevance,
holographic QCD has been employed
extensively to explore the phase diagram of QCD where
conventional models of nuclear matter cease to be applicable.

The holographic description of baryons as described in Sec.~\ref{baryon_as_soliton}
goes beyond the traditional Skyrme model because it includes interactions
with (axial) vector mesons, which indeed improves their phenomenology
\citep{Sutcliffe:2011ig}. Some efforts have been made to investigate
the consequences for holographic nuclear matter constructed from
individual solitons \citep{Preis:2016fsp,BitaghsirFadafan:2018uzs,Bartolini:2026dbr}, where increasing
density will eventually force 4-dimensional crystalline layers of solitonic
solutions to populate also the holographic direction (which
has been termed ``popcorn'' transition \citep{Kaplunovsky:2012gb}),
or lead to transitions with dyonic solitons \citep{Rho:2009ym,Gorsky:2015pra}.

Most of the studies of dense holographic matter have, however,
drastically simplified the prohibitively difficult task of
analyzing multi-soliton configurations by a homogeneous ansatz \citep{Rozali:2007rx},
which is expected to become a good approximation at sufficiently high densities.
In the \WSS model with non-antipodal branes, this ansatz leads to a rich
phase diagram, see the review of \citet{Kovensky:2026zdd}; for
bottom-up models like the V-QCD model
the state of the art has been reviewed by \citet{Jarvinen:2021jbd,Hoyos:2021uff}.

The phase diagram of strongly coupled gauge theories with fundamental matter
at nonzero chemical potential has also been studied
using D3-D7 brane configurations
by \citet{Kobayashi:2006sb,Mateos:2007vc},
also including strong magnetic fields by \citet{Evans:2010iy,Evans:2011mu},
and more recently with a view of applications to neutron stars by \citet{Hoyos:2016zke,BitaghsirFadafan:2025rpm}.

\subsection{Higher-spin mesons as spinning open strings}\label{sec:higherspinmesons}

A $q\bar q$ meson is associated with an open confining flux tube which, at large orbital angular
momentum, is long compared to its thickness; at large $N_c$ string splitting is suppressed, so the
rotating string can first be studied as a stable classical object. In top-down holography the
flavor-brane supergravity sector contains only $J\le1$ towers ($\pi$, $\rho$, $a_1$, \dots\ in WSS),
while flavored states with $J\ge2$ ($a_2$, $\rho_3$, \dots) are massive open-string excitations,
with masses set by the effective tension at the bottom of the geometry.

Three descriptions are available, with related motivations but different inputs and predictive
content: a specified holographic background and brane system, an effective string model fitted to
hadronic data, and a systematic long-string EFT
\citep{Kruczenski:2004me,Imoto:2010ef,Sonnenschein:2016pim,Cuomo:2024gek}.

The action of a string with massive endpoints in flat 4D spacetime is
\begin{equation}
 S=-T\int d^2\sigma\sqrt{-h}-\sum_{i=1}^{2}m_i\int ds_i,
 \qquad T=T_{\rm eff},\qquad \alpha'_{\rm eff}=\frac{1}{2\pi T_{\rm eff}} .
 \label{eq:c-action}
\end{equation}
For the rigidly rotating configuration $X^0=t$, $X^1+iX^2=r\,e^{i\omega t}$, $-\ell_1\leq r\leq\ell_2$,
with massless endpoints, the equations of motion give
\begin{equation}
 E=\frac{\pi T}{\omega},\qquad
 \mathcal J=\frac{\pi T}{2\omega^2},\qquad
 L_{\rm phys}=\frac{2}{\omega},\qquad
 \mathcal J=\alpha'_{\rm eff}E^2 ,
 \label{eq:c-massless}
\end{equation}
the classical origin of the open-string Regge slope. Massive endpoints move with velocities
$\beta_i=\omega\ell_i<1$ and correct these relations.

\begin{figure}[htbp]
 \centering
 \includegraphics[width=.75\linewidth]{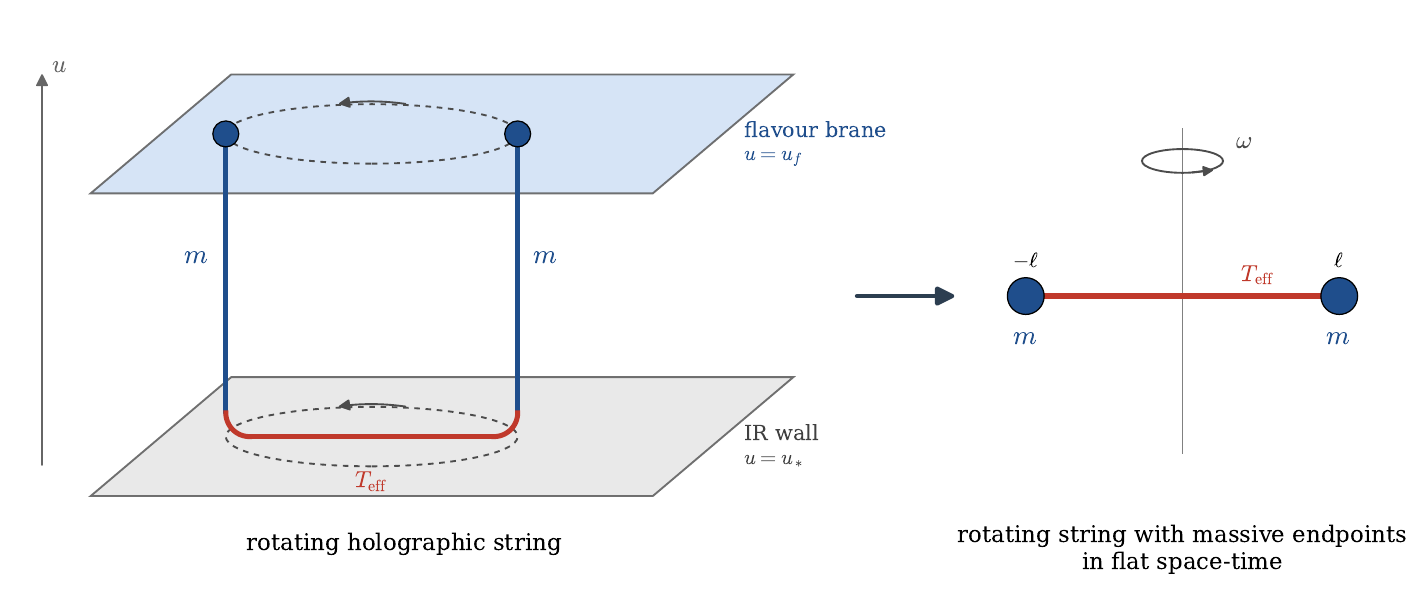}
 \caption{Left: rotating string in a confining holographic background, with both endpoints on the
 same flavor brane. The vertical segments (blue) from the brane to the IR wall carry the endpoint
 mass $m$; the horizontal segment (red) along the wall has the effective tension $T_{\rm eff}$.
 Right: the equivalent rotating string with two equal massive endpoints in flat space-time.}
 \label{fig:holo-string-map}
\end{figure}

Consider now a rotating string in a background
\begin{equation}
 ds_s^2=-A(u)dt^2+B(u)d\boldsymbol{x}^{2}+C(u)du^2+ds_{\rm int}^2,
 \qquad T_s=\frac{1}{2\pi\ell_s^2},
 \label{eq:g-metric}
\end{equation}
with $ds_{\rm int}^2$ a compact internal metric and $A=B$ at zero temperature, governed by the
Nambu--Goto action
\begin{equation}
 S_{\rm NG}=-T_s\int d^2\sigma\sqrt{-\det h},\qquad
 h_{ab}=G^{(s)}_{MN}\partial_aX^M\partial_bX^N .
 \label{eq:g-ng}
\end{equation}
With the ansatz $u=u(r)$, $X^1+iX^2=r\,e^{i\omega t}$, the solutions in a confining background take
the form sketched in Fig.~\ref{fig:holo-string-map}.

The long central segment lies near the infrared point $u_*$ where the redshifted tension is
minimal and nonzero \citep{Witten:1998zw,Kinar:1998vq}, so that
\begin{equation}
 T_{\rm eff}=T_s\sqrt{A(u_*)B(u_*)} .
 \label{eq:g-tension}
\end{equation}
Flavor is introduced by probe branes \citep{Karch:2002sh}, and a vertical segment from $u_*$ to the
flavor brane has rest energy
\begin{equation}
 m_i=T_s\int_{u_*}^{u_{f,i}}du\,\sqrt{A(u)C(u)} .
 \label{eq:g-endpoint-mass}
\end{equation}
The holographic string thus reduces to the 4D rotating string with massive end points, described by the action \eqref{eq:c-action}  \citep{Kruczenski:2004me} (see Fig.~\ref{fig:holo-string-map}); identifying
$m_i$ with current or constituent quark masses requires additional matching \citep{Kruczenski:2004me,Sonnenschein:2016pim}.

\subsubsection{Excited open strings in the \WSS model}\label{ISS}

Since the \WSS model is constructed in string theory, it naturally contains higher-spin mesons as excited open string states. 
As discussed in section \ref{holoYM}, the effective tension near the tip of the cigar geometry  $u_*=\ukk$ is
\begin{equation}
 T_{\rm eff}
 =\frac{1}{2\pi\ell_s^2}\Big(\frac{\ukk^3}{Q_4}\Big)^{1/2}
 =\frac{2\lambda}{27\pi}\,M_{\rm KK}^2,
 \qquad \alpha'_{\rm eff}=\frac{27}{4\lambda}\,M_{\rm KK}^{-2} .
 \label{eq:WSS-T}
\end{equation}
Using $M_{\rm KK}\simeq949$~MeV, $\lambda\simeq16.6$, obtained by fitting the $\rho$-meson mass and the pion decay constant, yields
$\alpha'_{\rm eff}\simeq0.45~{\rm GeV}^{-2}$ and $T_{\rm eff}\simeq 0.35~{\rm GeV}^2$. The latter is roughly twice
the empirical tension $\simeq0.18~{\rm GeV}^2$. Alternatively, if we use the empirical Regge slope $0.884~{\rm GeV}^{-2}$ to fix $\lambda$, we obtain $\lambda\simeq 8.5$. 

For $\lambda\gg 1$, strings are short, as the effective string length near the tip is $\tilde\ell_s=(\alpha'_{\rm eff})^{1/2}\propto \lambda^{-1/2}$, and are localized in the region $|Z|\lesssim \lambda^{-1/2}$. Therefore, they can be quantized in a flat spacetime approximation, with curvature effects treated perturbatively. 
Then, after eliminating the $\SO(5)$ non-invariant states, \footnote{See section \ref{D4D8} for the discussion. In addition to the $\SO(5)$ symmetry, $\tau$-parity can be used to further constrain the spectrum. See \citet{Imoto:2010ef} for details.}
the meson masses are labeled by two integers $N$ and $n$, where $N=0,1,2,\dots$ is the excitation level of the string oscillation, while $n=0,1,2,\dots$ labels the excitation number associated with the motion in the $Z$ direction. States with the same $(N,n)$ are degenerate at this order.
For example, the states with $(N,n)=(0,n)$ correspond to $\varphi^{(0)}$ for $n=0$ and 
$v^{(n)}_\mu(x)$ for $n=1,2,\dots$ in the expansion (\ref{expansion}). 

The spectrum of the excited open string states with $N=1,2$ was analyzed by \citet{Imoto:2010ef}.
The states in the first excited level $(N,n)=(1,0)$ are ($i,j,k$ spatial rest-frame indices, $h_{ij}$ traceless):
\begin{center}
\begin{tabular}{@{}cccccccc@{}}
\hline
Component & $h_{ij}$ & $h_{iz}$ & $h_{zz}$ & $h_{yy}$ & $A_{ijk}$ & $A_{ijz}$ & $\varphi$\\
\hline
$J^{PC}$ & $2^{++}$ & $1^{+-}$ & $0^{++}$ & $0^{++}$ & $0^{-+}$ & $1^{--}$ & $0^{++}$\\
\hline
\end{tabular}
\end{center}
For $(N,n)=(2,0)$, one finds $3^{--}$, $2^{++}$, $2^{-+}(\times2)$, $2^{--}$, $1^{--}(\times7)$,
$1^{+-}(\times4)$, $1^{++}(\times3)$, $1^{-+}$, $0^{++}(\times2)$, $0^{-+}(\times6)$. 

Let us compare these states with the mesons listed in the PDG meson summary table \citep{ParticleDataGroup:2026mpi}. (See Table \ref{excitedmeson}.)
Here, we focus on the $I=1$ states to avoid the analysis of mixing with glueballs.
Since $N=0$ states consist of $J^{PC}=0^{-+}$ (pion), $1^{--}$ ($\rho$-meson) and $1^{++}$ ($a_1$-meson), the mesons with the other quantum numbers should be identified with the excited states with $N\ge 1$. The $2^{++}$ state with $N=1$ is naturally identified as the lightest $2^{++}$ meson, $a_2(1320)$. Then, we expect that all the $(N,n)=(1,0)$ states found above are nearly degenerate around 1300 MeV.
Possible candidates for 
$1^{+-}$, $0^{++}$ and $0^{-+}$ states with $N=1$ are $b_1(1235)$, $a_0(1450)$ and $\pi(1300)$, respectively.
Note that the lightest $0^{++}$ state is $a_0(980)$, but this looks rather too light to be identified with an $N=1$ state. As a candidate for the $1^{--}$ state with $N=1$, we find $\rho(1450)$. 
However, since $1^{--}$ states are also contained in the spectrum of the $N=0$ states this identification is less clear. If the $1^{--}$ state with $N=1$ is identified with $\rho(1450)$, the $(N,n)=(0,2)$ vector meson with 1606 MeV listed in Table \ref{tab:vmasses} could be interpreted as $\rho(1570)$ or $\rho(1700)$.
Similarly, the $3^{--}$ state with $N=2$ is identified with the lightest $3^{--}$ meson, $\rho_3(1690)$. Thus, the $(N,n)=(2,0)$ states found above are expected to be nearly degenerate around 1700 MeV. 
For example, the spectrum of the $N=2$ states obtained above predicts the existence of a $2^{-+}$ state around 1700 MeV, which can be naturally identified with the observed $\pi_2(1670)$ meson.
See Table \ref{excitedmeson} and \citep{Imoto:2010ef} for more discussion.
\begin{table}
    \centering
    \begin{tabular}{l|ccccccc}
    \hline
    $0^{-+}$$(\pi)$ & \color{red}{\underline{$\pi$}}&\color{blue}{\underline{$\pi(1300)$}} &\color{green}{$\pi(1800)$}\\
    $0^{++}$$(a_0)$ & $a_0(980)$ &\color{blue}{$a_0(1450)$}&\color{green}{$a_0(1710)^\Delta$}&$a_0(1950)^\Delta$\\
    $1^{--}$$(\rho)$& $\color{red}{\underline{\rho(770)}}$&\color{blue}{$\rho(1450)$}&\color{red}{$\rho(1570)^\Delta$}&\color{green}{$\rho(1700)$}&$\rho(1900)^\Delta$&$\rho(2150)^\Delta$\\
    $1^{++}$$(a_1)$&\color{red}{\underline{$a_1(1260)$}}&\color{green}{$a_1(1640)$} \\
    $1^{+-}$$(b_1)$&\color{blue}{\underline{$b_1(1235)$}}&\color{green}{?}\\
    $1^{-+}$$(\pi_1)$&$\pi_1(1400)^\Delta$&\color{green}{$\pi_1(1600)$}\\
    $2^{--}$$(\rho_2)$&\color{green}{?} \\
    $2^{++}$$(a_2)$&\color{blue}{\underline{$a_2(1320)$}}&\color{green}{$a_2(1700)$} \\
    $2^{-+}$$(\pi_2)$&\color{green}{\underline{$\pi_2(1670)$}}&$\pi_2(1880)$&$\pi_2(2005)^\Delta$&$\pi_2(2100)^\Delta$\\
    $3^{--}$$(\rho_3)$&\color{green}{\underline{$\rho_3(1690)$}} &$\rho_3(1990)^\Delta$&$\rho_3(2250)^\Delta$\\
    $4^{++}$$(a_4)$&$a_4(1970)$\\    
    $5^{--}$$(\rho_5)$&$\rho_5(2350)^\Delta$\\
    \hline
    \end{tabular}
    \caption{The light unflavored mesons with $I=1$ listed in the PDG meson summary table \citep{ParticleDataGroup:2026mpi} and a possible interpretation as open string states with $N=0,1,2$. The mesons marked with $\Delta$ are experimentally not well-established.
    The mesons shown in red correspond to the states with $N=0$ and $n=0,1,2,3$, while those shown in blue and green are considered to be states with $(N,n)=(1,0)$ and $(N,n)=(2,0)$, respectively. The underlines indicate that the interpretations are relatively more reliable than the others,  because they cannot be obtained as excitations of lower $N$ states.
    }
    \label{excitedmeson}
\end{table}

As we see in Table \ref{excitedmeson}, the spectrum predicted by string theory does not contain a state that can be naturally identified with $a_0(980)$ or $\pi_1(1400)$. This may suggest that these mesons are tetraquark states.
In the large-$N_c$ limit, interactions among $q\bar q$ bound states are suppressed, which may explain why these states are absent in the analysis based on large-$N_c$ approximation.
The highest-spin state at each level has $N=J-1$; for $n=0$
\begin{equation}
 M^2=\frac{N}{\alpha'_{\rm eff}}+\sqrt{\frac{N}{2\alpha'_{\rm eff}}}\,M_{\rm KK}
 +\mathcal O(\lambda^{0}) ,
 \label{eq:iss-regge}
\end{equation}
i.e.\ $J=1+\alpha'_{\rm eff}M^2-\sqrt{\alpha'_{\rm eff}(J-1)/2}\,M_{\rm KK}$ up to
the same order.
With the Regge slope $\alpha'_{\rm eff}\simeq0.45~{\rm GeV}^{-2}$ estimated by
\eqref{eq:WSS-T}, the formula \eqref{eq:iss-regge} implies that the mass of the $N=1$ states lies at $M\simeq1.8$~GeV, which is too heavy;
with a fitted $\alpha'_{\rm eff}\simeq1$--$1.1~{\rm GeV}^{-2}$, \eqref{eq:iss-regge} reproduces
$a_2(1320)$, $\rho_3(1690)$, $a_4(1970)$, $\rho_5(2350)$, $a_6(2450)$ to within about $5\%$ (see Table~\ref{tab:num}).
However, at strong coupling, the masses of the KK modes of the D8 gauge fields are $\mathcal O(M_{\rm KK})$, while
massive string modes have $m_N\sim\sqrt\lambda\,M_{\rm KK}$. This gap, together with unwanted
compactification modes, limits quantitative comparison with QCD: the construction organizes string
states in its own regime, and precision spectroscopy requires a justified extrapolation beyond it.

\subsubsection{The Holography Inspired Stringy Hadron model}\label{sec:HISH}

The Holography Inspired Stringy Hadron (HISH) framework retains the string and its endpoint masses
as effective 4D degrees of freedom, fitting tension, masses and intercepts to hadronic spectra
\citep{Sonnenschein:2016pim}. It is a phenomenological continuation of the holographic picture, not a
controlled expansion from large $\lambda,N_c$ to QCD. Each trajectory is described by the classical
curve plus an intercept,
\begin{equation}
 J=\mathcal J_{\rm cl}(M;m_1,m_2,T)+a_J .
 \label{eq:h-fit}
\end{equation}
The intercept depends on the convention: fitting $L_{\rm orb}$ instead of $J$ gives $a_J=a_L+S$ for
an aligned family. Quark-spin operators are absent from the basic endpoint action.

The universal fit uses
\begin{equation}
 m_{u/d}=0.060\,\mathrm{GeV},\quad m_s=0.220\,\mathrm{GeV},\quad
 m_c=1.500\,\mathrm{GeV},\quad \alpha'_{\rm eff}=0.884\,\mathrm{GeV}^{-2},
 \label{eq:h-parameters}
\end{equation}
with separate intercepts for nine trajectories ($a_\pi=-0.33$, $a_\rho=0.52$, $a_\eta=-0.22$,
$a_\omega=0.53$, $a_{K^*}=0.50$, $a_\phi=0.46$, $a_D=-0.19$, $a_{D_s^*}=-0.39$, $a_\psi=-0.06$),
excluding bottomonium; heavy-quark radial trajectories also prefer different slopes
\citep{Sonnenschein:2014jwa}. Individual (non-universal) fits in the
orbital plane \citep{Sonnenschein:2014jwa} are shown in Table \ref{tab:hish-individual-fits}.
\begin{table}[!ht]
\centering
\setlength{\tabcolsep}{5pt}
\begin{tabular*}{\textwidth}{@{\extracolsep{\fill}}lcccc@{}}
\toprule
Trajectory & $m$ [MeV] & $\alpha'$ [GeV$^{-2}$] & $a$ & $N_{\mathrm{states}}$ \\
\midrule
$\rho/a$
  & $m_{u/d}=0\text{--}180$
  & $0.883\text{--}0.933$ & $0.47\text{--}0.66$ & 6 \\
$\pi/b$
  & $m_{u/d}=90\text{--}185$
  & $0.808\text{--}0.863$ & $-0.23\text{ to }0.00$ & 4 \\
$\omega/f$
  & $m_{u/d}=0\text{--}60$
  & $0.910\text{--}0.918$ & $0.45\text{--}0.50$ & 6 \\
$K^*$
  & $\begin{gathered}
       m_{u/d}=0\text{--}240\\[-2pt]
       m_s=0\text{--}390
     \end{gathered}$
  & $0.848\text{--}0.927$ & $0.32\text{--}0.62$ & 5 \\
$\phi/f'$
  & $m_s=400$
  & $1.078$ & $0.82$ & 3 \\
$\Upsilon$ (angular)
  & $m_b=4730$
  & $0.635$ & $1.00$ & 3 \\
$\Upsilon$ (radial)
  & $m_b=4725\text{--}4740$
  & $0.455\text{--}0.471$ & $1.00$ & 6 \\
\bottomrule
\end{tabular*}
\caption[Individual massive-endpoint rotating-string fits]{%
Selected individual meson-trajectory fits from \citep{Sonnenschein:2014jwa}. The first six rows concern angular trajectories; the last
concerns radial excitations. Here $m_q$ denotes an effective string-endpoint
mass, $\alpha'=(2\pi T)^{-1}$, and the intercept convention is
$J+n-a=J_{\mathrm{cl}}(M)$.  We refer the reader to the original reference for a discussion of the displayed ranges.}
\label{tab:hish-individual-fits}
\end{table}

Light endpoint masses are poorly constrained ($\chi^2$ flat up to $\sim200$~MeV); $m_s\sim200$--$500$~MeV
depending on the trajectory, $m_c\simeq1.4$--$1.7$~GeV, $m_b\simeq4.73$~GeV. A string with a quark
at one end and a diquark at the other describes baryons, with $\alpha'_{\rm baryon}\simeq0.95~{\rm
GeV}^{-2}$, which disfavors the Y-shaped configuration (slope $\tfrac23\alpha'$) \citep{Sonnenschein:2014bia};
glueballs are folded closed strings with half the slope \citep{Sonnenschein:2015zaa}.

The first quantum correction around the classical trajectory, with the endpoint boundary conditions,
follows from fluctuation quantization and the effective theory of long strings
\citep{Baker:2002km,Zahn:2013yma,Aharony:2013ipa,Zahn:2016bam}; in noncritical dimension the action
needs the term restoring worldsheet consistency \citep{Polchinski:1991ax,Hellerman:2013kba}. The
resulting intercept belongs to this effective string theory, and matching it to a given meson
trajectory is a separate question. The results of \citet{Imoto:2010ef} (ISS) and \citep{Sonnenschein:2014jwa} (HISH)  are compared in Table~\ref{tab:num}.
\begin{table}[tbp]
\centering 
\setlength\tabcolsep{4.5pt}
\begin{tabular}{lccccccc}
\toprule
State & $J^{PC}$ & ISS $(N,n)$ & PDG [MeV] & ISS $\alpha'{=}0.45$ & ISS $\alpha'{=}1.1$ & HISH univ. & trajectory\\
\midrule
$\rho(770)$     & $1^{--}$ & $(0,1)$ & 775  & input & input & 776  & $\rho/a$ \\
$b_1(1235)$     & $1^{+-}$ & $(1,0)$ & 1230 & 1795 & 1245 & 1257 & $\pi/b$ \\
$a_2(1320)$     & $2^{++}$ & $(1,0)$ & 1318 & 1795 & 1245 & 1324 & $\rho/a$ \\
$\pi(1300)$     & $0^{-+}$ & $(1,0)$ & 1300 & 1795 & 1245 & --   & -- \\
$a_0(1450)$     & $0^{++}$ & $(1,0)$ & 1439 & 1795 & 1245 & --   & -- \\
$\pi_2(1670)$   & $2^{-+}$ & $(2,0)$ & 1671 & 2421 & 1650 & 1650 & $\pi/b$ \\
$\rho_3(1690)$  & $3^{--}$ & $(2,0)$ & 1689 & 2421 & 1650 & 1701 & $\rho/a$ \\
$a_4(1970)$     & $4^{++}$ & $(3,0)$ & 1967 & 2898 & 1958 & 2008 & $\rho/a$ \\
$\rho_5(2350)$  & $5^{--}$ & $(4,0)$ & 2330 & 3300 & 2217 & 2274 & $\rho/a$ \\
$a_6(2450)$     & $6^{++}$ & $(5,0)$ & 2450 & 3653 & 2445 & 2512 & $\rho/a$ \\
\bottomrule
\end{tabular}
\caption{Masses (MeV) from Eq.~\eqref{eq:iss-regge} with $M_{\rm KK}=949$~MeV (top-down and
fitted slope) and from the HISH universal fit \eqref{eq:h-parameters} ($a_\rho=0.52$, $a_\pi=-0.33$). ISS \citep{Imoto:2010ef} states at fixed $(N,n)$ are degenerate at this order; the $\rho(770)$
belongs to the massless gauge sector and fixes $M_{\rm KK}$.
}
\label{tab:num}
\end{table}

The long-string effective field theory of \citet{Cuomo:2024gek} organizes the light-quark spectrum as a $1/J$
expansion,
\begin{equation}
 M^2=\frac{J}{\bar\ell_s^2}+\frac{2\mu_1}{\bar\ell_s}J^{1/4}-\frac{1}{\bar\ell_s^2}
 +\frac{2\mu_2}{\bar\ell_s}J^{-1/4}+\frac34\mu_1^2J^{-1/2}+\dots,\qquad \bar\ell_s^2=\alpha'_{\rm eff},
 \label{eq:q-eft}
\end{equation}
where the $J^{1/4}$ term reproduces the classical $m^{3/2}$ correction from massive endpoints
($\mu_1\supset\tfrac43\sqrt\pi\,m^{3/2}\bar\ell_s^{1/2}$), the $-1/\bar\ell_s^2$ term is the intercept
$a=1$, and the boundary Wilson coefficients $\mu_{1,2}$ are free parameters. The same work treats
heavy quarks, quark spin and daughter trajectories from the worldsheet axion.

Decays proceed by creation of a new endpoint pair, splitting the open string in two. In a confining
holographic geometry, a horizontal segment fluctuates toward a flavor brane and splits on reaching
it \citep{Peeters:2005fq,Cotrone:2005fr}; worldsheet instantons reproduce the Casher--Neuberger--Nussinov rate for
long strings \citep{Peeters:2018xei}. The phenomenological HISH width is \citep{Sonnenschein:2017ylo}
\begin{equation}
 \Gamma=\frac{\pi}{2}\,A\,T\,L(M,m_1,m_2,T),\qquad A\simeq0.095\pm0.015,\qquad
 \frac{\Gamma_i}{\Gamma}=\Phi_i\exp\!\Big[-2\pi C\,\frac{m_{\rm sep}^2}{T}\Big],\quad C\simeq1,
 \label{eq:d-width}
\end{equation}
linear in the string length, with heavy-pair creation exponentially suppressed ($s\bar s$
suppression $\simeq0.3$--$0.4$). A specific channel still requires its kinematics and spin and
flavor factors.

The literature on this subject is vast; we refer the reader to the review \citep{Sonnenschein:2016pim} for a comprehensive list of references.

\subsection{Hadronic light-by-light scattering}
\label{sec:hlbl}

\begin{figure}[htbp]
 \centering \includegraphics[width=1.\linewidth]{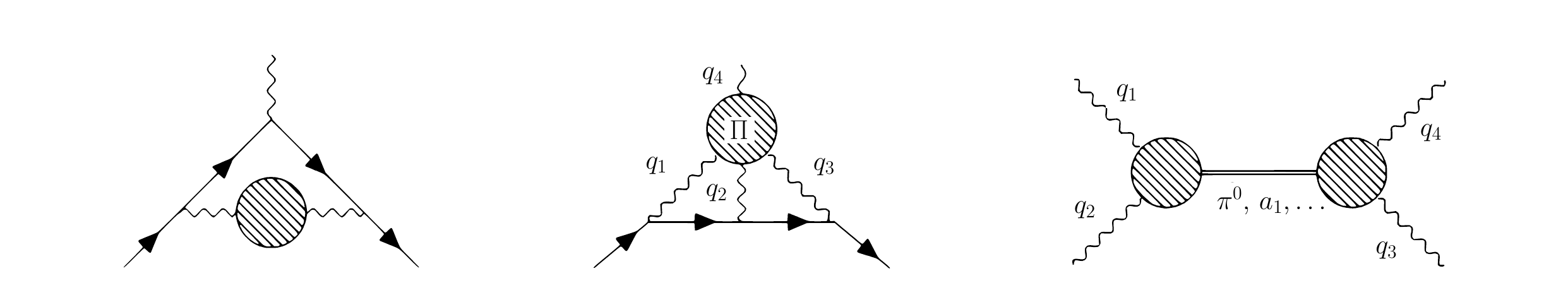}
     \caption{HVP and HLbL contributions to the muon-photon vertex diagram, with 
    single-particle exchanges in the HLbL amplitude (only one of the possible permutations of the photon lines is shown)
    }
    \label{fig:muong-2}
\end{figure}

In recent years, the experimental efforts to determine with high precision
the anomalous magnetic moment of the muon, which culminated in the
Fermilab Muon $g-2$ experiment \citep{Muong-2:2025xyk}, have triggered a large amount of
studies of the interactions of hadrons with photons, which give rise to
the radiative corrections shown in Fig.~\ref{fig:muong-2} and which are
responsible for virtually all of the uncertainties in the Standard Model (SM)
prediction; see
the White Papers of the International Muon $g-2$ Theory Initiative
\citep{Aoyama:2020ynm,Aliberti:2025beg}. To match the experimental precision, the contributions from
hadronic vacuum polarization (HVP) need to be known at the per-mille level,
which is the goal of both data-driven and lattice QCD studies;
their combination is indeed capable of reaching the required
level as demonstrated by \citet{Boccaletti:2024guq}, and will hopefully be further
improved in the future. A smaller contribution that is also crucial for the SM prediction 
arises from the 4-point amplitude of hadronic light-by-light scattering (HLbL)
with an uncertainty
of (currently) about $10\%$ \citep{Aliberti:2025beg}.
This is evaluated also by data-driven studies using dispersion theory
as well as by lattice QCD simulations, which yield compatible
results with similar precision, albeit with a certain tension between them.
Holographic QCD has turned out to be a valuable complementary approach, since
it provides a purely hadronic model in which short-distance constraints
of QCD can be realized in situations where conventional hadronic models
with a finite number of resonances are bound to fail. Moreover, it
could rather accurately reproduce and in some cases predict the
experimental and dispersion-theoretic results for transition form factors that determine
the HLbL amplitude in single-meson exchange contributions.

\subsubsection{Pseudoscalar exchange}
The largest contribution to the HLbL amplitude (see Fig.~\ref{fig:muong-2}) is the
exchange of a neutral pion which couples anomalously to two photons.
For virtual photons, the momentum dependence of this coupling is 
described by the transition form factor \eqref{TFFpi},
which involves
the vector bulk-to-boundary propagator encoding the
coupling to the infinite tower of vector-meson resonances; see e.g.\ \eqref{HWVF}. 
Its implications for the muon $g-2$ anomaly have been initially considered
by \citet{Hong:2009zw,Cappiello:2010uy} with a first complete
evaluation by \citet{Leutgeb:2019zpq}.

The short-distance behavior of a pseudoscalar transition form factor
has been derived in pQCD by means of a light-cone expansion by Brodsky and
Lepage and others
\citep{Lepage:1979zb,Lepage:1980fj,Brodsky:1981rp,Efremov:1979qk,Hoferichter:2020lap}
with the result
\be\label{Fpiggasymptotics}
F_{P\gamma^*\gamma^*}(Q_1^2,Q_2^2)\propto \frac{1}{Q^2}
\left[\frac1{w^2}-\frac{1-w^2}{2w^3}\ln\frac{1+w}{1-w}\right],
\ee
for $Q^2=\frac12 (Q_1^2+Q_2^2)\to\infty$ and $w=(Q_1^2-Q_2^2)/(Q_1^2+Q_2^2)$.

While ordinary hadronic models with VMD and a finite number of vector-meson resonances
cannot reproduce the above dependence on $w$ in the doubly virtual case, 
the asymptotic behavior \eqref{Fpiggasymptotics} is a universal result for holographic
QCD models with a bulk geometry that is asymptotically AdS$_5$
\citep{Grigoryan:2008up,Leutgeb:2021mpu} (this excludes the \WSS
model, which has a different asymptotic dependence on $Q$ and $w$
\citep{Leutgeb:2019gbz}).

\begin{figure}
\centering
\includegraphics[width=0.33\textwidth,clip]{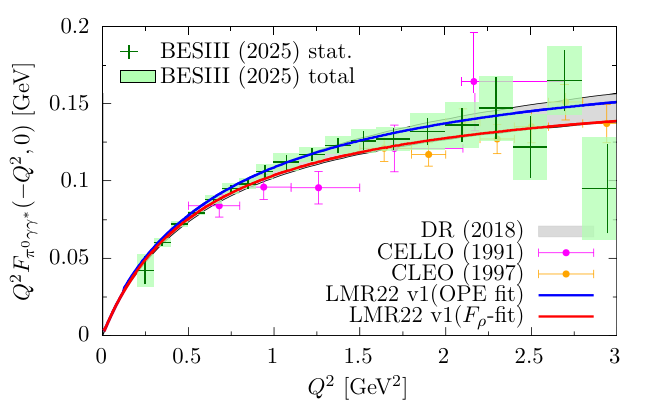}\includegraphics[width=0.33\textwidth,clip]{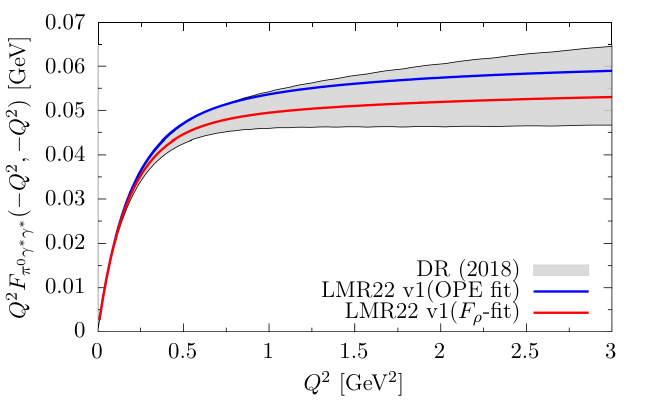}
\includegraphics[width=0.33\textwidth,clip]{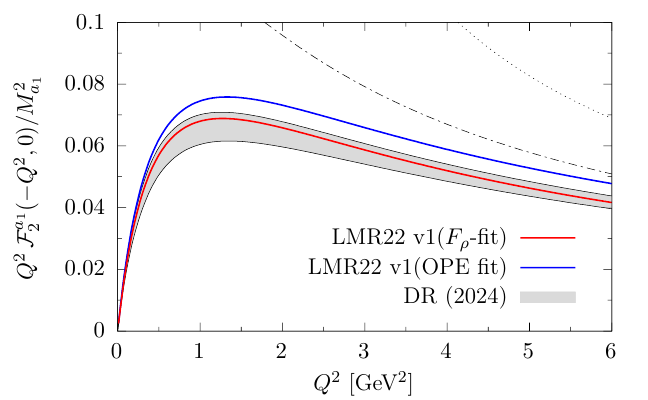}
\caption{Results for the singly and doubly virtual
$\pi^0$ transition form factors, and for the singly-virtual $a_1(1260)$ transition form factor
obtained in the hard-wall model of \citet{Leutgeb:2022lqw} (LMR22) with
OPE fit of $g_5$ according to Eq.~\eqref{eq:g5} (blue lines) and  
an alternative fit through the $\rho$ decay constant $F_\rho$ (red lines) compared to the dispersive results of
\citep{Hoferichter:2018kwz,Ludtke:2024ase} (DR)
and experimental data, the latter including the recent data from BESIII \citep{BESIII:2025zjx}).
In the plot for $a_1$, the dash-dotted and dotted lines show
the LCE limit \citep{Hoferichter:2020lap}  with and without mass corrections,
respectively.
(Figures taken from \citep{Aliberti:2025beg} and \citep{Mager:2026ddo}.)}
\label{fig:pi0a1TFF}       
\end{figure}

In the hard-wall model\footnote{In the conventional soft-wall model,
the corresponding results have been obtained by \citet{Colangelo:2023een,Colangelo:2024xfh}.
However, as found by \citet{Leutgeb:2025jmv},
this model leads to an ill-defined total HLbL amplitude
due to divergences and needs modification.}
with transition form factor \eqref{TFFpi},
the numerical results for $\pi^0$ 
and the resulting contribution to the muon $g-2$
turn out to agree surprisingly accurately with
both the data-driven results and the lattice QCD results, as shown
in Fig.~\ref{fig:pi0a1TFF}; see \citet{Leutgeb:2021mpu,Leutgeb:2022lqw,Leutgeb:2024rfs,Aliberti:2025beg} for
further details and results for the $\eta$ and $\eta'$ mesons.

\subsubsection{Axial-vector mesons and the Melnikov-Vainshtein constraint}

After pseudoscalar meson exchange, the second most important contribution
to the HLbL amplitude is due to three towers of axial-vector mesons
with the ground-state representatives $a_1(1260)$, $f_1(1285)$, and $f_1'(1420)$.
In holographic QCD, their couplings to photons is derived
exclusively from the CS action, as is the case for pseudoscalars.

According to the Landau-Yang theorem \citep{Landau:1948kw,Yang:1950rg}, axial-vector mesons cannot decay into two real photons, and this is
automatically realized by the structure of the amplitude $A^{(n)a}\to\gamma^*\gamma^*$ in hQCD\footnote{Compared to the most general expression possible for this
amplitude \citep{Pascalutsa:2012pr,Roig:2019reh,Zanke:2021wiq}, the holographic result
(\ref{calMa}) has one asymmetric structure function (denoted $C(Q_1^2,Q_2^2)$
by \citet{Roig:2019reh}) set to zero; this is in fact broadly
consistent with phenomenology \citep{Zanke:2021wiq}.}
\be\label{calMa}
\mathcal{M}(A^{(n)a}\to\gamma^*\gamma^*)=i\frac{N_c}{4\pi^2}\mathrm{tr}(\mathcal{Q}^2 t^a)\,\epsilon_{(1)}^\mu \epsilon_{(2)}^\nu
\epsilon_{A}^{*\rho} \epsilon_{\mu\nu\rho\sigma}
\left[q_{(2)}^\sigma Q_1^2 A_n(Q_1^2,Q_2^2)-q_{(1)}^\sigma Q_2^2 A_n(Q_2^2,Q_1^2)\right],
\ee
with transition form factor
\begin{align}\label{TFFAn} 
A_n(Q_1^2,Q_2^2) &= \frac{2g_5}{Q_1^2} \int_0^{z_0} dz \left[ \partial_z \mathcal{J}(Q_1,z) \right]
\mathcal{J}(Q_2,z) \,\psi^A_n(z),
\end{align}
because $\partial_z \mathcal{J}(Q,z)\sim Q^2$ as $Q^2\to 0$.

Again, there is a universal result for the short-distance behavior in all
hQCD models with a bulk geometry that is asymptotically AdS$_5$, obtained
first by \citet{Leutgeb:2019gbz}
\be
A_n(Q_1^2,Q_2^2) \to 
\frac{g_5^2 F^A_{n}}{Q^4}
\frac1{w^4}\left[
w(3-2w)+\tfrac12 (w+3)(1-w)\ln\frac{1-w}{1+w}
\right],
\ee
which anticipated the result derived
from light-cone expansions in pQCD in
\citep{Hoferichter:2020lap}.
Moreover, the quantitative results for the transition form factor
and the contribution to the muon $g-2$ were confirmed
by the recent dispersive results of \citet{Ludtke:2024ase,Hoferichter:2024bae}
(see Fig.~\ref{fig:pi0a1TFF}) adopted by the White Paper 2025 \citep{Aliberti:2025beg}
for the data-driven estimate of the HLbL contribution.

The one-loop exactness of the chiral anomaly together with the OPE
lead to the short-distance constraint of Melnikov and Vainshtein (MV) \citep{Melnikov:2003xd} in the region
$Q_1^2 \sim Q_2^2\gg Q_3^2 \gg \Lambda^2_\mathrm{QCD}$
and $Q_4=0$, which
in the Bardeen-Tung-Tarrach basis of the HLbL four-point function \citep{Colangelo:2015ama}
reads \citep{Colangelo:2019lpu,Colangelo:2019uex}
\be\label{MVSDC}
\lim_{Q_3\to\infty} \lim_{Q\to\infty} Q^2 Q_3^2 \bar\Pi_1(Q,Q,Q_3)=-\frac{2}{3\pi^2}
\ee
for $N_c=N_f=3$. 
The short-distance behavior of the transition form factors implies that each individual 
pseudoscalar or axial-vector meson 
gives a vanishing contribution $\bar\Pi_1(Q,Q,Q_3)\sim Q^{-2} Q_3^{-4}$.
However, \citet{Leutgeb:2019gbz,Cappiello:2019hwh} have shown that
the infinite tower of axial-vector mesons in hQCD does naturally saturate this constraint,
providing the first hadronic model consistent with this requirement (in the chiral
limit \footnote{Away from the chiral limit, also excited pseudoscalars couple to the axial
vector current and thus to the axial anomaly. \citet{Colangelo:2019lpu,Colangelo:2019uex} have constructed a Regge model 
for excited pseudoscalars
in order to satisfy the MV constraint and to estimate
the corresponding contributions to the muon $g-2$.}).
Moreover, the quantitative results for short-distance contributions to the
muon $g-2$ in the hQCD model \citep{Leutgeb:2022lqw} turned out to agree
remarkably well with the 
dispersive analysis of \citet{Hoferichter:2024bae}, where axial-vector
contributions were included fully for the first time.

\subsubsection{Tensor meson contributions}\label{sec:hlbltensor}

It was found recently that the contribution of tensor mesons to HLbL and the muon $g-2$
had been underestimated in the past, which is related to overlooked problems of kinematical singularities in the formalism used
\citep{Hoferichter:2024bae}. With a simple dipole ansatz for only one of five
independent transition form factors that tensor mesons can have, the
previous evaluation of \citet{Danilkin:2016hnh} for
the contribution of the tensor mesons $f_2(1270)$ and $a_2(1320)$ gave
a tiny positive contribution $\approx 0.64\times 10^{-11}$ to $a_\mu=(g-2)_\mu/2$.
With the new optimized basis of \citet{Hoferichter:2024fsj}, \citet{Hoferichter:2024bae}
obtain instead a much larger and negative contribution $\approx -2.5(3)\times 10^{-11}$.

In hQCD, superseding an expressly incomplete calculation of \citet{Colangelo:2024xfh},
the transition form factors and contributions to $a_\mu$ of tensor mesons
in the minimal model of \citet{Katz:2005ir} (see Sec.~\ref{sec:tensormesons})
were worked out by \citet{Mager:2025pvz,Cappiello:2025fyf}.
In this model, two out of the five transition form factors appear, and the
additional one has the effect of changing not only the magnitude but also
the sign of the total contribution, implying again a positive one.

The infinite tower of tensor mesons in hQCD also contributes to the
asymptotic limits of the HLbL amplitude, again despite vanishing contributions
from individual resonances. The MV limit \eqref{MVSDC} is unaffected, but
the symmetric longitudinal short-distance constraint \citep{Melnikov:2003xd,Colangelo:2019lpu,Bijnens:2020xnl,Bijnens:2021jqo}
\begin{align}\label{symLSDC}
    \lim_{Q\to\infty}
    Q^4 \bar\Pi_1(Q,Q,Q)=-\frac4{9\pi^2},
\end{align}
where the tower of axial-vector mesons contributes only $81\%$ of this result, can be
completed by the tensor contribution. Tuning the strength of 
tensor interaction such that
\eqref{symLSDC} is matched
fixes also the two-photon width of tensor mesons, which
turns out to be fully consistent with the observed
experimental values for the combined ground-state multiplet.

In the
minimal model considered by \citet{Mager:2025pvz,Cappiello:2025fyf}, the
tower of tensor mesons contributes in total $\approx +11\times10^{-11}$
to $a_\mu$, with $\approx +8.5\times10^{-11}$ coming from
the low-energy region $Q_i<1.5\,\mathrm{GeV}$, while the
contributions from larger $Q_i$ actually improve the agreement
with the amount of short-distance contributions estimated
in the dispersive approach \citep{Aliberti:2025beg}.
A positive contribution of this size from tensor mesons might
in fact remove the tension observed by \citet{Aliberti:2025beg} between
the dispersive and the lattice QCD results for the total HLbL contribution
\citep{Mager:2025pvz}.

An important caveat in the treatment of \citet{Mager:2025pvz,Cappiello:2025fyf}
is that
only terms with the lowest order of derivatives
in the action have been kept, namely an Einstein-Hilbert action
and a minimal coupling of the metric to the Yang-Mills action
following from five-dimensional diffeomorphism invariance.
A full description (already to leading order in $N_c$) would
certainly require the inclusion of higher-derivative terms,
because those generically contribute to all five tensor
transition form factors, and all five have nontrivial short-distance
limits 
\citep{Hoferichter:2020lap} of the same order in $N_c$
(the two transition form factors of the minimal model
have the correct scaling in $Q$, but wrong asymmetry functions).
In the minimal model, one simply assumes that 
the two nonzero tensor transition form factors
obtained in the minimal model are the most important ones
phenomenologically.
A direct experimental test will become possible when
new experimental studies of tensor mesons probe
also the doubly virtual regime; the singly virtual data
available \citep{Belle:2015oin,Lellmann:2026ajj} agree well with the holographic prediction but
do not constrain the additional tensor
transition form factor obtained by \citet{Mager:2025pvz,Cappiello:2025fyf}.

\subsection{Gravitational form factors}
\label{GFF}

The gravitational form factors are scalar functions that characterize the internal structure of hadrons through matrix elements of the energy-momentum tensor.
From the symmetry and the conservation law of the energy-momentum tensor, the matrix elements are written as
\be
\langle p'|T_{\mu\nu}(x)|p\rangle
=\left[2P_\mu P_\nu A(t)
+\tfrac{1}{2}(k_\mu k_\nu-\eta_{\mu\nu}k^2)D(t)
\right]
e^{ikx} \ ,
\ee
for spin-0 states and
\begin{align}
\langle p',s'|T_{\mu\nu}(x)|p,s\rangle
&=\bar u(p',s')\left[
\frac{P_\mu P_\nu}{M}A(t)
+\frac{iP_{(\mu}\sigma_{\nu)\rho} k^\rho}{2M}J(t)+\frac{k_\mu k_\nu-\eta_{\mu\nu}k^2}{4M}D(t)\right]
u(p,s)\,e^{ikx}
\end{align}
for spin-1/2 states, where $p$ and $p'$ ($s$ and $s'$) are the momenta (spin) of the hadrons, $P:=\frac12(p'+p)$, $k:=p'-p$, $t:=k^2$ and $M$ is the mass of the hadron. Here, the normalization of the state $|p\rangle$ and the spinor $u(p,s)$ are chosen such that $\langle p'|p\rangle=2p^0(2\pi)^3\delta^{(3)}(\boldsymbol{p'}-\boldsymbol{p})$ and $\bar u(p,s)u(p,s)=2M$.
The functions $A(t)$, $D(t)$ and $J(t)$ are called gravitational form factors, which cannot be fixed by symmetries alone, but reflect the internal dynamics of the hadron. Roughly speaking, $A(t)$ and $J(t)$ are related to the distributions of mass and angular momentum, and $D(t)$ describes the pressure and shear force inside the hadron.\footnote{See, e.g., \citep{Polyakov:2018zvc,Burkert:2023wzr} for reviews.} 
The values of $A(t)$ and $J(t)$ (for spin-1/2 states) in the $t\ra0$ limit are $1$ and $1/2$, respectively. In contrast, $D(0)$ is not fixed by symmetries, except for the Nambu-Goldstone bosons for which $D(0)=-1$. The quantity $D(0)$, called the D-term, has not yet been precisely determined experimentally and has attracted considerable interest.

Because the energy-momentum tensor couples to the gravitational field, the gravitational form factors encode the coupling of hadrons to gravitons. However, since the gravitational interaction is extremely weak, it is currently impractical to determine the 
gravitational form factors through the scattering process involving graviton exchange. Instead, these form factors have recently been extracted indirectly through experimentally accessible processes \citep{Kumano:2017lhr,Burkert:2018bqq,Burkert:2021ith,Duran:2022xag}.
In light of these developments, theoretical predictions for gravitational form factors have become increasingly important.

Holographic QCD provides a useful framework for computing the gravitational form factors, since it naturally contains a gravitational field. The energy-momentum tensor can be obtained by the variation of the action with respect to the boundary value of the metric, as we did in (\ref{varS}) for the currents associated with the chiral symmetry, and its matrix elements can be calculated by sandwiching it between the hadron states. Calculations of the gravitational form factors in holographic QCD were first performed by Abidin and Carlson around 2008: \citep{Abidin:2008ku,Abidin:2008hn} for pion and vector/axial vector mesons, and \citep{Abidin:2009hr} for nucleons in HW and SW models.
In these works, the gravitational form factor $A(t)$ of mesons and nucleons was calculated.
Since around 2020, the study of gravitational form factors in holographic QCD has attracted renewed attention and become an active area of research.

In the \WSS model, a general discussion of the gravitational form factors for baryons and an attempt to obtain an approximate analytical estimate of the D-term was presented by \citet{Fujita:2022jus}. It was pointed out that the gravitational form factors can be expressed as a sum of contributions from glueball exchange diagrams. This property was referred to as `glueball dominance' in direct analogy with vector meson dominance for electromagnetic form factors.
The calculation of the D-term is rather involved, and its magnitude was actually underestimated in this work, because the flat space instanton approximation discussed in section \ref{baryon} is not sufficiently accurate for this quantity. This estimate was subsequently improved by \citet{Sugimoto:2025btn} by numerically solving the equations of motion for a baryon configuration, yielding $D(0)\approx -2.05$. However, note that the calculation in this work was done in a classical approximation, and the quantization of the fluctuations around the solution was not taken into account. In particular, the $J(t)$ form factor was not calculated.
The calculation of the gravitational form factors for the pion in the \WSS model was given in
\citep{Fujii:2024rqd,Fujii:2025eug,Tanaka:2025znc}. The behavior of both $A(t)$ and $D(t)$ was consistent, at least qualitatively, with the recent lattice results \citep{Hackett:2023nkr}, and the condition $D(0)=-1$ in the chiral limit was reproduced. Gravitational form factors have also been studied in bottom-up models. See
\citep{Allahverdiyeva:2023fhn,Liu:2025vfe,Deng:2026suw,Fornetti:2026yup,Liu:2026isp}
for mesons, including pions, kaons, $\rho$ and $\eta$ mesons;
\citep{Mamo:2019mka,Mamo:2021krl,Mamo:2022eui,Deng:2025fpq,Nascimento:2025smf,Hippelainen:2026izm}
for nucleons, and \citep{Liu:2025qja} for the baryon octet.

\section{Conclusion and outlook }

Holographic QCD provides a complementary framework for investigating the nonperturbative dynamics of the strong interaction and the structure of hadrons. By relating four-dimensional quantum gauge theories in the limit of large coupling and color number to higher-dimensional gravitational theories with (mostly) classical field dynamics, it offers an intuitive, geometric perspective on phenomena that are difficult to describe analytically using the fundamental degrees of freedom of QCD. The applications reviewed in this article illustrate the potential of this approach, from the realization of confinement and spontaneous chiral symmetry breaking to the calculation of hadronic spectra, interaction strengths, and form factors. At the same time, they demonstrate that the predictive power of holographic QCD depends crucially on the choice of model and on the approximations used in its construction.

An important distinction is that between top-down and bottom-up approaches. In top-down constructions, particularly the Witten–Sakai–Sugimoto model, the dynamics of mesons, baryons, and glueballs can be related to a common string-theoretic description. The geometry provides an intuitive realization of confinement and chiral symmetry breaking, while the effective action for the flavor sector yields a systematic description of meson interactions. Baryons emerge either as wrapped D-branes or as solitonic configurations of an effective five-dimensional gauge theory. These connections illustrate how holography can relate different classes of hadrons within a common framework rather than treating each as an independent phenomenological degree of freedom.

Bottom-up models complement this perspective by allowing the effective five-dimensional action and background geometry to be adapted to specific features of QCD. Hard-wall and soft-wall models, together with their improved extensions, provide economical descriptions of hadron spectra and interactions and can incorporate constraints from the ultraviolet behavior of QCD. The resulting diversity of approaches is a strength of holographic QCD, but it also means that quantitative predictions must be assessed in the context of the particular assumptions and observables involved.

The phenomenology of mesons illustrates both the achievements and the limitations of these constructions. Holographic models provide a common description of vector and axial-vector mesons, pseudoscalar modes, and their interactions, while incorporating chiral symmetry, its spontaneous breaking, and the associated anomalies. They also offer a natural setting for vector meson dominance and for the analysis of electromagnetic transition form factors. The description of scalar and tensor mesons, however, is more model-dependent, and reproducing the observed spectrum does not by itself guarantee that the corresponding interactions and decay patterns are accurately described. The glueball sector provides a particularly clear example: although the Witten–Sakai–Sugimoto model reproduces some qualitative features of the glueball spectrum, discrepancies in predicted masses and decay widths leave important questions about the identification of particular modes and the role of model-specific states unresolved.

A central strength of holographic QCD is its ability, in suitable implementations, to connect hadronic descriptions with short-distance constraints. This is particularly relevant for hadronic light-by-light scattering and the anomalous magnetic moment of the muon. Conventional hadronic models containing only a finite number of vector-meson resonances cannot, in general, reproduce the full short-distance behavior of doubly virtual meson transition form factors; in contrast to bottom-up QCD models with asymptotic conformal symmetry, they fail to implement important short-distance constraints on hadronic light-by-light scattering such as the Melnikov-Vainshtein constraint. Holographic models incorporate this constraint naturally through its infinite tower of axial-vector resonances. The encouraging agreement obtained for several pseudoscalar and axial-vector contributions with dispersive analyses and lattice calculations illustrates the practical value of this structure. Recent studies of tensor-meson contributions further indicate that additional resonance sectors may be important. However, the results also show that simplified interactions can omit relevant form factors and short-distance contributions, so further improvements are necessary before the full phenomenological implications can be established reliably.

Holography also provides a useful framework for exploring hadronic structure beyond electromagnetic observables. Gravitational form factors offer information about the distributions of mass and angular momentum, as well as the pressure and shear forces inside hadrons. Their calculation is particularly natural in the holographic framework that incorporates gravitational fields, and the qualitative agreement of some predictions with lattice results is encouraging. At finite temperature, holographic constructions provide a geometrical description of deconfinement, chiral symmetry restoration, and the behavior of hadronic excitations in a thermal medium. At finite baryon density, they offer access to strongly coupled regimes for which conventional numerical methods face severe difficulties. These applications broaden the scope of the approach beyond vacuum spectroscopy, although quantitative predictions remain sensitive to model-dependent features of the background and to approximations in the treatment of dense matter.

Further progress will depend on developing more realistic holographic models and on testing them systematically against experiment and independent nonperturbative calculations. Improved ultraviolet matching, more consistent treatments of chiral dynamics and quark masses, and a better understanding of higher-spin and scalar or tensor excitations are important directions. In applications to precision observables, complete short-distance constraints and the role of higher-derivative interactions must be examined together with the phenomenology of individual resonances. In the study of dense matter, more controlled treatments of baryonic configurations will be essential for assessing the reliability of predicted phase structures. Lattice QCD, dispersive methods, and experimental measurements will remain indispensable in distinguishing robust holographic results from features specific to individual models.

The broader significance of holographic QCD lies not in replacing these complementary approaches, but in providing a framework in which nonperturbative dynamics, hadronic degrees of freedom, and short-distance constraints can be investigated together. Its successes demonstrate that gauge/gravity duality can yield useful insights into theories without the supersymmetry and conformal symmetry of the original AdS/CFT correspondence. Its limitations, in turn, identify where more complete constructions and more stringent phenomenological tests are required. Continued comparison with QCD calculations and data will determine how far this framework can be developed into a quantitatively reliable description of hadron structure, while clarifying which aspects of the strong interaction are most naturally understood through holography.

\begin{ack}[Acknowledgments]\1
 We wish to thank Jonas Mager for many useful comments on the manuscript.
This work was funded in part by the Austrian Science Fund (FWF), grant-DOI \url{https://www.doi.org/10.55776/PAT7221623}.  L.C.\ acknowledges the support of the INFN research project
ENP (Exploring New Physics).
The work of S.S.\ is supported by the JSPS KAKENHI (Grant-in-Aid for Scientific Research (B)) grant number JP24K00628.
\end{ack}

\appendix

\section{Dimensional deconstruction of  QCD }\label{appDD}
\citet{Son:2003et} proposed a bridge between three
otherwise distinct research lines: phenomenological models of \emph{hidden local
symmetry} \citep{Bando:1984ej,Bando:1987br}, which promote the vector mesons $\rho,\omega,\ldots$ to gauge bosons of dynamically generated local symmetries, 
\emph{dimensional deconstruction} \citep{Arkani-Hamed:2001kyx, Hill:2000mu}, a technique that reconstructs a higher-dimensional gauge theory from a (1D)  lattice of four-dimensional gauge groups connected by link fields; and \emph{gauge/gravity duality} \citep{Maldacena:1997re, Gubser:1998bc,Witten:1998qj}.

The central observation  is that the continuum limit of a 4D 
\emph{open moose} theory is a 4+1-dimensional gauge theory in a curved
background with two boundaries, offering an explicit
bottom-up holographic model for low-energy QCD.

\begin{figure}[tbp]
 \centering
 \includegraphics[width=.75\linewidth]{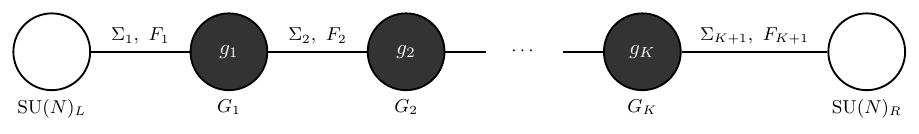}
 \caption{Open-moose diagram for generic $K$.
Open circles denote \emph{global} (ungauged) symmetry groups; filled
circles denote \emph{gauged} (dynamical) groups.}
\label{fig:moose}
\end{figure}

The model is defined by a \emph{moose diagram} \citep{Georgi:1985hf},
see Fig.~\ref{fig:moose}, with
$K$ internal nodes and two dangling ends, containing 
$K$ hidden symmetry gauge groups $G_i = \mathrm{SU}(N)_i$,
$i = 1,\ldots,K$, each with gauge coupling $g_i$,  $K+1$  link fields $\Sigma_i$,
each transforming as a sigma-model bi-fundamental field under adjacent groups, and each with a decay constant $F_i$. The two endpoint groups $G_0 = \mathrm{SU}(N)_L$ and $G_{K+1}=\mathrm{SU}(N)_R$ remain ungauged and are identified with the approximate global chiral symmetry of QCD.
The Lagrangian is
\begin{equation}
  \mathcal{L} = -\frac{1}{2}\sum_{i=1}^{K}
\frac{1}{g_i^2}\,\mathrm{tr}\,\left(F_{\mu\nu}^{(i)}\right)^2
 + \sum_{i=1}^{K+1}
\frac{F_i^2}{4}\,\mathrm{tr}
 \bigl|D_\mu \Sigma_i\bigr|^2.
\end{equation}
The link fields are scalars  $\Sigma_i \in \mathrm{SU}(N)$ transforming as
\begin{equation}
  \Sigma_i \;\longmapsto\; U_{i-1}(x)\,\Sigma_i(x)\,U_i^\dagger(x),
  \qquad U_0 \in \mathrm{SU}(N)_L,\quad U_{K+1} \in \mathrm{SU}(N)_R,
\end{equation}
where $U_0$ and $U_{K+1}$ act globally (ungauged endpoints). 
 $D_\mu\Sigma_i = \partial_\mu\Sigma_i - iA_\mu^{(i-1)}\Sigma_i
+i\Sigma_i A_\mu^{(i)}$, where the $A_\mu^{(i)}$ are the gauge fields of the $i$-th node group $G_i$. Parity invariance is imposed by the symmetry
$g_i = g_{K+1-i}$ and $F_i = F_{K+2-i}$.

Labeling the sites $i = 0, 1, \ldots, K+1$, the continuum limit of the model is  obtained 
by assigning a lattice spacing $a$
such that $z_i = i\,a$ with $L = (K+1)a$ fixed.    Writing
$\Sigma_i = \exp(i\,a\,A_z(z_i))$, and identifying
$g_i\,A_\mu^{(i)} \equiv A_\mu(z_i)$, the covariant derivative reconstructs the $({\mu z})$ components
of the 5D non-Abelian field strength.  Identifying $a\, f_i^2 \equiv f^2(z_i)$ and 
$a\, g_i^2\equiv g^2(z_i)$, in the continuous limit the full 5D non-Abelian action reads
\begin{equation}
\label{SSAction}
S_5 = \int d^4x\int_0^L dz\;
\mathrm{Tr}\Bigl[-\frac{1}{2g^2(z)}\,F_{\mu\nu}F^{\mu\nu}+ f^2(z)\,F_{\mu z}F^{\mu z}\Bigr].
\end{equation}
The two
boundaries at $z=0$ and $z=L$ implement
$\mathrm{SU}(N)_L\times\mathrm{SU}(N)_R$ as independent global
symmetries, broken to the diagonal $\mathrm{SU}(N)_V$  because the two boundaries are connected by a single bulk field, so that only
$z$-independent transformations, $g_L=g_R$, preserve the vacuum—the holographic realization of chiral symmetry breaking.

The $A_z$ component of the 5D gauge field plays the role of a Higgs field 
in unitary gauge (it has no potential and no vacuum expectation value; the
chiral symmetry is broken by the boundaries). To make this explicit, use the
$z$-dependent gauge freedom $g(x,z)$ with $\partial_z g=-i\,g A_z$ to set
$A_z(x,z)=0$; this absorbs $A_z$ into a Wilson line
\begin{equation}\label{WilsonLine}
W(x,z)=\mathcal{P}\exp\!\left(i\int_z^L A_z(x,z')\,dz'\right)\in\mathrm{SU}(N),
\end{equation}
whose value at $z=0$ is the chiral field $U(x)=W(x,0)$, which contains the
Goldstone bosons of spontaneous chiral symmetry breaking.

As we have seen, the expression \eqref{SSAction} is the typical action of hQCD models; moreover,  hQCD models like  WSS and HW2 realize the chiral field as a Wilson line as in Eq.~\eqref{WilsonLine}.

\section{Migdal's meromorphic approach to QCD and its connection to hQCD}

Pad\'e approximants are a powerful technique widely used in hadronic physics (for a comprehensive review, see, e.g., \citet{Masjuan:2007ay}. In the mid-1970s, \citet{Migdal:1977nu,Migdal:1977ut} proposed a systematic program to approximate QCD correlation functions at large $N_c$ by \emph{meromorphic} (rational, Pad\'e) functions whose poles are identified with physical hadron resonances. In particular, as summarized by \citet{Shifman:2005zn}, for the vector two-point function he asked how accurately the partonic $\log Q^2$ can be approximated by an infinite sum of infinitely narrow resonances and what  masses and decay constants such an optimal approximation implies. He found exponential accuracy at large Euclidean $Q^2$ when the resonance masses are proportional to the zeros of a Bessel function. As noted by \citet{Shifman:2005zn}, this is exactly what happens in the hard-wall (HW) hQCD model. Here we sketch the derivation of this result given by \citet{Erlich:2006hq} (see also \citet{Cata:2006ak,Falkowski:2006uy}).

Neglecting condensates, the UV behavior of the vector two-point function is $\Pi(q^2)\sim\log(-q^2/\mu^2)$, with Minkowskian momentum $q$. Setting $t=q^2/\mu^2$, its $[N,N]$ Pad\'e approximant $R_N(t)/Q_N(t)$ is defined by matching $\log t$ up to order $(t+1)^{2N+1}$ around the Euclidean point $t=-1$, i.e.\ $q^2=-\mu^2$, far from the branch point. Using complex analysis, one finds that the denominator is given in terms of Legendre polynomials,
\begin{equation}
Q_N(t)=(t+1)^N P_N\!\left(\frac{1-t}{1+t}\right),
\end{equation}
while the numerator follows from a dispersive integral over the discontinuity of the logarithm across its cut.

Migdal's crucial step is to go beyond the Pad\'e approximation by taking simultaneously the limits $N\to\infty$ and $q^2\ll\mu^2$. By the Mehler--Heine formula \citep{Szego:1939}, $P_N(1-z^2/2N^2)\to J_0(z)$, so that $Q_N(t)\to J_0(2N\sqrt{t})$, and the dispersive integral produces the second-kind Bessel function $Y_0$. The approximant then becomes
\begin{equation}
\Pi(q^2)\simeq \log\frac{q^2}{\mu^2}-\pi\,\frac{Y_0(2Nq/\mu)}{J_0(2Nq/\mu)} .
\end{equation}
This is Migdal's basic result: the logarithmic branch cut is replaced by an infinite tower of poles at the zeros of $J_0$, i.e.\ at the resonance masses
\begin{equation}
m_n=\frac{\mu}{2N}\,\gamma_{0,n},
\end{equation}
where $\gamma_{0,n}$ is the $n$-th zero of $J_0$.

In the double-scaling limit $\mu\to\infty$, $N\to\infty$ with $\mu/N$ fixed, a new infrared scale $\mu_{\rm IR}\sim\mu/N$ emerges. This scale is not predicted by the Pad\'e approximation, but is an external IR input. With the identification
\begin{equation}
\epsilon=\frac{1}{\mu},\qquad z_0=\frac{2N}{\mu},
\end{equation}
Migdal's result reproduces the HW expression of the vector correlator, with UV and IR cutoffs $\epsilon$ and $z_0$ respectively. Like $\mu_{\rm IR}$ in the Migdal limit, $z_0$ is an external input of the HW model.

Despite its elegance, Migdal's approach left several questions open. It gave no mechanism to distinguish the vector and axial-vector channels at low energies, nor to implement chiral symmetry breaking. It assumed a meromorphic structure without deriving it from a confinement mechanism. It was difficult to improve systematically beyond the two-point function and, providing no wave functions, it could not address form factors, electromagnetic radii or three-point functions.

The same period saw the development of the SVZ QCD sum rules \citep{Shifman:1978bx,Shifman:1978by}, which offered a systematic incorporation of the full OPE, including condensates and radiative corrections, and a well-defined approximation scheme suppressing both UV and IR uncertainties, allowing reliable error estimates of phenomenological quantities.

hQCD models, as well as the Son--Stephanov approach \citep{Son:2003et}, can be regarded as dynamical completions of Migdal's purely analytic ansatz, with the missing ingredients, namely confinement, chiral symmetry breaking and wave functions, provided by the 5D (or 4D deconstruction) setting.

\bibliographystyle{Harvard}
\bibliography{references}

\end{document}